\newif\ifpdf
\ifx\pdfoutput\undefined
\pdffalse 
\else
\pdfoutput=1 
\pdftrue \fi

\documentclass[cmp,final,10pt]{svjour-MB}
\usepackage{cite}
\usepackage{amsmath}
\usepackage{amsfonts}
\usepackage{amssymb}
\usepackage{epsf}
\usepackage{graphicx}
\usepackage{graphics}
\usepackage{verbatim}

\IfFileExists{myowntimes.sty}
	{\usepackage{myowntimes}}
	{\usepackage{times}\usepackage{mathrsfs}}

\DeclareFontFamily{OT1}{eusb}{} \DeclareFontShape{OT1}{eusb}{m}{n}
{<5> <6> <7> <8> <9> <10> <11> <12> <14.4> eusb10}{}
\DeclareMathAlphabet{\eusb}{OT1}{eusb}{m}{n}

\DeclareFontFamily{OT1}{eusm}{} \DeclareFontShape{OT1}{eusm}{m}{n}
{<5> <6> <7> <8> <9> <10> <11> <12> <14.4> eusm10}{}
\DeclareMathAlphabet{\eusm}{OT1}{eusm}{m}{n}

\DeclareFontFamily{OT1}{eufm}{} \DeclareFontShape{OT1}{eufm}{m}{n}
{<5> <6> <7> <8> <9> <10> <11> <12> <14.4> eufm10}{}
\DeclareMathAlphabet{\mathfrak}{OT1}{eufm}{m}{n}

\newenvironment{proofsect}[1]{\vskip0.1cm\noindent{\rmfamily\itshape #1.}}{\qed\vspace{0.15cm}}

\spnewtheorem*{Main Theorem}{Main Theorem}{\bf}{\it}
\spnewtheorem*{Key Estimate}{Key Estimate}{\bf}{\it}
\spnewtheorem*{Uniformity Property}{Uniformity Property}{\bf}{\it}

\spnewtheorem{mylemma}[theorem]{Lemma}{\bf}{\it}
\spnewtheorem{myproposition}[theorem]{Proposition}{\bf}{\it}
\spnewtheorem{mycorollary}[theorem]{Corollary}{\bf}{\it}

\numberwithin{equation}{section}
\numberwithin{theorem}{section}

\newcounter{obrazek}

\newcommand{\dist}{\operatorname{dist}}

\newcommand{\per}{{\text{\rm per}}}

\newcommand{\textd}{\text{\rm d}}

\newcommand{\Arg}{\operatorname{Arg}}

\newcommand{\QQ}{\mathcal Q}
\newcommand{\RR}{\mathcal R}

\newcommand{\C}{\mathbb C}
\newcommand{\D}{\mathbb D}

\newcommand{\BbbL}{\mathbb L}
\newcommand{\M}{\mathbb M}
\newcommand{\N}{\mathbb N}
\newcommand{\BbbO}{\mathbb O}

\newcommand{\R}{\mathbb R}

\newcommand{\Z}{\mathbb Z}

\newcommand{\abs}[1]{{\lvert #1\rvert}}

\newcommand{\RE}{\text{\rm Re}}
\newcommand{\IM}{\text{\rm Im}}
\renewcommand{\eqref}[1]{(\ref{#1})}
\newcommand{\twoeqref}[2]{(\ref{#1}--\ref{#2})}

\newcommand{\texti}{\text{\rm i}}

\newcommand{\zM}{z_{\text{\rm M}}}
\newcommand{\hate}{{\text{\rm \^e}}}
\newcommand{\LB}{\tilde L_0}
\newcommand{\hatv}{{\text{\rm \^v}}}
\newcommand{\hatw}{{\text{\rm \^w}}}
\newcommand{\frakz}{\mathfrak z}
\newcommand{\gammaL}{{\gamma_L}}
\newcommand{\barz}{\bar z}

\newcommand{\UU}{\mathscr{U}}
\newcommand{\CalS}{\mathscr{S}}
\newcommand{\GG}{\mathscr{G}}
\newcommand{\KK}{\mathscr{K}}
\newcommand{\OO}{\mathscr{O}}
\newcommand{\DD}{\mathscr{D}}
\renewcommand{\AA}{\mathscr{A}}
\newcommand{\CC}{\mathscr{C}}
\newcommand{\WW}{\mathscr{W}}
\newcommand{\PP}{\mathscr{P}}
\newcommand{\MM}{\mathscr{M}}

\begin{document}

\title{Partition function zeros at first-order phase transitions: A general analysis}
\titlerunning{Partition function zeros at first-order phase transitions}
\author{M.~Biskup\inst{1} \and C.~Borgs\inst{2} \and J.T.~Chayes\inst{2} \and L.J.~Kleinwaks\inst{3} \and R.~Koteck\'y\inst{4}}
\institute{Department of Mathematics, UCLA, Los Angeles CA 90095-1555, U.S.A.
\and Microsoft Research, One Microsoft Way, Redmond WA 98052, U.S.A.
\and Department of Physics, Princeton University, Princeton NJ 08544, U.S.A.
\and Center for Theoretical Study, Charles University, Prague 110 00, Czech Republic}
\authorrunning{Biskup et al}
\date{Received: April 3, 2003 / Accepted: March 31, 2004}
\communicated{M. Aizenman}
\maketitle
\renewcommand{\thefootnote}{}
\footnotetext{\hglue-5.3mm\copyright\,\,\,2003 M.~Biskup, C.~Borgs, J.T.~Chayes, L.J.~Kleinwaks, R.~Koteck\'y. Reproduction of the entire article for non-commercial purposes is permitted without charge.}
\renewcommand{\thefootnote}{\arabic{footnote}}
\begin{abstract}
We present a general, rigorous theory of partition function zeros for
lattice spin models depending on one complex parameter.
First, we formulate a set of natural assumptions which are verified
for a large class of spin models in a
companion~paper~\cite{BBCKK2}. Under these assumptions, we derive
equations whose solutions give the location of the zeros of the
partition function with periodic boundary conditions, up to an
error which
we prove is (generically) exponentially small in the linear
size of the system. For asymptotically large systems, the zeros
concentrate  on phase boundaries which are simple curves
ending in multiple points. For models with
an Ising-like plus-minus symmetry, we also establish a local
version of the Lee-Yang Circle Theorem. This result allows us to control situations when
in one region of the complex plane
the zeros lie precisely on the unit circle, while in
the complement of this region
the zeros concentrate on less symmetric~curves.
\end{abstract}

\setcounter{tocdepth}{3}
\tableofcontents

\section{Introduction}
\label{sec1}
\subsection{Motivation}
One of the cornerstones of equilibrium statistical mechanics is
the notion that macroscopic systems undergo phase transitions as
the external parameters change. A mathematical description of
phase transitions was given by Gibbs~\cite{Gibbs} who
characterized a phase transition as a point of non-analyticity in
thermodynamic functions, e.g., the pressure. This definition was
originally somewhat puzzling since actual physical systems are
finite, and therefore their thermodynamic functions are manifestly
real-analytic. A solution to this contradiction came in two
seminal papers by Yang and Lee~\cite{YL,LY}, where it was argued
that non-analyticities develop in physical quantities because, as
the system passes to the thermodynamic limit, complex
singularities of the pressure pinch the physical (i.e., real)
domain of the system parameters. Since the pressure is
proportional to the logarithm of the partition function, these
singularities correspond exactly to the zeros of the partition
function.

In their second paper~\cite{LY}, Lee and Yang demonstrated the
validity of their theory in a particular example of the Ising
model in a complex magnetic field~$h$. Using an induction
argument, they proved the celebrated Lee-Yang Circle Theorem which
states that, in this model, the complex-$e^h$ zeros of the
partition function on any finite graph with free boundary
conditions lie on the unit circle. The subject has been further
pursued by a number of authors in the following fifty years.
Generalizations  of the Lee-Yang theorem have been
developed~\cite{Ruelle,Newman,Lieb-Sokal,Nashimori-Griffiths} and
extensions to other complex parameters have been derived (for
instance, the Fisher zeros~\cite{Fisher} in the complex
temperature plane and  the zeros of the~$q$-state Potts model in
the complex-$q$ plane~\cite{Sokal1,Sokal2}). Numerous papers have
appeared studying the partition function zeros using various techniques
including
computer simulations~\cite{Chen-Hu-Wu,Kim-Creswick,Janke-Kenna},
approximate analyses~\cite{Kenna-Lang,Lee,Matveev-Shrock2} and
exact solutions of 1D and 2D lattice
systems~\cite{Glumac-Uzelac,Matveev-Shrock1,Lu-Wu,
Shrock-Tsai,Shrock,Chang-Shrock1,Chang-Shrock2,Dolan-Johnston}.
However, in spite of this progress, it seems fair to say that much
of the original Lee-Yang program---namely, to learn about the
transitions in physical systems by studying the zeros of partition
functions---had remained unfulfilled.

In \cite{BBCKK}, we outlined a general program, based on
Pirogov-Sinai theory~\cite{PSa,PSb,Zahradnik1,Borgs-Imbrie}, to
determine the partition function zeros for a large class of
lattice models depending on one complex parameter~$z$. The present
paper, and its companion~\cite{BBCKK2}, give the mathematical
details of that program. Our results apply to a host of systems
with first-order phase transitions; among others, they can be
applied to field-driven transitions in many low-temperature spin
systems as well as temperature-driven transitions---for instance,
the order-disorder transition in the~$q$-state Potts model with
large~$q$ or the confinement Higgs transition in lattice gauge
theories. We consider lattice models with a finite number of
equilibrium states that satisfy several general assumptions
(formulated in detail below). The validity of the assumptions
follows whenever a model can be analyzed using
 a convergent contour expansion based on Pirogov-Sinai theory,
even in the complex domain. In the present work, we
study only models with periodic boundary conditions,
although---with some technically involved modifications---our
techniques should allow us to treat also other boundary
conditions.

Under our  general assumptions, we derive a set of model-specific
equations; the solutions of these equations yield the locations of
the partition function zeros, up to rigorously controlled errors
which are typically exponentially small in the linear size of the
system. It turns out that, as the system size tends to infinity,
the partition function zeros concentrate on the union of a
countable number of simple smooth curves in the complex~$z$-plane.
Another outcome of our analysis is a local version of the Lee-Yang
Circle Theorem. Whereas the global theorem says that, for models
with the full Ising interaction, all partition function zeros lie
on the unit circle, our local theorem says that if the model has
an Ising-like symmetry in a restricted region of the
complex~$z$-plane, the corresponding portion of the zeros lies on
a piece of the unit circle. In particular, there are natural
examples (see the discussion of the Blume-Capel model in
\cite{BBCKK}) where only some of the partition function zeros lie
on the unit circle, and others lie on less symmetric curves. Our
proof  indicates that it is just the Ising plus-minus
symmetry (and a natural non-degeneracy condition) that makes the
Lee-Yang theorem true, which is a fact not entirely apparent in
the original derivations of this result.

In addition to being of interest for the foundations of
statistical mechanics, our results can often be useful on a
practical level---even when the parameters of the model are such
that we cannot rigorously verify all of our assumptions. We have
found that our equations seem to give accurate locations of
finite-volume partition function zeros for system sizes well beyond
what can be currently achieved using, e.g., computer-assisted
evaluations of these partition functions (see~\cite{BBCKK} for the
example of the three dimensional~$25$-state Potts model on~$1000$
sites). Our techniques are also capable of handling  situations
with more than one complex parameter in the system. However, the
actual analysis of the manifolds of partition function zeros may
be technically rather involved. Finally, we remark that, in one respect, our
program falls short of the ultimate goal of the original
Lee-Young program---namely, to describe the phase structure of
any statistical-mechanical
system directly on the basis of its partition function zeros.
Instead, we show that both the location of the partition function zeros
and the phase structure are consequences of an even
more fundamental property: the ability to represent the partition function as
a sum of terms corresponding to different metastable phases.
This representation is described in the next section.

\subsection{Basic ideas}
Here we will discuss the main ideas of our program, its
technical difficulties and our assumptions in more detail. We
consider spin models on~$\Z^d$, with~$d\ge2$, whose interaction
depends on a complex parameter~$z$. Our program is based on the
fact that, for a large class of such models, the partition
function~$Z_L^\per$ in a box of side~$L$ and with periodic
boundary conditions can be written as
\begin{equation}
\label{intro.1}
Z_L^\per(z) = \sum_{m=1}^r q_m e^{-f_m(z) L^d} +
O(e^{-\text{const}\,L}e^{-f(z) L^d}).
\end{equation}
Here~$q_1,\dots,q_r$ are positive integers describing the
degeneracies of the phases~$1,\dots,r$, the
quantities~$f_1,\dots,f_r$ are smooth (but not in general analytic) complex functions of the
parameter~$z$ which play the role of \emph{metastable free energies} of
the corresponding phases, and~$f(z)=\min_{1\le m\le r}\RE f_m(z)$.
The real version of the formula~\eqref{intro.1} was instrumental
for the theory of finite-size scaling near first-order phase
transitions~\cite{Borgs-Kotecky}; the original derivation goes
back to~\cite{Borgs-Imbrie}.

It follows immediately from  \eqref{intro.1} that, asymptotically
as~$L$ tends to infinity, $Z_L^\per = 0$ requires that~$\RE
f_m(z)=\RE f_{\widetilde m}(z)=f(z)$ for at least two distinct
indices~$m$ and~$\widetilde m$. (Indeed, otherwise the sum in
\eqref{intro.1} would be dominated by a single, non-vanishing
term.) Therefore, asymptotically, all zeros of~$Z_L^\per$
concentrate on the set
\begin{equation}
\label{intro.G}
\GG=\bigl\{z\colon
\text{ there exist }m\neq\widetilde m \text{ with }
\RE f_m(z)=\RE f_{\widetilde m}(z)=f(z)\bigr\}.
\end{equation}
Our first concern is the topological structure of~$\GG$. Let us
call a point where~$\RE f_m(z)=f(z)$ for at least three
different~$m$ a \emph{multiple point}; the points~$z\in\GG$ that
are not multiple points are called \emph{points of two-phase
coexistence}. Under suitable assumptions on the
functions~$f_1,\dots,f_r$, we show that~$\GG$ is a countable union
of non-intersecting simple smooth curves that begin and end at
multiple points. Moreover,  there are only a finite number of
multiple points inside any compact subset of~$\C$. See
Theorem~\ref{T:PD} for details.

The relative interior of each curve comprising~$\GG$ consists entirely of
the points of two-phase coexistence, i.e.,
we have~$\RE f_m(z)=\RE f_{\widetilde m}(z)=f(z)$ for exactly two
indices~$m$ and~$\widetilde m$. In particular, the sum in
\eqref{intro.1} is dominated by two terms. Supposing for a moment
that we can neglect all the remaining contributions, we would have
\begin{equation}
\label{intro.1a} Z_L^\per(z) = q_m e^{-f_m(z) L^d}+ q_{\widetilde m}
e^{-f_{\widetilde m}(z) L^d},
\end{equation}
and the zeros of~$Z_L^\per$ would be determined by the equations
\begin{equation}
\label{intro.1b}
\begin{aligned}
\RE f_m(z) &=\RE f_{\widetilde m}(z)+ L^{-d}\log
(q_m/q_{\widetilde m})
\\
\IM f_m(z)
&=\IM f_{\widetilde m}(z)+ (2\ell+1)\pi L^{-d},
\end{aligned}
\end{equation}
where~$\ell$ is an integer.  The presence of additional  terms of
course makes the actual zeros only approximate solutions to
\eqref{intro.1b}; the main technical problem is to give a
reasonable estimate of the distance between the solutions of
\eqref{intro.1b} and the zeros of~$Z_L^\per$. In a neighborhood of
multiple points, the situation is even more complicated because
there the equations \eqref{intro.1b} will not be even approximately
correct.

It turns out that the above heuristic argument cannot possibly be
converted into a rigorous proof without making serious adjustments
to the initial formula~\eqref{intro.1}. This is a consequence of subtle analytic
properties of the functions~$f_m$.  For typical physical systems,
the metastable free energy~$f_m$ is known to be analytic only in
the interior of the region
\begin{equation}
\label{1.5}
\CalS_m=\bigl\{z\colon\RE
f_m(z)=f(z)\bigr\}.
\end{equation}
On the boundary of~$\CalS_m$, one expects---and in some cases
proves \cite{I,FP}---the existence of essential singularities.
Thus \eqref{intro.1} describes an approximation of an analytic
function, the function~$Z_L^\per$, by a sum of non-analytic
functions, with singularities appearing precisely in the region
where we expect to find the zeros of~$Z_L^\per$! It is easy to
construct examples where an arbitrarily small non-analytic
perturbation of a complex polynomial \emph{with a degenerate zero}
produces extraneous roots. This would not be an issue along the
two-phase coexistence lines, where the roots of~$Z_L^\per$ turn
out to be non-degenerate, but we would not be able to say much
about the roots near the multiple points. In short, we need an
approximation that respects the analytic structure of our model.

\begin{figure}[t]
\refstepcounter{obrazek}
\ifpdf
\centerline{\includegraphics[width=0.9\textwidth]{decomposition.pdf}}
\else \centerline{\epsfxsize=0.9\textwidth\epsffile{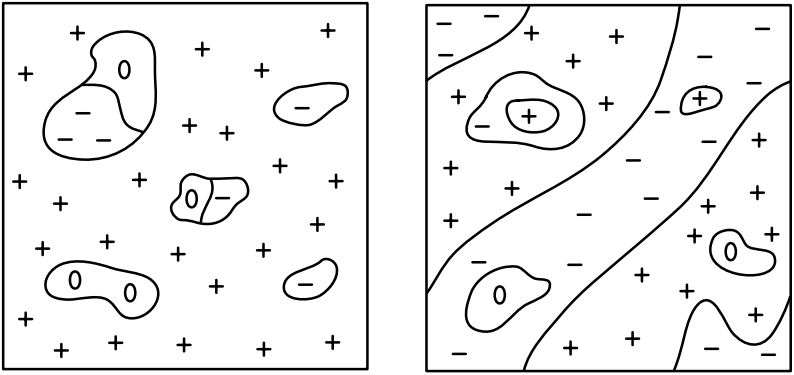}}
\fi
\bigskip
\label{fig1}
\caption{Schematic examples of configurations, along with their associated contours,
which contribute to different terms in the decomposition in
\eqref{intro.2}. Here we have a spin model with~$r=3$ equilibrium
phases denoted by~$+$,~$-$ and~$0$. The configuration on the left
has all contours smaller than the cutoff---which we set to~$L/3$
where~$L$ is the side of the box---and will thus contribute
to~$Z_+^{(L)}$ because~$+$ is the external phase for all external
contours. The configuration on the right has long contours and
will be assigned to the error term.
}
\end{figure}

Fortunately, we do not need to look far to get the desirable
analytic counterpart of \eqref{intro.1}. In fact, it suffices to
modify slightly  the derivation of the original formula. For the
benefit of the reader, we will recall the main steps of this
derivation: First we use a contour representation of the
model---the class of models we consider is characterized by the
property of having such a contour reformulation---to rewrite the
partition function as a sum over the collections of contours. Then
we divide the configurations contributing to~$Z_L^\per$ into~$r+1$
categories: Those in which all contours are of diameter smaller than,
say,~$L/3$ and in which the dominant phase is~$m$, where~$m=1,\dots,r$,
and those not falling into the preceding categories.
Let~$Z_m^{(L)}$ be the partial partition function obtained by
summing the contributions corresponding to the configurations in
the~$m$-th category, see Fig.~\ref{fig1}. It turns out that the
error term is still uniformly bounded as in \eqref{intro.1}, so we
have
\begin{equation}
\label{intro.2} Z_L^\per(z) = \sum_{m=1}^r Z_m^{(L)}(z) +
O(e^{-\text{const}\,L}e^{-f(z) L^d}),
\end{equation}
but now the functions~$Z_m^{(L)}(z)$ are analytic, and non-zero in a small
neighborhood of~$\CalS_m$. (However, the size of the neighborhood
shrinks with~$L\to\infty$, and one of the challenges of using the
formula \eqref{intro.2} is to cope with this restriction of
analyticity.) Moreover, writing
\begin{equation}
Z_m^{(L)}(z)=q_m e^{-f_m^{(L)}(z) L^d}
\end{equation}
and using the contour representation, the functions~$f_m^{(L)}$
can be expressed by means of convergent cluster
expansions~\cite{Kotecky-Preiss,Dobrushin}. In particular, they
can be shown to converge quickly to the functions~$f_m$
as~$L\to\infty$.

In this paper, we carry out the analysis of the partition
function zeros starting from the representation~\eqref{intro.2}.
In particular, we formulate minimal conditions (see Assumptions~A
and~B in Sect.~\ref{sec2}) on the functions~$f_m^{(L)}$ and the
error terms that allow us to analyze the roots of~$Z_L^\per$ in
great detail. The actual construction of the functions~$f_m^{(L)}$
and the proof that they satisfy the required conditions is
presented in \cite{BBCK1,BBCK2} for the~$q$-state Potts model with
one complex external field and~$q$ sufficiently large,  and in
\cite{BBCKK2} for a general class lattice models with finite
number of equilibrium states.

\subsection{Discussion of assumptions and results}
Here we will describe our main assumptions and indicate how they
feed into the proofs of our main theorems. For consistency with
the previous sections, we will keep using the functions~$f_m$
and~${f_m^{(L)}}$ even though the assumptions will actually be
stated in terms of the associated exponential variables
\begin{equation}
\zeta_m(z)=e^{-f_m(z)}
\quad\text{and}\quad
\zeta_m^{(L)}(z)=e^{-f_m^{(L)}(z)}.
\end{equation}
The first set of assumptions (Assumption~A, see Sect.~\ref{sec2.1}) 
concerns the
infinite-volume quantities~$f_m$, and is important for the
description of the set of coexistence points~$\GG$. The
functions~$f_m$ are taken to be twice differentiable in
the variables~$x=\RE z$ and~$y=\IM z$, and analytic in the
interior of the set~$\CalS_m$. If, in addition,~$f(z)=\min_m\RE
f_m$ is uniformly bounded from above, good control of the
two-phase coexistence curves is obtained by assuming that, for any
distinct~$m$ and~$\widetilde m$, the difference of the first derivatives
of~$f_m$ and~$f_{\widetilde m}$ is uniformly bounded from below
on~$\CalS_m\cap\CalS_{\widetilde m}$. Finally, in order to discuss
multiple coexistence points, we need an additional non-degeneracy
assumption on the derivatives of the functions~$f_m$ for the
coexisting phases. Given these assumptions, we are able to give a
very precise characterization of the topology of the coexistence
set~$\GG$, see Theorem~\ref{T:PD}.

The second set of assumptions (Assumption~B, see Sect.~\ref{sec2.2}) is crucial for our
results on the partition function zeros, and is formulated in
terms of the functions~${f_m^{(L)}}$. These will be taken to be
analytic with a uniform upper bound on the first~$r$ derivatives
in an order-$(1/L)$ neighborhood of the sets~$\CalS_m$.  In this
neighborhood,~${f_m^{(L)}}$ is also assumed to be exponentially
close to~$f_m$, with a lower bound on the difference of the first
derivatives for any pair~${f_m^{(L)}}$ and~${f_{\widetilde
m}^{(L)}}$ in the intersection of the corresponding order-$(1/L)$
neighborhoods. Finally, we need a bound on the error term and its
derivatives in an approximation of the form \eqref{intro.2} where
the sum runs only over the dominating terms, i.e., those~$m$ for
which~$z$ lies in the order-$(1/L)$ neighborhood of~$\CalS_m$.

Combining Assumptions~A and~B, we are able to prove several
statements on the location of the partition function zeros.
We will start by covering the set of available $z$-values by
sets with a given number of stable (or
``almost stable'')
phases. The covering involves three scale functions, $\omega_L$,
$\gamma_L$ and $\rho_L$
which give rise to three classes of
sets: the region where one phase is decisively dominating the others
(more precisely, the complement of
an~$L^{-d}\omega_L$-neighborhood of the set~$\GG$),
a~$\gamma_L$-neighborhood of sets
with two stable phases,
excluding a~$\gamma_L$-neighborhood of multiple points, and
the~$\rho_L$-
neighborhoods of multiple points.
As is shown in
Proposition~\ref{prop2.6},
for a suitable choice of
sequences~$\omega_L$,~$\gamma_L$,
and~$\rho_L$, these three sets cover
all~possibilities.

In each part of the cover, we will control the zeros by a
different method. The results of our analysis can be summarized as
follows: First, there are no zeros of~$Z_L^\per$ outside an
$L^{-d}\omega_L$-neighborhood of the set~$\GG$.  This claim,
together with a statement on the maximal possible degeneracy of
zeros, is the content of Theorem~\ref{T:deg}. The next theorem,
Theorem~\ref{T:2ph}, states that in a $\gamma_L$-neighborhood of
the two-phase coexistence points, excluding a neighborhood of
multiple  points, the zeros of~$Z_L^\per$ are exponentially close
to the solutions of \eqref{intro.1b}.  In particular, this implies
that the zeros are spaced in intervals of order-$L^{-d}$ along the
two-phase coexistence curves with the asymptotic density expressed
in terms of the difference of the derivatives of the corresponding
free energies---a result known in a special case already to Yang
and Lee~\cite{YL}; see Proposition~\ref{cor2.1}. The control of
the zeros in the vicinity of multiple points is more difficult and
the results are less detailed. Specifically, in the
$\rho_L$-neighborhood of a multiple point with~$q$ coexisting
phases, the zeros of~$Z_L^\per$ are shown to be located within
a~$L^{-d-d/q}$ neighborhood of the solutions of an explicitly
specified equation.

\smallskip
We finish our discussion with a remark concerning the positions of zeros of complex functions of the form:
\begin{equation}
Z_N(z)=\sum_{m=1}^r\alpha_m(z)\zeta_m(z)^N,
\end{equation}
where~$\alpha_1,\dots,\alpha_r$ and~$\zeta_1,\dots,\zeta_r$ are analytic functions of~$z$. Here there is a  general theorem, due to Beraha, Kahane and Weiss~\cite{BKW} (generalized recently by Sokal~\cite{Sokal2}), that the set of zeros of~$Z_N$ asymptotically concentrates on the set of~$z$ such that either~$\alpha_m(z)=0$ and~$|\zeta_m(z)|=\max_k|\zeta_k(z)|$ for some~$m=1,\dots,r$ or~$|\zeta_m(z)|=|\zeta_n(z)|=\max_k|\zeta_k(z)|$ for two distinct indices~$m$ and~$n$. The present paper provides a substantial extension of this result to situations when analyticity of~$\zeta_m(z)$ can be guaranteed only in a shrinking neighborhood of the sets where~$m$ is the ``dominant'' index. In addition, we also provide detailed control of the rate of convergence.

\section{Main results}
\label{sec2}

\subsection{Complex phase diagram}
\label{sec2.1}
We begin by abstracting the assumptions on the meta\-stable free
energies of the contour model and showing what kind of complex
phase diagram they can yield. Throughout the paper, we will assume
that a domain~$\OO\subset\C$ and a positive integer~$r$ are given,
and use~$\RR$ to denote the set~$\RR=\{1,\dots,r\}$. For each~$z\in\OO$,
we let~$x=\RE z$ and~$y=\IM z$ and define, as usual,
\begin{equation}
\label{2.1}
\textstyle
\partial_z=
\frac12\bigl(\frac{\partial}{\partial x}
-\texti \frac{\partial}{\partial y}\bigr)
\quad\text{and}\quad
\partial_{\bar z}=
\frac12\bigl(\frac{\partial}{\partial x}
+\texti \frac{\partial}{\partial y}\bigr).
\end{equation}

\medskip\noindent
\textbf{Assumption~A.\ }
There exists a constant~$\alpha>0$ and, for
each~$m\in\RR$, a function $\zeta_m\colon\OO\to\C$, such that the
following conditions are satisfied:
\settowidth{\leftmargini}{(11)}
\begin{enumerate}
\item[(1)] The quantity~$\zeta(z)=\max_{m\in\RR}\abs{\zeta_m(z)}$
is uniformly positive in~$\OO$, i.e., we have $\inf_{z\in\OO}\zeta(z)>0$.
\item[(2)] Each function~$\zeta_m$, viewed as a function of two
real variables~$x=\RE z$ and~$y=\IM z$, is twice
continuously differentiable on~$\OO$ and it satisfies the
Cauchy-Riemann  equations~$\partial_{\bar z}\zeta_m(z)=0$ for
all~$z\in\CalS_m$, where
\begin{equation}
\label{betterSm}
\CalS_m=\bigl\{z\in\OO\colon |\zeta_m(z)|=\zeta(z)\bigr\}.
\end{equation}
In particular,~$\zeta_m$ is analytic on the interior of~$\CalS_m$.
\item[(3)] For any pair of distinct indices~$m,n\in\RR$ and
any~$z\in\CalS_m\cap \CalS_n$ we have
\begin{equation}
\label{nondeg} \biggr|\frac{\partial_z\zeta_m(z)}{\zeta_m(z)}-
\frac{\partial_z\zeta_n(z)}{\zeta_n(z)} \biggl|\ge\alpha.
\end{equation}
\item[(4)] If~$\QQ\subset \RR$ is such that~$|\QQ|\ge 3$, then for
any~$z\in\bigcap_{m\in\QQ}\CalS_m$,
\begin{equation}
\label{tripledeg}
v_m(z)=\frac{\partial_z\zeta_m(z)}{\zeta_m(z)},\quad m\in\QQ,
\end{equation}
are the vertices of a strictly convex polygon in~$\C\simeq\R^2$.
\end{enumerate}

\begin{remark}
In~(1), we assumed uniform positivity in order to simplify some of
our later arguments. However, uniformity in~$\OO$ can easily be replaced by
uniformity on compact sets.
Note that Assumptions~A3--4 are invariant with respect to
conformal transformations of~$\OO$ because the functions involved
in \eqref{nondeg} and \eqref{tripledeg} satisfy the Cauchy-Riemann
conditions.
Also note that, by Assumption~A3, the length of each side of the polygon
from Assumption~A4 is at least~$\alpha$;~cf~Fig.~\ref{fig3}.
\end{remark}

The indices $m\in\RR$ will be often referred to as \emph{phases}.
We call a phase~$m$ \emph{stable} at~$z$ if~$z\in\CalS_m$, i.e.,
if~$|\zeta_m(z)|=\zeta(z)$. For each~$z\in\OO$ we define
\begin{equation}
\label{2.4}
\QQ(z)=\bigl\{m\in\RR\colon|\zeta_m(z)|=\zeta(z)\bigr\}
\end{equation}
to be the set of phases \emph{stable at~$z$}. If~$m,n\in\QQ(z)$,
then we say that the phases~$m$ and~$n$ \emph{coexist at~$z$}. The
phase diagram is determined by the \emph{set of coexistence
points}:
\begin{equation}
\label{GG}
\GG=\bigcup_{m,n\in\RR\colon m\neq n}\GG(m,n)
\quad\text{ with }\quad \GG(m,n)=\CalS_m\cap\CalS_n.
\end{equation}
If~$|\zeta_m(z)|=\zeta(z)$ for at least three distinct~$m\in\RR$,
we call such~$z\in\OO$ a \emph{multiple point}.

In the following, the phrase \emph{simple arc} denotes the image of~$(0,1)$
under a continuous and injective map while \emph{simple closed curve} denotes
a corresponding image of the unit circle $\{z\in\C\colon|z|=1\}$. 
A curve will be called \emph{smooth} if it can be parametrized using
twice continuously differentiable functions.

\smallskip
Our main result concerning the topology of~$\GG$ is then as follows.

\begin{theorem}
\label{T:PD}
Suppose that Assumption~A holds and let~$\DD\subset\OO$ be a
compact set. Then there exists a finite set
of open discs $\D_1,\D_2,\dots,\D_\ell\subset\OO$
covering~$\DD$, such that
for each~$k=1,\dots,\ell$, the set~$\AA_k=\GG\cap\D_k$
satisfies exactly one of the following properties:
\settowidth{\leftmargini}{(111)}
\begin{enumerate}
\item[(1)]~$\AA_k=\emptyset$.
\item[(2)]~$\AA_k$ is a smooth simple arc with both
endpoints on~$\partial \D_k$. Exactly two distinct
phases coexist along the arc constituting~$\AA_k$.
\item[(3)]~$\AA_k$ contains a single multiple
point~$z_k$ with~$s_k=|\QQ(z_k)|\ge3$ coexisting
phases, and~$\AA_k\setminus\{z_k\}$ is a collection of~$s_k$ 
smooth, non-intersecting, simple arcs connecting~$z_k$
to~$\partial \D_k$. Each pair of distinct
curves from~$\AA_k\setminus\{z_k\}$ intersects at a positive angle
at~$z_k$. Exactly two distinct phases coexist along each component
of~$\AA_k\setminus\{z_k\}$.
\end{enumerate}
In particular,~$\GG=\bigcup_{\CC\in\eusb C}\CC$, where~$\eusb C$
is a finite or countably-infinite collection of smooth simple closed curves
and simple arcs which intersect each other only at the endpoints.
\end{theorem}

Theorem~\ref{T:PD} is proved in Sect.~\ref{sec3.2}. Further
discussion is provided in Sect.~\ref{sec2.4}.

\subsection{Partition function zeros}
\label{sec2.2}
\begin{figure}[t]
\refstepcounter{obrazek}
\ifpdf
\centerline{\includegraphics[width=0.9\textwidth]{diag.pdf}}
\else \centerline{\epsfxsize=0.9\textwidth\epsffile{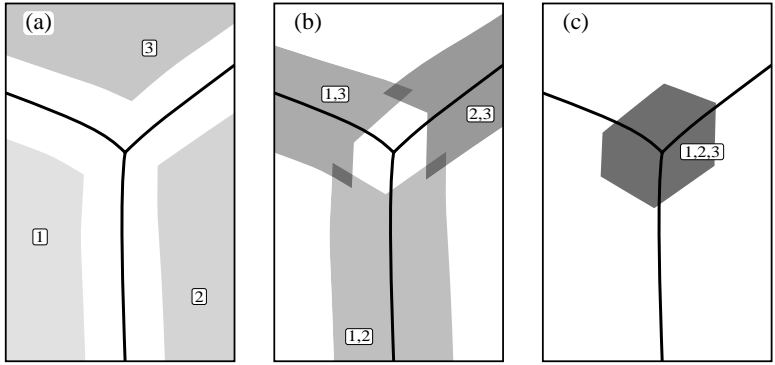}}
\fi
\bigskip
\label{fig2}
\caption{An illustration of the
sets $\UU_\epsilon(\QQ)$ in the vicinity of a
multiple point. The thick lines indicate the visible portion of
the set of coexistence points~$\GG$. Three phases, here labeled
$1$, $2$ and $3$, are stable at the multiple point. In~(a), the
three shaded domains represent the sets $\UU_\epsilon(\{1\})$,
$\UU_\epsilon(\{2\})$ and $\UU_\epsilon(\{3\})$, with the label
indicated by the number in the box. Similarly, in~(b) the three
regions represent the sets $\UU_\epsilon(\{1,2\})$,
$\UU_\epsilon(\{2,3\})$ and $\UU_\epsilon(\{1,3\})$. Finally,~(c)
contains only one shaded region, representing the
set~$\UU_\epsilon(\{1,2,3\})$. The various
regions~$\UU_\epsilon(\QQ)$ generously overlap so that their union
covers the entire box.
}
\end{figure}

\noindent
Next we will discuss our assumptions and results concerning the
zeros of the partition function. We assume that the
functions~$Z_L^\per\colon\OO \to \C$, playing the role of the
partition function in a box of side~$L$ with periodic boundary
conditions, are defined for each integer~$L$, or, more generally,
for any~$L\in \BbbL$, where~$\BbbL\subset \N$ is a fixed infinite
set. Given any~$m\in\RR$ and~$\epsilon>0$, we
use~$\CalS_\epsilon(m)$ to denote the region where the phase~$m$
is ``almost stable,''
\begin{equation}
\CalS_\epsilon(m)= \bigl\{z\in\OO\colon \abs{\zeta_m(z)}>
e^{-\epsilon} \zeta(z)\bigr\}.
\end{equation}
For any~$\QQ\subset\RR$, we also introduce the region where all
phases from~$\QQ$ are ``almost stable'' while the remaining ones
are not,
\begin{equation}
\UU_\epsilon(\QQ)=\bigcap_{m\in\QQ} \CalS_\epsilon(m)\setminus
\bigcup_{n\in\QQ^{\text{\rm c}}}\overline{\CalS_{\epsilon/2}(n)},
\label{2.7}
\end{equation}
with the bar denoting the set closure. Notice that the
function~$\zeta_m$ is non-vanishing on~$\CalS_\epsilon(m)$ and
that~$\bigcup_{\QQ\subset\RR}\UU_\epsilon(\QQ)=\OO$,
see~Fig.~\ref{fig2}. Note also that
$\UU_\epsilon(\emptyset)=\emptyset$, so we may assume that
$\QQ\ne\emptyset$ for the rest of this paper.

\medskip\noindent
\textbf{Assumption~B.\ } There exist constants
$\kappa,\tau\in(0,\infty)$ and, for each $m\in\RR$, a positive
integer~$q_m$ and a
function~$\zeta_m^{(L)}\colon\CalS_{\kappa/L}(m)\to\C$ such that
for any~$L\in\BbbL$
the following is true:
\settowidth{\leftmargini}{(11)}
\begin{enumerate}
\item[(1)]
The function~$Z_L^\per$ is analytic in~$\OO$.
\item[(2)]
Each~$\zeta_m^{(L)}$ is non-vanishing and analytic
in~$\CalS_{\kappa/L}(m)$. Furthermore,
\begin{equation}
\label{zetamL}
\biggl|\log\frac{\zeta_m^{(L)}(z)}{\zeta_m(z)}\biggr| \le
e^{-\tau L}
\end{equation}
and
\begin{equation}
\label{der-zetamL}
\biggl|\partial_z\log\frac{\zeta_m^{(L)}(z)}{\zeta_m(z)}\biggr| +
\biggl|\partial_{\bar z}\log\frac{\zeta_m^{(L)}(z)}{\zeta_m(z)}\biggr|
\le e^{-\tau L}
\end{equation}
for all~$m\in\RR$ and all~$z\in\CalS_{\kappa/L}(m)$. (Here
``$\log$'' denotes the principal branch of the complex logarithm.)
\item[(3)]
There exist constants~$\tilde\alpha>0$,~$M<\infty$
and~$\LB<\infty$ such that for any~$L\ge \LB$ we have
\begin{equation}
\label{uplogderL}
\biggl|\frac{\partial^\ell_z\zeta_m^{(L)}(z)}{\zeta_m^{(L)}(z)}
\biggr|\le M,
\end{equation}
whenever~$m\in\RR$,~$\ell=1,\dots,r$, and $z\in\CalS_{\kappa/L}(m)$.
In addition,
\begin{equation}
\label{nondegL}
\biggr|\frac{\partial_z\zeta_m^{(L)}(z)}{\zeta_m^{(L)}(z)}-
\frac{\partial_z\zeta_n^{(L)}(z)}{\zeta_n^{(L)}(z)}
\biggl|\ge{\tilde \alpha}
\end{equation}
whenever~$m,n\in\RR$ are distinct and~$z\in\CalS_{\kappa/L}(m)\cap
\CalS_{\kappa/L}(n)$.
\item[(4)]
There exist constants~$C_\ell<\infty$,~$\ell=0,1,\dots,r+1$,
such that for any~$\QQ\subset\RR$, the difference
\begin{equation}
\label{ZLper} \varXi_{\QQ,L}(z)=Z_L^\per(z)-\sum_{m\in\QQ} q_m
\bigl[\zeta_m^{(L)}(z)\bigr]^{L^d}
\end{equation}
satisfies the bound
\begin{equation}
\label{error}
\Bigl|\partial_z^\ell \varXi_{\QQ,L}(z)\Bigr|\le C_\ell
L^{d(\ell+1)}\zeta(z)^{L^d}\biggl(\sum_{m\in\RR}q_m\biggr)
e^{-\tau L} ,
\end{equation}
for all~$\ell=0,1,\dots,r+1$, uniformly in~$z\in\UU_{\kappa/L}(\QQ)$.
\end{enumerate}

\begin{remark}
\label{rem3}
In applications,~$q_m$ will represent the degeneracy
of the phase~$m$; thus we have taken it to be a positive
integer. However, our arguments would go through even if
we assumed only that all~$q_m$'s are real and positive.
It is also worth noting that in many physical models 
the partition function is not directly
of the form required by Assumption~B; but it can be brought into
this form by extracting a multiplicative ``fudge'' factor~$F(z)^{L^d}$,
where~$F(z)\ne0$ in the region of interest. For instance, in the
Ising model with~$z$ related to the complex external field~$h$ by~$z=e^h$ 
we will have to take~$F(z)=z^{-1/2}$ to make the partition function
analytic in the neighborhood of~$z=0$.
\end{remark}

Our first theorem in this section states that the zeros
of~$Z_L^\per(z)$ are concentrated in a narrow strip along the
phase boundaries. In addition, their maximal degeneracy near the
multiple points of the phase diagram can be evaluated. In accord
with the standard terminology, we will call a point~$z_0$ a
\emph{$k$-times degenerate root} of an analytic function~$h(z)$
if~$h(z)=g(z)(z-z_0)^k$ for some~$g(z)$ that is finite and
non-zero in a neighborhood of~$z_0$. Recalling the definition
\eqref{2.7} of the set~$\UU_\epsilon(\QQ)$, we introduce
the shorthand
\begin{equation}
\label{UU_epsilon}
\GG_\epsilon=\bigcup_{m\neq n}\Big(\overline{\CalS_{\epsilon/2}(n)}
\cap\overline{\CalS_{\epsilon/2}(m)}\Big)=
\OO\setminus\bigcup_{m\in\RR} \UU_\epsilon\bigl(\{m\}\bigr).
\end{equation}
An easy way to check the
second equality in \eqref{UU_epsilon} is by noting that
$\OO\setminus\UU_\epsilon(\{m\})$ can be written as the union
$\bigcup_{n:n\neq m}\overline{\CalS_{\epsilon/2}(n)}$. Then we
have the following result.

\begin{theorem}
\label{T:deg}
Suppose that Assumptions~A1-3 and~B hold and let~$\kappa>0$ be as
in Assumption~B. Let~$(\omega_L)$ be a sequence of positive
numbers such that~$\omega_L\to\infty$. Then there exists
a constant~$L_0<\infty$
such that for~$L\ge L_0$ all roots of~$Z_L^\per$ lie
in~$\GG_{L^{-d}\omega_L}$ and are at most~$|\RR|-1$ times degenerate.
For each~$\QQ\subset\RR$, the roots of~$Z_L^\per$
in~$\UU_{\kappa/L}(\QQ)$ are at most~$|\QQ|-1$ times
degenerate.
\end{theorem}

In other words, as $L\to\infty$, the zeros of $Z_L^\per$
asymptotically concentrate on the set of coexistence points~$\GG$.
Notice that we explicitly do \emph{not} require Assumption~A4 to
hold; see Sect.~\ref{sec2.4} for further discussion.
Theorem~\ref{T:deg} is proved in Sect.~\ref{sec4.1}.

\smallskip
Our next theorem deals with the zeros of~$Z_L^\per$ in the regions
where at most two phases from~$\RR$ are ``almost stable.'' It
turns out that we have a much better control on the location of
zeros in regions that are sufficiently far from multiple points.
To quantify the meaning of ``sufficiently far,'' we let $\gammaL$
be a sequence of positive numbers (to be specified below) and, for
any $\QQ\subset\RR$ with $|\QQ|=2$ and any $L\ge0$,
let~$\delta_L\colon\UU_\gammaL(\QQ)\to(0,\infty)$ be a function
defined by
\begin{equation}
\label{deltaLz}
\delta_L(z)=\begin{cases}
e^{-\tau L},\qquad&\text{if }
z\in\UU_\gammaL(\QQ)\cap\UU_{2\kappa/L}(\QQ),
\\
L^de^{-\frac 12\gammaL L^d},\qquad&\text{otherwise}.
\end{cases}
\end{equation}
(Clearly, $\delta_L(z)$ depends on the index set~$\QQ$. However,
this set will always be clear from the context and so we will not
make it notationally explicit.)
Finally, given $\epsilon>0$ and $z\in\OO$, let
$\D_\epsilon(z)$ denote the open disc of radius~$\epsilon$
centered at $z$.

\smallskip
The exact control of the roots in two-phase regions is then as follows.

\begin{theorem}
\label{T:2ph} Suppose that Assumptions~A and~B hold and
let~$\Omega_L^\star$ be the set of all zeros of the
function~$Z_L^\per(z)$ in~$\OO$, including multiplicity.
If~$m,n\in\RR$ are distinct indices, let~$\QQ=\{m,n\}$, and
let~$\Omega_L(\QQ)$ be the set of the solutions of the system of
equations
\begin{align}
\label{Req}
{}&q_m^{1/L^d}|\zeta_m(z)|=q_n^{1/L^d}|\zeta_n(z)|,
\\
\label{Imq} {}&L^d\Arg \bigl(\zeta_m(z)/\zeta_n(z)\bigr)
=\pi\operatorname{ mod }2\pi.
\end{align}
Let $\gammaL$ be such that
\begin{equation}
\label{gLass}
\liminf_{L\to\infty}\frac{L^d\gammaL}{\log L}>4d
\quad\text{and}\quad
\limsup_{L\to\infty}L^{d-1}\gammaL<2\tau,
\end{equation}
and let $\delta_L\colon\UU_\gammaL(\QQ)\to(0,\infty)$ be as defined in
\eqref{deltaLz}.
Then there exist finite positive constants $B$, $C$, $D$, and~$L_0$
such that for any~$\QQ\subset\RR$ with~$|\QQ|=2$
and any $L\ge L_0$ we~have:
\settowidth{\leftmargini}{(111)}
\begin{enumerate}
\item[(1)]
For all~$z\in\GG\cap\UU_\gammaL(\QQ)$ with
$\D_{DL^{-d}}(z)\subset\OO$,
the disc~$\D_{DL^{-d}}(z)$ contains at least one root
from~$\Omega_L^\star$.
\item[(2)]
For all~$z\in\Omega_L^\star\cap\UU_\gammaL(\QQ)$ with
$\D_{C\delta_L(z)}(z)\subset\OO$,
the disc~$\D_{C\delta_L(z)}(z)$ contains exactly one point
from~$\Omega_L(\QQ)$.
\item[(3)]
For all~$z\in\Omega_L(\QQ)\cap\UU_\gammaL(\QQ)$ with
$\D_{C\delta_L(z)}(z)\subset\OO$,
the disc~$\D_{C\delta_L(z)}(z)$ contains exactly one root
from~$\Omega_L^\star$.
\item[(4)]
Any two distinct
roots of~$Z_L^\per$
in the set $\{z\in\UU_\gammaL(\QQ)\colon\D_{BL^{-d}}(z)\subset\OO\}$
are at least $BL^{-d}$
apart.
\end{enumerate}
\end{theorem}

\smallskip
Note that the first limit in \eqref{gLass} ensures
that $L^d\delta_L(z)\to 0$ as
$L\to\infty$ throughout $\UU_\gammaL(\QQ)$ (for any
$\QQ\subset\RR$ with $|\QQ|=2$). Thus $\delta_L(z)$
is much smaller than the distance of
the ``neighboring'' roots of \twoeqref{Req}{Imq}.
Theorem~\ref{T:2ph} is proved in Sect.~\ref{sec4.2}.

\smallskip
Theorem~\ref{T:2ph}
allows us to describe the asymptotic density of the roots
of~$Z_L^\per$
along the arcs of the complex phase diagram. Let~$m,n\in\RR$ be
distinct and
let~$\GG(m,n)$ be as in \eqref{GG}.
For each~$\epsilon>0$ and each~$z\in\GG(m,n)$,
let~$\rho_{m,n}^{(L,\epsilon)}(z)$ be defined by
\begin{equation}
\label{adens} \rho_{m,n}^{(L,\epsilon)}(z)=\frac1{2\epsilon
L^d}\bigl|\Omega_L^\star\cap\D_\epsilon(z)\bigr|,
\end{equation}
where~$|\Omega_L^\star\cap\D_\epsilon(z)|$ is the number of roots
of~$Z_L^\per$ in~$\D_\epsilon(z)$ including multiplicity.
Since~$\GG(m,n)$ is a union of simple arcs and closed curves, and
since the roots of (\ref{Req}-\ref{Imq}) are spaced within $O(L^{-d})$
from each other,~$\rho_{m,n}^{(L,\epsilon)}(z)$ has the
natural interpretation of the approximate \emph{line density of
zeros} of~$Z_L^\per$ along~$\GG(m,n)$. As can be expected from
Theorem~\ref{T:2ph}, the approximate
density~$\rho_{m,n}^{(L,\epsilon)}(z)$ tends to an explicitly
computable limit.

\begin{myproposition}
\label{cor2.1}
Let~$m,n\in\RR$ be distinct and let~$\rho_{m,n}^{(L,\epsilon)}(z)$ be
as in \eqref{adens}.
Then the limit
\begin{equation}
\rho_{m,n}(z)=\lim_{\epsilon\downarrow0}\lim_{L\to\infty}
\rho_{m,n}^{(L,\epsilon)}(z)
\end{equation}
exists for all~$z\in\GG(m,n)$ such that~$|\QQ(z)|=2$, and
\begin{equation}
\label{dens}
\rho_{m,n}(z)=\frac1{2\pi}\Biggl|
\frac{\partial_z\zeta_m(z)}{\zeta_m(z)}
-\frac{\partial_z\zeta_n(z)}{\zeta_n(z)}\Biggr|.
\end{equation}
\end{myproposition}

\begin{remark}
Note that, on the basis of Assumption~A3, we have that
$\rho_{m,n}(z)\ge\alpha/(2\pi)$.
In particular, the density of zeros is always positive.
This is directly related to the fact that all points $z\in\GG$
will exhibit a first-order phase transition (defined in
an appropriate sense, once $\IM z\ne0$ or $\RE z<0$)---hence the
title of the paper.
The observation that the (positive) density of zeros and the order
of the transition are closely related goes back to~\cite{YL}.
\end{remark}

In order to complete the description of the roots of~$Z_L^\per$,
we also need to cover the regions with more than two ``almost
stable'' phases. This is done in the following theorem.

\begin{theorem}
\label{T:Mph}
Suppose that Assumptions~A and~B are satisfied. Let~$\zM$
be a multiple point and let~$\QQ=\QQ(\zM)$ with~$q=|\QQ|\ge 3$.
For each~$m\in\QQ$, let
\begin{equation}
\label{phi-v}
\phi_m(L)=L^d\Arg \zeta_m(\zM)\;(\operatorname{ mod
}2\pi)\quad \text{ and } \quad
v_m=\frac{\partial_z\zeta_m(\zM)}{\zeta_m(\zM)}.
\end{equation}
Consider the set~$\Omega_L(\QQ)$ of all solutions of the
equation
\begin{equation}
\label{Meq}
\sum_{m\in\QQ} q_m\,e^{\texti\phi_m(L)+L^d(z-\zM)v_m}=0,
\end{equation}
including multiplicity,
and let~$(\rho_L)$ be a sequence of positive numbers such that
\begin{equation}
\label{rholim}
\lim_{L\to\infty}L^d\rho_L=\infty\qquad\text{but}\qquad
\lim_{L\to\infty}L^{d-d/(2q)}\rho_L=0.
\end{equation}
Define $\rho_L'=\rho_L+L^{-d(1+1/q)}$. Then there exists a
constant~$L_0<\infty$ and, for any~$L\ge L_0$, an open, connected
and simply connected set~$\UU$ satisfying
$\D_{\rho_L}(\zM)\subset\UU\subset\D_{\rho_L'}(\zM)$ such that the
zeros in $\Omega\cap\UU$ are in one-to-one correspondence with the
solutions in $\Omega(\QQ)\cap\UU$ and the corresponding points are
not farther apart than~$L^{- d(1+1/q)}$.
\end{theorem}

Theorem~\ref{T:Mph} is proved in Sect.~\ref{sec4.3}.
Sect.~\ref{sec2.4} contains a discussion of the role of
Assumption~A4 in this theorem; some information will also be provided
concerning the actual form of the solutions of \eqref{Meq}.

\smallskip
To finish the exposition of our results, we will need to show that
the results of Theorems~\ref{T:deg},~\ref{T:2ph} and~\ref{T:Mph}
can be patched together to provide
complete control of
the roots of~$Z_L^\per$, at least in any compact subset of~$\OO$. This
is done in the following claim,
the proof of which
essentially relies
only on Assumption~A and compactness arguments:

\begin{myproposition}
\label{prop2.6}
Suppose that Assumption~A  holds and let
$\omega_L$, $\gamma_L$ and $\rho_L$ be sequences of positive
numbers such that $\omega_L\le \gamma_L L^d$, $\gamma_L\to 0$,
and~$\rho_L\to 0$.
 For each compact set $\DD\subset\OO$, there exist constants
$\chi=\chi(\DD)>0$ and $L_0=L_0(\DD)<\infty$ such that, if
$\rho_L\ge\chi\gamma_L$, we have
\begin{equation}
\label{inkluze}
\GG_{L^{-d}\omega_L}\cap\DD\subset
\bigcup_{\begin{subarray}{c}
\QQ\subset\RR\\|\QQ|=2
\end{subarray}}
\UU_\gammaL(\QQ)
\cup
\bigcup_{\begin{subarray}{c}
\zM\in\DD\\|\QQ(\zM)|\ge3
\end{subarray}}
\D_{\rho_L}(\zM)
\end{equation}
for any $L\ge L_0$.
\end{myproposition}

Note that in \eqref{inkluze} we consider only that portion of~$\DD$
in~$\GG_{L^{-d}\omega_L}$, since by Theorem~\ref{T:deg} the roots
of $Z_L^\per$ are contained in this set.  Note also that the
conditions we impose on the sequences $\omega_L$, $\gamma_L$ and
$\rho_L$ in Theorems~\ref{T:PD}, \ref{T:2ph} and \ref{T:Mph} and
Proposition~\ref{prop2.6} are not very restrictive. In particular,
it is very easy to verify the existence of these sequences.
(For example, one can take both~$\gamma_L$ and~$\rho_L$ to be
proportional to $L^{- d}\log L$ with suitable prefactors and then
let $\omega_L=L^d\gamma_L$.)

\subsection{Local Lee-Yang theorem}
As our last result, we state a generalized version of the classic
Lee-Yang Circle Theorem~\cite{LY}, the proof of which is based entirely
on the exact symmetries of the model.

\begin{theorem}
\label{T:locLY}
Suppose that Assumptions~A and~B hold. Let~$+$ and~$-$ be
two selected indices from~$\RR$ and let
$\UU$ be an open set with compact closure
$\DD\subset \OO$ such that $\UU\cap\{z\colon |z|=1\}\neq\emptyset$.
Assume that $\DD$ is invariant under circle inversion~$z\mapsto 1/z^*$,~and
\settowidth{\leftmargini}{(111)}
\begin{enumerate}
\item[(1)]~$Z_L^\per(z)=Z_L^\per(1/z^*)^*$,
\item[(2)]~$\zeta_+(z)=\zeta_-(1/z^*)^*$ and~$q_+=q_-$
\end{enumerate}
hold for all~$z\in\DD$ and all~$L\in\BbbL$.
Then there exists a constant~$L_0$ such that
the following holds for all $L\geq L_0$:
If the intersection of~$\DD$
with the set of coexistence points~$\GG$ is connected
and if $+$ and $-$ are the only stable phases in~$\DD$,
then all
zeros in~$\DD$ lie on the unit circle, and the number of
zeros on any segment of $\DD\cap\{z\colon |z|=1\}$
is proportional to~$L^d$ as $L\to\infty$.
\end{theorem}

Condition (2) is the rigorous formulation of the statement
that the~$+$ and~$-$ phases are related by~$z\leftrightarrow 1/z^*$
(or~$h\leftrightarrow-h$, when~$z=e^h$) symmetry. Condition~(1)
then stipulates that this 
symmetry is actually respected by
the remaining phases and, in particular, by~$Z_L^\per$ itself.

\begin{remark}
As discussed in Remark~\ref{rem3}, in order to satisfy Assumption~B it may 
be necessary to extract a multiplicative ``fudge'' factor from the partition function,
perform the analysis of partition function zeros in various restricted regions in~$\C$ 
and patch the results appropriately. A similar manipulation may be required in order to apply Theorem~\ref{T:locLY}.
\end{remark}

Here are the main steps of the proof of Theorem~\ref{T:locLY}:
First we show that
the phase diagram in~$\DD$ falls  exactly on the unit circle,
i.e.,
\begin{equation}
\label{nakruhu}
\DD\cap\GG=\{z\in\DD\colon |z|=1\}.
\end{equation}
This fact is essentially an immediate  consequence of the symmetry
between ``$+$'' and ``$-$.'' \emph{A priori} one would then expect
that the zeros are close to, but not necessarily on, the unit
circle. However, the symmetry of $Z_L^{\per}$
combined with
the fact that distinct zeros are at least $BL^{-d}$ apart
is not compatible with the existence of zeros away
from the unit circle.
Indeed, if $z$ is a root of $Z_L^\per$,
it is bound to be within
a distance
$O(e^{-\tau L})$
of the unit circle.
If, in addition, $|z|\ne1$, then the $z\leftrightarrow 1/z^*$ symmetry
implies that $1/z^*$ is also a root of $Z_L^\per$, again within
$O(e^{-\tau L})$
of the unit circle. But then the distance
between~$z$ and~$1/z^*$ is
of the order $e^{-\tau L}$ which is forbidden by
claim (4) of Theorem~\ref{T:2ph}.

\smallskip
This argument is made precise in the following proof.

\begin{proofsect}{Proof of Theorem~\ref{T:locLY}}
We start with the proof of~\eqref{nakruhu}.
Let us suppose that $\DD\subset\OO$
and ${\QQ(z)}\subset\{+,-\}$ for all $z\in\DD$.
Invoking the continuity of $\zeta_\pm$
and condition~(2) above, we have
${\QQ(z)}=\{+,-\}$ for all
$z\in\DD\cap\{z\colon|z|=1\}$ and thus
${\DD}\cap\{z\colon|z|=1\}\subset\GG$.
Assume now that $\GG\cap\DD\setminus\{z\colon|z|=1\}\neq
\emptyset$. By
the fact that $\GG\cap\DD$ is connected and the assumption that
$\UU\cap\{z\colon |z|=1\}\ne\emptyset$, we can find
a path $z_t\in\GG\cap\DD$, $t\in [-1,1]$, such that
$z_t\in\DD\cap\{z\colon |z|=1\}$ if $t\leq 0$ and $z_t\in
\GG\cap\DD\setminus\{z\colon|z|=1\}$ if $t>0$.  Since
$\QQ(z_0)=\{+,-\}$, we know that there is a disc
$\D_\epsilon(z_0)\subset \OO$ that contains no multiple points.
Applying Theorem~\ref{T:PD} to this disc, we conclude that there
is an open disc $\D$ with $z_0\in\D\subset\D_\epsilon(z_0)$, such
that $\GG\cap\D$ is a simple curve which ends at $\partial\D$.
However, using condition (2) above, we note that
as with~$z_t$, also the curve $t\mapsto 1/z_t^*$ lies in $\GG\cap\DD$,
contradicting the fact that $\GG\cap\D$ is a simple curve. This
completes the proof of~\eqref{nakruhu}.

Next, we will show that for any $z_0\in\DD\cap\{z\colon|z|=1\}$,
and any $\delta>0$, there exists an open disc
$\D_\epsilon(z_0)\subset\OO$ such that
the set $\GG\cap\D_\epsilon(z_0)$ is a smooth curve with the property
that for any $z\in\D_\epsilon(z_0)$ with $|z|\neq 1$, the line
connecting $z$ and $1/z^*$
intersects the curve $\GG\cap\D_\epsilon(z_0)$ exactly once,
and at an angle that
lies between $\pi/2-\delta$ and $\pi/2+\delta$.  If $z_0$ lies in
the interior of~$\DD$, this statement (with $\delta=0$)
follows trivially from~\eqref{nakruhu}.
If~$z_0$ is a boundary point
of~$\DD$, we first choose a sufficiently small
disc
$\D\ni z_0$ so that
$\D\subset\OO$ and, for all
points in~$\D$,
only the phases~$+$ and~$-$ are stable.  Then we
use Theorem~\ref{T:2ph} and
\eqref{nakruhu}
to infer that $\epsilon$ can be chosen small enough to
guarantee the above statement about intersection angles.

Furthermore, we claim
that given $z_0\in\DD\cap\{z\colon|z|=1\}$
and  $\epsilon>0$ such that
$\D_{3\epsilon}(z_0)\subset\OO$ and
$\QQ(z)\subset\{+,-\}$ for all $z\in\D_{3\epsilon}(z_0)$,
one can choose~$L$ sufficiently large so
that
\begin{equation}
\label{LY-cond}
\D_{2\epsilon}(z_0)\cap\GG_{L^{-d}\omega_L}
\subset\UU_\gammaL(\{+,-\})\cap\UU_{2\kappa/L}(\{+,-\}).
\end{equation}
To
prove this,
let us first note that,
for $\gamma_L\le2\kappa/L$, the right
hand side can be rewritten~as
\begin{equation}
\UU_\gammaL(\{+,-\})
\setminus\bigcup_{m\neq-,+}\overline{\CalS_{\kappa/L}(m)}.
\end{equation}
Next, by the compactness of $\overline{\D_{2\epsilon}(z_0)}$
and the fact that no~$m\in\RR$ different
from~$\pm$ is stable anywhere in~${\D_{3\epsilon}}
(z_0)$, we can
choose~$L_0$  so large that
$\overline{\CalS_{\kappa/L}(m)}
\cap\overline{\D_{2\epsilon}(z_0)}=\emptyset$ for
all $L\ge L_0$ and all~$m\ne\pm$.  Using the closure of
$\D_{2\epsilon}(z_0)$
in place of the set~$\DD$
in \eqref{inkluze}, we get \eqref{LY-cond}.

We are now ready to prove that for any
$z_0\in\DD\cap\{z\colon|z|=1\}$,
there exist constants $\epsilon>0$ and~$L_0$ such that all roots
of $Z_L^\per$in $\D_\epsilon(z_0)\cap\DD$ lie on the unit circle.
To this end, let us first assume that $\epsilon$ has been chosen
small enough to guarantee that
$(1-\epsilon)^{-1}<1+2\epsilon$,
$\D_{3\epsilon}(z_0)\subset\OO$, $\QQ(z)\subset\{+,-\}$ for all
$z\in\D_{3\epsilon}(z_0)$, and
$\GG\cap\D_{3\epsilon}(z_0)$ is a smooth curve with the above
property about the intersections angles, with, say, $\delta=\pi/4$.
Assume further that~$L$ is chosen so
that \eqref{LY-cond} holds and
$\epsilon>\max(C\delta_L(z_0), BL^{-d})$,
where~$C$ and~$B$ are the constants from
Theorem~\ref{T:2ph}.

Let~$z\in\D_\epsilon(z_0)\cap\DD$ be a root of~$Z_L^\per$.
If~$L$ is so large that Theorem~\ref{T:deg} applies, we have
$z\in\GG_{L^{-d}\omega_L}$ and thus $\delta_L(z)=e^{-\tau L}$
in view of \eqref{LY-cond}.  By Theorem~\ref{T:2ph},
there exists a solution~$\tilde z$ to \twoeqref{Req}{Imq} that lies in a
$C\delta_L(z)$-neighborhood of~$z$, implying that
$z$ has distance less than $C\delta_L(z)$ from
$\D_{2\epsilon}(z_0)\cap\GG$.
(Here we need that $q_+=q_-$ to conclude that $\tilde z\in\GG$.)
Suppose now that $|z|\ne1$. Then the condition~(1) above implies
that~$z'=(z^*)^{-1}$ is a \emph{distinct} root of~$Z_L^\per$
in~$\DD$. Moreover, if $\epsilon$ is so small that
$(1-\epsilon)^{-1}<1+2\epsilon$, then
$z'\in\GG_{L^{-d}\omega_L}\cap\D_{2\epsilon}(z_0)$ and
$\delta_L(z')$ also equals~$e^{-\tau L}$,
implying that $z'$ has distance less than
$C\delta_L(z)$ from
$\D_{3\epsilon}(z_0)\cap\GG$.  Since both $z$ and
$z'$ have distance less than $C\delta_L(z)$ from
$\D_{3\epsilon}(z_0)\cap\GG$, and the curve
$\D_{3\epsilon}(z_0)\cap\GG$ intersects the line
through~$z$ and~$z'$ in an angle that is near~$\pi/2$,
we conclude that $|z-z'|\leq 2\sqrt 2 Ce^{-\tau L}$ which for~$L$
sufficiently large contradicts
the last claim of Theorem~\ref{T:2ph}. Hence,~$z$ must have been
on the unit circle after all.

The rest of the argument is based on compactness. The set
$\DD\cap\{z\colon|z|=1\}$ is compact, and can thus be
covered by a finite number of such discs. Picking one such cover,
let~$\DD'$ be the complement of these disc in~$\DD$. Then the
set~$\DD'$ is a finite distance away from~$\GG$ and thus
$\DD'\cap\GG_{L^{-d}\omega_L}=\emptyset$ for~$L$ sufficiently
large. From here it follows that for some finite $L_0<\infty$
(which has to exceed the maximum of the corresponding quantity for
the discs
that constitute the covering
of $\DD\cap\{z\colon|z|=1\}$), all roots of~$Z_L^\per$
in~$\UU$ lie on the unit circle.
\end{proofsect}

\subsection{Discussion}
\label{sec2.4}\noindent
We finish with a brief discussion of the
results stated in the previous three sections. We will also
mention the role of (and possible exceptions to) our assumptions,
as well as extensions to more general situations.

We begin with the results on the complex phase diagram.
Theorem~\ref{T:PD} describes the situation in the generic cases
when Assumptions~A1-A4 hold. We note that Assumption~A3 is crucial
for the fact that the set~$\GG$ is a collection of \emph{curves}.
A consequence of this is also that the zeros of~$Z_L^\per$
asymptotically concentrate on curves---exceptions to this ``rule''
are known, see, e.g.,~\cite{Saarloos-Kurtze}. Assumption~A4 prevents
the phase coexistence curves from merging in a tangential fashion and,
as a result of that, guarantees that multiple points do not proliferate
throughout~$\OO$. Unfortunately, in several models of interest
(e.g., the Potts and Blume-Capel model) Assumption~A4 happens to
be violated at some~$\tilde z$ for one or two ``critical'' values
of the model parameters. In such cases, the region~$\OO$ has to be
restricted to the complement of some neighborhood of~$\tilde z$
and, inside the neighborhood, the claim has to be verified using a
refined and often model-specific analysis. (It often suffices to
show that the phase coexistence curves meeting at $\tilde z$ have
different curvatures, which amounts to a statement about the
second derivatives of the functions~$\log\zeta_m(z)$.) Examples of
such analysis have appeared in~\cite{BBCKK} for the Blume-Capel
model and in~\cite{BBCK2} for the Potts model in a complex external field.

Next we will look at the results of Theorems~\ref{T:deg}
and~\ref{T:2ph}. The fact that the roots of~$Z_L^\per$ are only
finitely degenerate is again independent of Assumption~A4. (This
is of some relevance in
view of the aforementioned exceptions
to this assumption.) The fact that, in the cases when all~$q_m$'s
are the same, the zeros shift only by an
exponentially small amount away from the two-phase coexistence
lines is a direct consequence of our choice of the boundary
conditions. Indeed, the factor~$e^{-\tau L}$ in \eqref{deltaLz}
can be traced to the similar factors in \eqref{zetamL} and
\eqref{error}. For strong (e.g., fixed-spin) boundary conditions,
we expect the corresponding terms in \eqref{zetamL} and
\eqref{error} to be replaced by~$1/L$. In particular, in these
cases, the lateral shift of the partition function zeros away from
the phase-coexistence lines should be of the order~$1/L$.
See~\cite{Zahradnik2} for some results on this problem.

\begin{figure}[t]
\refstepcounter{obrazek}
\ifpdf
\centerline{\includegraphics[width=0.75\textwidth]{kite.pdf}}
\else
\centerline{\epsfxsize=0.75\textwidth\epsffile{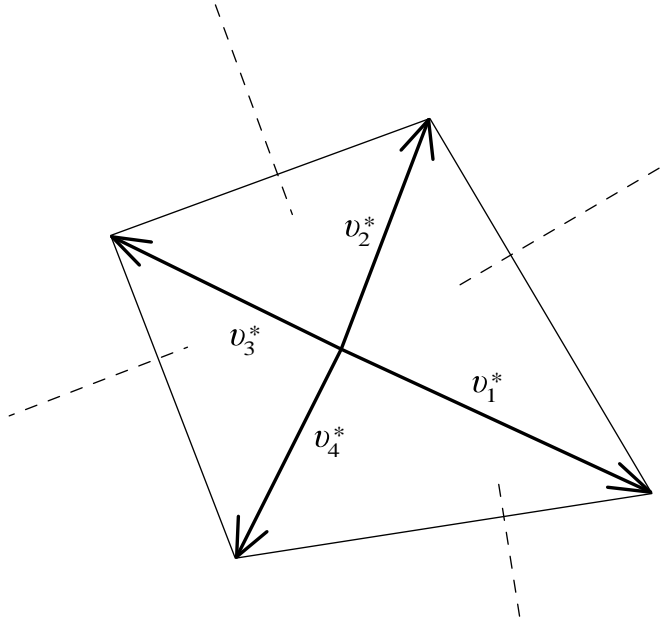}}
\fi
\bigskip
\label{fig3}
\caption{An illustration of the situation around a quadruple point.
Here~$v_1^*,\dots,v_4^*$ are the complex conjugates of the
quantities from \eqref{tripledeg} and $q_1=q_2=q_3<q_4$. (The
quadruple point lies at the common tail point of
the vectors~$v_1^*,\dots,v_4^*$.) The dashed lines indicate the
asymptotes of the ``strings'' of zeros sufficiently far---on the
scale~$L^{-d}$---from the quadruple point. Note the lateral shift
of these lines due to the fact that $q_4>q_1,q_3$. The picture
seems to suggest that, on the scale~$L^{-d}$, the quadruple point
splits into two~triple~points.
}
\end{figure}

Finally, let us examine the situation around multiple points in
some detail. Theorem~\ref{T:Mph} can be given the following
geometrical interpretation: Let~$\zM$ be a multiple point.
Introducing the parametrization~$\frakz=(z-\zM)L^d$, we
effectively zoom in on the scale~$L^{-d}$, where the zeros
of~$Z_L^\per$ are well approximated by the roots of the linearized
problem \eqref{Meq} with~$\QQ=\QQ(\zM)$. Let us plot the complex
conjugates~$v_m^*$ of the logarithmic derivatives $v_m$ (see
\eqref{phi-v}),~$m\in\QQ$, as vectors in~$\R^2$. By Assumption~A4,
the vectors~$v_m^*$ are the endpoints of a convex set
in~$\C\simeq\R^2$. Let~$v_1^*,\dots,v_q^*$ be the ordering of~$\QQ$
in the counterclockwise direction, see Fig.~\ref{fig3}.
Noting that the real part~$\RE(v_m\frakz)$ can be written in terms
of the dot product,
$\RE(v_m\frakz)=v_m^*\cdot \frakz$,
\eqref{Meq} can be recast as
\begin{equation}
\label{Meq1}
\sum_{m\in\QQ(\zM)} q_m\,e^{\texti\phi'_m(L)+v_m^*\cdot \frakz}=0,
\end{equation}
where~$\phi'_m(L)=\phi_m(L)+\IM(v_m\frakz)$.

On the basis of \eqref{Meq1}, it is easy to verify the following
facts: Let $\frakz=|\frakz|\hate$, with $\hate$ a unit vector in
$\C$. An inspection of \eqref{Meq1} shows that, for
$|\frakz|\gg1$, the roots of \eqref{Meq1} will concentrate along
the ``directions'' for which the projection of $\hate$ on at least
two $v_n^*$'s is the same. Invoking the convexity assumption
(Assumption~A4), this can only happen when $v_n^*\cdot\hate =
v_{n+1}^*\cdot\hate$ for some~$n$. In such cases, the
contributions of the terms with indices $m\ne n,n+1$ in
\eqref{Meq1} are negligible---at least once $|\frakz|\gg1$---and
the zeros will thus asymptotically lie along the half-lines given
in the parametric form by
\begin{equation}
\frakz=\frakz(t)=\frac{v_n^*-v_{n+1}^*}{|v_n-v_{n+1}|^2}\log\Bigl(
\frac{q_{n+1}}{q_n}\Bigr)+\texti t(v_n^*-v_{n+1}^*),\quad
t\in[0,\infty).
\end{equation}
Clearly, the latter is a line perpendicular to the~$(n,n+1)$-st
side of the convex set with vertices~$v_1^*,\dots,v_q^*$, which is
shifted (away from the origin) along the corresponding side by a
factor proportional to~$\log(q_{n+1}/q_n)$, see Fig.~\ref{fig3}.

Sufficiently far away from~$\zM$ (on the scale~$L^{-d}$), the
zeros resume the pattern established around the two-phase
coexistence curves. In particular, the zeros are asymptotically
equally spaced but their overall shift along the asymptote is
determined by the factor~$\phi_m(L)$---which we note depends very
sensitively on~$L$. Computer simulations show that, at least in
generic cases, this pattern will persists all the way down to the
multiple point. Thus, even
on the ``microscopic'' level, the zeros seem to form a ``phase
diagram.'' However, due to the lateral shifts caused by
$q_{m+1}\ne q_m$,
a ``macroscopic'' quadruple point may resolve into
two ``microscopic'' triple points, and similarly for higher-order multiple points.

\section{Characterization of phase diagrams}
\noindent
The  goal of this section is to give the proof
of Theorem~\ref{T:PD}. We begin by proving a series of auxiliary
lemmas
whose purpose is to elevate the pointwise Assumptions~A3-A4 into
statements extending over a small neighborhood of each coexistence
point.

\subsection{Auxiliary claims}
\noindent
Recall the definitions of $\CalS_m$, $\QQ(z)$ and $v_m(z)$, in
\eqref{betterSm}, \eqref{2.4} and \eqref{phi-v}, respectively. The
first lemma gives a limiting characterization of stability of
phases around coexistence points.

\begin{mylemma}
\label{lemma3.1} Let Assumption~A1--A2 hold and let $\barz\in\OO$
be such that $|\QQ(\barz)|\ge2$. Let $(z_k)$ be a sequence of
numbers $z_k\in\OO$ such that $z_k\to\barz$ but $z_k\ne\barz$
for~all~$k$. Suppose~that
\begin{equation}
\label{eitheta}
e^{\texti\theta}=\lim_{k\to\infty} \frac{z_k-\barz}{|z_k-\barz|}
\end{equation}
exists and let $m\in\QQ(\barz)$. If $z_k\in\CalS_m$ for
infinitely many $k\ge1$, then
\begin{equation}
\label{ineq}
\RE(e^{\texti\theta}v_m)\ge\RE(e^{\texti\theta}v_n)
\quad\text{ for all }n\in\QQ(\barz),
\end{equation}
where
$v_n=v_n(\barz)$.
Conversely, if the inequality in \eqref{ineq} fails for at least
one $n\in\QQ(\barz)$, then there is an $\epsilon>0$ such that
\begin{equation}
\label{w-neighb}
\WW_{\epsilon,\theta}(\barz)=
\Bigl\{z\in\OO\colon|z-\barz|<\epsilon,\, z\neq\barz,\,
\bigl|\textstyle\frac{z-\barz}{|z-\barz|}
-e^{\texti\theta}\bigr|<\epsilon\Bigr\}
\end{equation}
has empty intersection with $\CalS_m$, i.e.,
$\CalS_m\cap\WW_{\epsilon,\theta}(\barz)=\emptyset$. In
particular, $z_k\not\in\CalS_m$ for~$k$ large~enough.
\end{mylemma}

\begin{remark}
\label{rem3a} 
In the following, it will be
useful to recall some simple facts about complex functions.
Let~$f$,~$g$ and~$h$ be functions~$\C\to\C$ and let~$\partial_z$
and~$\partial_{\bar z}$ be as in \eqref{2.1}. If $f$ satisfies
$\partial_{\bar z}f(z_0)=0$ (i.e., Cauchy-Riemann conditions),
then all directional derivatives of~$f$ at $z_0=x_0+\texti y_0$ can be
expressed using one complex number $A=\partial_z f(x_0+\texti y_0)$,
i.e., we have
\begin{equation}
f(x_0+\epsilon\cos\varphi+\texti y_0+\texti \epsilon\sin\varphi)
-f(x_0+\texti y_0)=\epsilon Ae^{\texti\varphi}+o(\epsilon),
\quad\epsilon\downarrow0,
\end{equation}
holds for every $\varphi\in[-\pi,\pi)$. Moreover, if~$g$ is
differentiable with respect to~$x$ and~$y$ at $z_0=x_0+\texti y_0$
and~$h$ satisfies $\partial_{\bar z}h(z')=0$ at $z'=g(z_0)$, then
the chain rule holds for $z\mapsto h(g(z))$ at $z=z_0$. In
particular, $\partial_z h(g(z_0))=(\partial_z
h)(g(z_0))\partial_zg(z_0)$.
\end{remark}

\begin{proofsect}{Proof of Lemma~\ref{lemma3.1}}
Let $m\in\QQ(\barz)$ be fixed.
Whenever $z_k\in\CalS_m$, we have
\begin{equation}
\label{coexrel}
\log\bigl|\zeta_m(z_k)\bigr|-\log\bigl|\zeta_m(\barz)\bigr|
\ge\log\bigl|\zeta_n(z_k)\bigr|-\log\bigl|\zeta_n(\barz)\bigr|,
\qquad n\in\QQ(\barz),
\end{equation}
because $|\zeta_m(\barz)|=|\zeta_n(\barz)|$, by our assumption
that $m,n\in\QQ(\barz)$. Using the notation
\begin{equation}\label{Fdef}
F_{m,n}(z)=\frac{\zeta_m(z)}{\zeta_n(z)}
\end{equation}
for $n\in\QQ(\barz)$ (which is
well defined and non-zero in a neighborhood of $\barz$),
the inequality \eqref{coexrel} becomes
\begin{equation}
\label{Fcoexrel}
\log\bigl|F_{m,n}(z_k)\bigr|-\log\bigl|F_{m,n}(\barz)\bigr|
\ge 0, \qquad n\in\QQ(\barz).
\end{equation}
Note that the complex derivative $\partial_z F_{m,n}(\barz)$ exists
for all $n\in\QQ(\barz)$. Our task is then to prove that
\begin{equation}
\label{Fineq}
\RE\Bigl(e^{\texti\theta}\,\frac{\partial_{\barz}
F_{m,n}(\barz)}{F_{m,n}(\barz)}\Bigr)\geq 0,
\qquad n\in\QQ(\barz).
\end{equation}
Fix $n\in\QQ(\barz)$. Viewing $z\mapsto F_{m,n}(z)$ as a function of
two real variables $x=\RE z$ and $y=\IM z$, we can expand $\log
\abs{F_{m,n}(z)}$ into a Taylor series around the point $\barz$ to
get
\begin{equation}
\label{Tay}
\log \bigl|F_{m,n}(z_k)\bigr|-\log \bigl|F_{m,n}(\barz)\bigr|
= \RE\biggl((z_k-\barz)
\frac{\partial_z F_{m,n}(\barz)}{F_{m,n}(\barz)}\biggr)
+O(\abs{z_k-\barz}^2).
\end{equation}
To derive \eqref{Tay} we recalled that $F_{m,n}$ is at least twice
continuously differentiable (hence the error bound) and then
applied the identity
\begin{equation}
\label{other}
\frac{\partial \log \abs{F_{m,n}(\barz)}}{\partial x}\,
\Delta x_k
+ \frac{\partial \log \abs{F_{m,n}(\barz)}}{\partial y}\,
\Delta y_k
=\RE\biggl((z_k-\barz)
\frac{\partial_z F_{m,n}(\barz)}{F_{m,n}(\barz)}\biggr),
\end{equation}
where $\Delta x_k=\RE (z_k-\barz)$ and
$\Delta y_k=\IM(z_k-\barz)$.
(To derive \eqref{other}, we just have to apply the chain
rule to the functions $z\mapsto\log F_{m,n}(z)$.
See Remark~\ref{rem3a}
for a discussion of this point.)
Using that $z_k\to\barz$, the
inequality \eqref{Fineq} and hence also \eqref{ineq} now follows
by combining \eqref{Tay} with \eqref{coexrel}, dividing by
$|z_k-\barz|$ and taking the limit $k\to\infty$.

If, on the contrary, the inequality \eqref{ineq} is violated for
some $n\in\QQ(\barz)$, then \eqref{Fineq} fails to hold as well
and hence \eqref{Fcoexrel} and \eqref{coexrel}, with $z_k$
replaced by $z$, must be wrong for
$z\in\WW_{\epsilon,\theta}(\barz)$ whenever~$\epsilon$
is small enough. But $m\in\QQ(\barz)$
implies that $|\zeta_m(\barz)|=|\zeta_n(\barz)|$ and thus
$|\zeta_m(z)|<|\zeta_n(z)|$ for all
$z\in\WW_{\epsilon,\theta}(\barz)$, proving that
$\CalS_m\cap\WW_{\epsilon,\theta}(\barz)=\emptyset$. By
\eqref{eitheta} and the fact that $z_k\to\barz$, we have
$z_k\in\WW_{\epsilon,\theta}(\barz)$ and hence
$z_k\not\in\CalS_m$ for all $k$ large enough.
\end{proofsect}

Lemma~\ref{lemma3.1} directly implies the following corollary.

\begin{mycorollary}
\label{cor3.1}
Let Assumption~A1--A2 hold and let $m,n\in\RR$ be
distinct. Let $(z_k)$ be a sequence of numbers
$z_k\in\CalS_m\cap\CalS_n$ such that $z_k\to\barz\in\OO$ but
$z_k\ne\barz$ for all $k$. Suppose that the
limit \eqref{eitheta}
exists and equals $e^{\texti\theta}$. Then
$\RE(e^{\texti\theta}v_m)=\RE(e^{\texti\theta}v_n)$.
\end{mycorollary}

\begin{proofsect}{Proof}
Follows immediately  applying \eqref{ineq} twice.
\end{proofsect}

The next lemma will ensure that multiple points do not cluster and
that the coexistence lines always intersect at positive angles.

\begin{mylemma}
\label{lemma3.2}
Suppose that Assumption~A holds and let $\barz\in\OO$.
Suppose there are two sequences $(z_k)$ and $(z_k')$ of
numbers from~$\OO$ such that $|z_k-\barz|=|z_k'-\barz|\ne 0$
for all~$k$ and $z_k,z_k'\to\barz$ as~$k\to\infty$.
Let $a,b,c\in\RR$ and suppose that $z_k\in\CalS_a\cap\CalS_b$ and $z_k'\in\CalS_a\cap\CalS_c$
for all $k$. Suppose the limit \eqref{eitheta} exists for both
sequences and let~$e^{\texti\theta}$ and~$e^{\texti\theta'}$ be the
corresponding limiting values.
\settowidth{\leftmargini}{(111)}
\begin{enumerate}
\item[(1)]
If $a,b,c$ are distinct, then
$e^{\texti\theta}\ne e^{\texti\theta'}$.
\item[(2)]
If $a\ne b=c$ and $z_k\ne z_k'$ for
infinitely many $k$, then $|\QQ(\barz)|=2$ and
$e^{\texti\theta}=-e^{\texti\theta'}$.
\end{enumerate}
\end{mylemma}

\begin{remark}
The conclusions of part~(2) have a very natural interpretation.
Indeed, in this case,  $\barz$ is a point on a two-phase coexistence
line (whose existence we have not established yet) and~$z_k$
and~$z_k'$ are the (eventually unique) intersections of this line
with a circle of radius $|z_k-\barz|=|z_k'-\barz|$ around $\barz$.
As the radius of this circle decreases, the intersections~$z_k$
and~$z_k'$ approach~$\barz$ from ``opposite'' sides, which
explains why we should expect to have $e^{\texti\theta}=-e^{\texti\theta'}$.
\end{remark}

\begin{proofsect}{Proof of Lemma~\ref{lemma3.2}}
Throughout the proof, we set $v_m=v_m(\barz)$.
We begin by proving~(1). Assume that $a,b,c\in\RR$ are distinct
and suppose that $e^{\texti\theta}=e^{\texti\theta'}$. Note that, since
$\QQ(\barz)\supset\{a,b,c\}$, the point $\barz$ is a multiple point.
Corollary~\ref{cor3.1} then implies that
\begin{equation}
\RE(e^{\texti\theta}v_a)=\RE(e^{\texti\theta}v_b)=\RE(e^{\texti\theta}v_c),
\end{equation}
and hence $v_a$, $v_b$ and $v_c$ lie on a straight line in $\C$.
But then $v_a$, $v_b$ and $v_c$ cannot simultaneously be vertices
of a strictly convex polygon, in contradiction with Assumption~A4.

In order to prove part (2), let $a\ne b=c$, suppose without loss
of generality that $z_k\ne z_k'$ for all $k$. If
$e^{\texti\theta}\ne\pm e^{\texti\theta'}$, then Corollary~\ref{cor3.1}
implies that
$\RE(e^{\texti\theta}(v_a-v_b))=0=\RE(e^{\texti\theta'}(v_a-v_b))$ and hence
$v_a=v_b$, in contradiction with Assumption~A3.
Next we will rule out the possibility that
$e^{\texti\theta}=e^{\texti\theta'}$, regardless of how many phases are
stable at $\barz$. Let $G(z)=\zeta_a(z)/\zeta_b(z)$ and note that
$|G(z_k)|=1=|G(z_k')|$ for all~$k$. Applying Taylor's theorem
(analogously to the derivation of \eqref{Tay}), dividing by
$|z_k-z_k'|$ and passing to the limit $k\to\infty$, we derive
\begin{equation}
\label{3.11}
\lim_{k\to\infty}
\RE\biggl(\frac{z_k-z_k'}{|z_k-z_k'|}
\frac{\partial_z G(z_k)}{G(z_k)}\biggr)=0.
\end{equation}
The second ratio on the left-hand side tends to $v_a-v_b$.
As for the first ratio, an easy computation reveals
that, since $|z_k-\barz|=|z_k'-\barz|\ne0$, we have
\begin{equation}
\label{3.12}
\frac{z_k-z_k'}{|z_k-z_k'|}=\texti e^{\texti\frac12(\theta_k+\theta_k')}
\frac{\sin((\theta_k-\theta_k')/2)}
{|\sin((\theta_k-\theta_k')/2)|},
\end{equation}
where
\begin{equation}
\label{thetak2}
e^{\texti\theta_k}=\frac{z_k-\barz}{|z_k-\barz|}
\quad\text{and}\quad e^{\texti\theta_k'}
=\frac{z_k'-\barz}{|z_k'-\barz|}.
\end{equation}
By our assumptions, we have $e^{\texti\theta_k}\to e^{\texti\theta}$ and
$e^{\texti\theta_k'}\to e^{\texti\theta'}$ as $k\to\infty$. Suppose now that
$e^{\texti\theta}=e^{\texti\theta'}$. Then, choosing a subsequence if
necessary, the left-hand side of \eqref{3.12} tends to a definite
sign times $\texti e^{\texti\theta}$. Inserting this into \eqref{3.11} and
using Corollary~\ref{cor3.1}, in addition to
$\RE(e^{\texti\theta}(v_a-v_b))=0$, we now get that also
$\RE(\texti e^{\texti\theta}(v_a-v_b))=\IM(e^{\texti\theta}(v_a-v_b))=0$.
Consequently, $v_a=v_b$, again contradicting Assumption~A3.

To finish the proof of the claim~(2), it remains to rule out the
possibility that $e^{\texti\theta'}=- e^{\texti\theta}$ in the case when
$\barz$ is a multiple point.
Let $n\in\QQ(\barz)$ be another phase stable at $\barz$,
i.e., $n\ne a,b$.
By Lemma~\ref{lemma3.1}, we have
\begin{equation}
\RE\bigl(e^{\texti\theta}(v_m-v_n)\bigr)\ge0\quad\text{and}\quad
\RE\bigl(e^{\texti\theta'}(v_m-v_n)\bigr)\ge0,\qquad m=a,b.
\end{equation}
But then  $e^{\texti\theta'}=- e^{\texti\theta}$ would imply that
$\RE(e^{\texti\theta}v_a)=\RE(e^{\texti\theta}v_n)=\RE(e^{\texti\theta}v_b)$, in
contradiction with Assumption~A4. Therefore, $|\QQ(\barz)|<3$, as
claimed.
\end{proofsect}

\begin{mycorollary}
\label{cor3.2}
Suppose that Assumption~A holds and let
$\barz\in\OO$ be a multiple point. Then there exists a constant
$\delta>0$ such that $|\QQ(z)|\le2$ for all
$z\in\{z'\in\OO\colon 0<|z'-\barz|<\delta\}$.
In particular, each multiple point in
$\OO$ is isolated.
\end{mycorollary}

\begin{proofsect}{Proof}
Suppose $\barz\in\OO$ is a non-isolated multiple point. Then there
is a sequence $z_k\in\OO$ such that $z_k\to\barz$ and, without
loss of generality, $\QQ(z_k)=\QQ_0$ with $|\QQ_0|\ge3$,
$z_k\ne\barz$ for all~$k$, and such that the limit \eqref{eitheta}
exists. Taking for $(z_k')$ the identical sequence, $z_k'=z_k$, we
get $e^{\texti\theta}=e^{\texti\theta'}$ in contradiction to
Lemma~\ref{lemma3.2}(1). Therefore, every multiple point in~$\OO$
is isolated.
\end{proofsect}

Our last auxiliary claim concerns the connectivity of sets of
$\theta$ such that \eqref{ineq} holds. As will be seen in the
proof of Lemma~\ref{lemma3.4}, this will be crucial for
characterizing the topology of the phase diagram in small
neighborhoods of multiple points.

\begin{mylemma}
\label{lemma3.3} Suppose that Assumption~A holds and let
$\barz\in\OO$ be a multiple point. For $m\in\QQ(\barz)$, let
$v_m=v_m(\barz)$.
Then, for
each $m\in\QQ(\barz)$, the set
\begin{equation}
\label{Im}
I_m=
\bigl\{
    e^{\texti\theta}
    \colon
    \theta\in[0,2\pi),\,
    \RE(e^{\texti\theta}v_m)>\RE(e^{\texti\theta}v_n),\,
    n\in\QQ(\barz)\setminus\{m\}
\bigr\}
\end{equation}
is connected and open as a subset of $\{z\in\OO\colon|z|=1\}$.
In particular, if~$e^{\texti\theta}$ is such that
\begin{equation}
\label{bdIm}
\RE(e^{\texti\theta}v_m)
=\max_{n\in\QQ(\barz)\smallsetminus\{m\}}\RE(e^{\texti\theta}v_n),
\end{equation}
then $e^{\texti\theta}$ is one of the two boundary points of~$I_m$.
\end{mylemma}

\begin{proofsect}{Proof}
By Assumption~A4, the numbers~$v_m$, $m\in\QQ(\barz)$, are the
vertices of a strictly convex polygon~$\PP$ in~$\C$. Let
$s=|\QQ(\barz)|$ and let $(v_1,\dots,v_s)$ be an ordering of the
vertices of~$\PP$ in the counterclockwise direction. For
$m=1,\dots,s$ define $\Delta v_m=v_m-v_{m-1}$, where we take
$v_0=v_s$. Note that, by strict convexity of~$\PP$, the
arguments~$\theta_m$ of~$\Delta v_m$, i.e., numbers~$\theta_m$
such that $\Delta v_m=|\Delta v_m|e^{\texti\theta_m}$, are such that
the vectors $e^{\texti\theta_1},\dots,e^{\texti\theta_s}$ are ordered
counterclockwise, with the angle between $e^{\texti\theta_m}$
and~$e^{\texti\theta_{m+1}}$ lying strictly between~$0$ and~$\pi$ for
all $m=1,\dots s$ (again, we identify $m=1$ and $m=s+1$).  In
other words, for each~$m$, the angles $\theta_1\dots,\theta_s$ can
be chosen in such a way that
$\theta_m<\theta_{m+1}<\dots<\theta_{m+s}$, with
$0<\theta_{m+k}-\theta_{m+k-1}<\pi$, $k=1,\dots, s$.  (Again, we
identified~$m+k$ with~$m+k-s$ whenever~$m+k>s$).

Using~$J_m$ to denote the set
$J_m=\bigl\{ \texti e^{-\texti\vartheta}\colon
\vartheta\in(\theta_m,\theta_{m+1})\bigr\}$,
we claim that
$I_m=J_m$ for all $m=1,\dots,s$. First, let us show that
$J_m\subset I_m$. Let thus $\vartheta\in(\theta_m,\theta_{m+1})$
and observe that
\begin{equation}
\RE(\texti e^{-\texti\vartheta}\Delta v_m)=|\Delta
v_m|\sin(\vartheta-\theta_m)>0,
\end{equation}
because $\theta_m<\vartheta<\theta_{m+1}<\theta_m+\pi$.
Similarly,
\begin{equation}
\RE(\texti e^{-\texti\vartheta}\Delta v_{m+1})= |\Delta
v_{m+1}|\sin(\vartheta-\theta_{m+1})<0,
\end{equation}
because $\theta_{m+1}-\pi<\theta_m<\vartheta<\theta_{m+1}$.
Consequently, $\RE(\texti e^{-\texti\vartheta}v_m)>\RE(\texti e^{-\texti\vartheta}v_n)$
holds for both~$n=m+1$ and~$n=m-1$.

It remains to show that
$\RE(\texti e^{-\texti\vartheta}v_m)>\RE(\texti e^{-\texti\vartheta}v_n)$ is true also
for all remaining $n\in\QQ(\barz)$.
Let~$n\in\QQ(\barz)\setminus\{m,m\pm 1\}$. We will separately
analyze the cases with $\theta_n-\theta_m\in(0,\pi]$ and
$\theta_n-\theta_m\in(-\pi,0)$. Suppose first that
$\theta_n-\theta_m\in(0,\pi]$. This allows us to write~$n=m+k$ for
some $k\in\{2,\dots, s-1\}$ and estimate
\begin{multline}
\qquad\RE(\texti e^{-\texti\vartheta}(v_n-v_m))
=\sum_{j=1}^k
\RE(\texti e^{-\texti\vartheta}\Delta v_{m+j})
\\
= \sum_{j=1}^k|\Delta v_{m+j}|\sin(\vartheta-\theta_{m+j})
<0.
\qquad
\end{multline}
The inequality holds since, in light of
$\vartheta<\theta_{m+1}<\dots<\theta_{m+k}\le\theta+\pi$,
each sine is negative
except perhaps for the last one which is
allowed to be zero. On the other hand, if
$\theta_n-\theta_m\in(-\pi,0)$, we write~$n=m-k$  instead, for
some $k\in\{2,\dots, s-1\}$, and estimate
\begin{multline}
\qquad
\RE(\texti e^{-\texti\vartheta}(v_m-v_n))=\sum_{j=-k+1}^0
\RE(\texti e^{-\texti\vartheta}\Delta v_{m+j})
\\=
\sum_{j=-k+1}^0|\Delta v_{m+j}|
\sin(\vartheta-\theta_{m+j})>0.
\qquad
\end{multline}
Here we invoked the inequalities
$\vartheta-\pi<\theta_{m-k}<\dots<\theta_m<\vartheta$ to show that
each sine on the right-hand side is strictly positive.

As a consequence of the previous estimates, we
conclude that
$J_m\subset I_m$ for all $m=1,\dots, s$. However,
the union of all~$J_m$'s covers the unit circle with the exception
of~$s$ points and, since the sets~$I_m$ are open and disjoint, we
must have~$I_m=J_m$ for all~$m\in\QQ(\barz)$.
Then, necessarily,~$I_m$ is connected and open.
Now the left-hand side of \eqref{bdIm} is strictly greater
than the right-hand side for $e^{\texti\theta}\in I_m$, and strictly
smaller than the right-hand side for $e^{\texti\theta}$ in the interior of
the complement of~$I_m$. By continuity of both sides,
\eqref{bdIm} can hold only on the boundary of~$I_m$.
\end{proofsect}

\subsection{Proof of Theorem~\ref{T:PD}}
\label{sec3.2}\noindent Having all the necessary tools ready, we
can start proving Theorem~\ref{T:PD}. First we will apply
Lemma~\ref{lemma3.3} to characterize the situation around multiple
points.

\vbox{
\begin{mylemma}
\label{lemma3.4}
Suppose that Assumption~A holds and let $\barz\in\OO$
be a multiple point. For $\delta>0$, let
\begin{equation}
I_m^{(\delta)}
=\bigl\{z\in \OO\colon |z-\barz|=\delta,\,Q(z)\ni m\bigr\}.
\end{equation}
Then the following is true once~$\delta$ is sufficiently small:
\settowidth{\leftmargini}{(111)}
\begin{enumerate}
\item[(1)] For each $m\in\QQ(\barz)$, the set $I_m^{(\delta)}$ is
connected and has a non-empty interior. \item[(2)]
$I_m^{(\delta)}=\emptyset$ whenever $m\notin\QQ(\barz)$.
\item[(3)] For distinct~$m$ and~$n$, the sets $I_m^{(\delta)}$ and
$I_n^{(\delta)}$ intersect in at most one point.
\end{enumerate}
\end{mylemma}
}

\begin{proofsect}{Proof}
The fact that $I_m^{(\delta)}=\emptyset$ for $m\notin\QQ(\barz)$
once $\delta>0$ is sufficiently small is a direct consequence
of the continuity of the functions~$\zeta_m$ and~$\zeta$.
Indeed, if there
were a sequence of points~$z_k$ tending to~$\barz$ such that a
phase~$m$ were stable at each~$z_k$, then~$m$ would be also stable
at~$\barz$.

We will proceed by proving that, as $\delta\downarrow0$, each set
$I_m^{(\delta)}$, $m\in\QQ(\barz)$, will eventually have a
non-empty interior. Let $m\in\QQ(\barz)$. Observe that,  by
Lemma~\ref{lemma3.3}, there is a value $e^{\texti\theta}$ (namely, a
number from~$I_m$) such that
$\RE(e^{\texti\theta}v_m)>\RE(e^{\texti\theta}v_n)$ for all
$n\in\QQ(\barz)\setminus\{m\}$. But then the second part of
Lemma~\ref{lemma3.1} guarantees the existence of an $\epsilon>0$
such that $\QQ(z)=\{m\}$ for all
$z\in\WW_{\epsilon,\theta}(\barz)$---see~\eqref{w-neighb}. In
particular, the intersection
$\WW_{\epsilon,\theta}(\barz)\cap\{z\colon |z-\barz|=\delta\}$,
which is non-empty and (relatively) open for $\delta<\epsilon$, is
a subset of $I_m^{(\delta)}$. It follows that the
set~$I_m^{(\delta)}$ has a nonempty interior once~$\delta$ is
sufficiently~small.

Next we will prove that each $I_m^{(\delta)}$, $m\in\QQ(\barz)$,
is eventually connected. Suppose that there exist a phase
$a\in\QQ(\barz)$ and a sequence $\delta_k\downarrow0$ such that
all sets $I_a^{(\delta_k)}$ are \emph{not} connected. Then, using
the fact that $I_a^{(\delta_k)}$ has nonempty interior and thus
cannot consist of just two separated points, we conclude that the
phase~$a$ coexists with some other phase at at least three
distinct points on each circle $\{z\colon|z-\barz|=\delta_k\}$.
Explicitly, there exist (not necessarily distinct) indices
$b_k^{(j)}\in\QQ(\barz)\setminus\{a\}$ and points $(z_k^{(j)})$,
$j=1,2,3$, with $|z_k^{(j)}-\barz|=\delta_k$ and $z_k^{(j)}\ne
z_k^{(\ell)}$ for $j\neq \ell$, such that
$a,b_k^{(j)}\in\QQ(z_k^{(j)})$. Moreover, (choosing subsequences
if needed) we can assume that $b_k^{(j)}=b^{(j)}$ for some
$b^{(j)}\in\QQ(\barz)\setminus\{a\}$ independent of~$k$. Resorting
again to subsequences, we also may assume that the limits in
\eqref{eitheta} exist for all three sequences.

Let us use $e^{\texti\theta_j}$ to denote the corresponding limits for
the three sequences. First we claim that the numbers
$e^{\texti\theta_j}$, $j=1,2,3$, are necessarily all distinct. Indeed,
suppose two of the $e^{\texti\theta_j}$'s are the same and let~$b$
and~$c$ be the phases coexisting with~$a$ along the corresponding
sequences. Then Lemma~\ref{lemma3.2}(1) forces  $b=c$, which
contradicts both conclusions of Lemma~\ref{lemma3.2}(2).
Therefore, all three $e^{\texti\theta_j}$ must be different. Applying
now Corollary~\ref{cor3.1} and Lemma~\ref{lemma3.1},  we  get
$\RE(e^{\texti\theta_j}v_a)=\max_{n\in\QQ(\barz)\setminus\{a\}}
\RE(e^{\texti\theta_j}v_n)$
for $j=1,2,3$.
According to Lemma~\ref{lemma3.3},
all three distinct numbers  $e^{\texti\theta_j}$, $j=1,2,3$, are
endpoints of~$I_a$, which is not possible since $I_a$ is a
connected subset of the unit circle. Thus, we can conclude
that~$I_a^{(\delta)}$ must be connected once~$\delta>0$ is
sufficiently small.

To finish the proof, we need to show that $I_a^{(\delta)}\cap
I_b^{(\delta)}$ contains at most one point for any $a\ne b$. First
note that we just ruled out the possibility that this intersection
contains \emph{three} distinct points for a sequence of $\delta$'s
tending to zero. (Indeed, then~$a$ would coexist with~$b$ along
three distinct sequences, which would in turn imply that~$a$ and
$b$ coexists along three distinct directions, in contradiction
with Lemma~\ref{lemma3.3}.) Suppose now that $I_a^{(\delta)}\cap
I_b^{(\delta)}$ contains two distinct points.
Since both~$I_a^{(\delta)}$ and~$I_b^{(\delta)}$ are connected
with open interior,
this would mean that $I_a^{(\delta)}$ and~$I_b^{(\delta)}$ cover
the entire circle of radius~$\delta$. Once again, applying the
fact that two $I_m^{(\delta)}$ have at most two points in common,
we then must have $I_c^{(\delta)}=\emptyset$ for all $c\ne a,b$.
But~$\QQ(\barz)$ contains at least three phases which necessitates
that $I_m^{(\delta)}\ne\emptyset$ for at least three distinct~$m$.
Hence $I_a^{(\delta)}\cap I_b^{(\delta)}$ cannot contain more than
one point.
\end{proofsect}

Next we will give a local characterization of two-phase
coexistence lines.

\begin{mylemma}
\label{lemma3.5}
Suppose that Assumption~A holds and let
$m,n\in\RR$ be distinct. Let $z\in\OO$ be such that
$z\in\CalS_m\cap\CalS_n$ and $\QQ(z')\subset\{m,n\}$ for
$z'\in\D_\delta(z)$. Then there exist numbers
$\delta'\in(0,\delta)$, $t_1<0$, $t_2>0$, and an  twice
continuously differentiable function
$\gamma_z\colon(t_1,t_2)\to\D_{\delta'}(z)$ such that
\settowidth{\leftmargini}{(111)}
\begin{enumerate}
\item[(1)]
$\gamma_z(0)=z$.
\item[(2)]
$|\zeta_m(\gamma_z(t))|=|\zeta_n(\gamma_z(t))|=\zeta(\gamma_z(t))$,
$t\in(t_1,t_2)$.
\item[(3)]
$\lim_{t\downarrow t_1}\gamma_z(t),
\,\lim_{t\uparrow t_2}\gamma_z(t)\in\partial \D_{\delta'}(z)$.
\end{enumerate}
The curve $t\mapsto \gamma_z(t)$ is unique up to
reparametrization. Moreover, the set
$\D_{\delta'}(z)\setminus\gamma_z(t_1,t_2)$ has two connected
components and~$m$ is the only stable phase in one of the
components while $n$ is the only stable phase in the other.
\end{mylemma}

\begin{proofsect}{Proof}
We begin by observing that by Assumption~A3, the function
\begin{equation}
\phi_{m,n}(x,y)=\log|\zeta_m(x+\texti y)|-
\log|\zeta_n(x+\texti y)|=\RE \log F_{m,n}(x+\texti y),
\end{equation}
has at least one of the derivatives
$\partial_x\phi_{m,n},\partial_y\phi_{m,n}$ non-vanishing at
$x+\texti y=z$. By continuity, there exists a constant $\eta>0$ such
that one of the derivatives is uniformly bounded away from zero
for all $z'=u+\texti v\in\D_\eta(z)$. Since
$z=x+\texti y\in\CalS_m\cap\CalS_n$, we have $\phi_{m,n}(x,y)=0$. By the
implicit function theorem, there exist
numbers~$t_0'$,~$t_1'$,~$x_0$,~$x_1$,~$y_0$ and~$y_1$
such that $t_0'<0<t_1'$,
$x_0<x<x_1$, $y_0<y<y_1$ and $(x_0,x_1)\times(y_0,y_1)
\subset\D_\eta(z)$, and twice continuously
differentiable functions $u\colon (t_0',t_1')\to(x_0,x_1)$ and
$v\colon (t_0',t_1')\to(y_0,y_1)$ such that
\begin{equation}
\label{phi=0} \phi_{m,n}\bigl(u(t),v(t)\bigr)=0, \qquad
t\in(t_0',t_1'),
\end{equation}
and
\begin{equation}
\label{inval}
u(0)=x,\quad\text{and}\quad v(0)=y.
\end{equation}
Moreover, since the second derivatives of~$\phi_{m,n}$ are
continuous in~$\OO$ and therefore bounded in $\D_\eta(z)$,
standard theorems on uniqueness of the solutions of ODEs
guarantee that the solution to \eqref{phi=0} and \eqref{inval} is
unique up to reparametrization. The construction of~$\gamma_z$ is
now finished by picking~$\delta'$ so small that
$\D_{\delta'}(z)\subset (x_0,x_1)\times(y_0,y_1)$, and
taking~$t_0$ and~$t_1$ to be the first backward and forward time,
respectively, when~$(u(t),v(t))$ leaves~$\D_{\delta'}(z)$.

The fact that $\D_{\delta'}(z)\setminus\gamma_z(t_1,t_2)$ splits
into two components is a consequence of the construction
of~$\gamma_z$. Moreover,~$\gamma_z$ is a (zero-)level curve of
function~$\phi_{m,n}$ which has a non-zero gradient. Hence,
$\phi_{m,n}<0$ on one component of
$\D_{\delta'}(z)\setminus\gamma_z(t_1,t_2)$, while $\phi_{m,n}>0$
on the other. Recalling the assumption that
$\QQ(z')\subset\{m,n\}$ for~$z'$ in a neighborhood of~$z$, the
claim follows.
\end{proofsect}

Now we can finally give the proof of Theorem~\ref{T:PD}.

\begin{proofsect}{Proof of Theorem~\ref{T:PD}}
Let~$\MM$ denote the set of all multiple points in~$\OO$, i.e., let

\begin{equation}
\MM=\bigl\{z\in\OO\colon |\QQ(z)|\ge3\bigr\}.
\end{equation}
By Corollary~\ref{cor3.2},
we know that~$\MM$  is relatively closed in~$\OO$ and so the set
$\OO'=\OO\setminus\MM$ is open. Moreover, the set $\GG\cap\OO'$
consists solely of points where exactly two phases coexist.
Lemma~\ref{lemma3.5} then shows that for each $z\in\GG\cap\OO'$,
there exists a disc~$\D_{\delta'}(z)$ and a unique, 
smooth~$\gamma_{z}$ in~$\D_{\delta'}(z)$ passing through~$z$ such
that $\QQ(z')=\QQ(z)$ for all~$z'$ on the curve~$\gamma_{z}$.
Let~$\tilde\gamma_{z}$ be a maximal extension of the
curve~$\gamma_{z}$ in~$\OO'$. We claim that~$\tilde\gamma_{z}$ is
either a closed curve or an arc with both endpoints
on~$\partial\OO'$. Indeed, if~$\tilde\gamma_{z}$ were open with an
end-point~$\tilde z\in\OO'$, then $\QQ(\tilde z)\supset\QQ(z)$, by
continuity of functions $\zeta_m$. But $\tilde z\in\OO'$ and so
$|\QQ(\tilde z)|\le2$, which implies that $\QQ(\tilde z)=\QQ(z)$.
By Lemma~\ref{lemma3.5}, there exists a non-trivial
curve~$\gamma_{\tilde z}$ along which the two phases
from~$\QQ(\tilde z)$ coexist in a neighborhood of~$\tilde z$. But
then $\gamma_{\tilde z}\cup\tilde\gamma_{z}$
would be a non-trivial extension of~$\tilde\gamma_{z}$, in
contradiction with the maximality of~$\tilde\gamma_{z}$. Thus we
can conclude that~$\tilde z\in\partial\OO'$.

Let~$\eusb C$ denote the set of maximal extensions of the curves
$\{\gamma_{z}\colon  z\in\GG\cap\OO'\}$. Let $\DD\subset\OO$ be a
compact set and note that Corollary~\ref{cor3.2} implies that
$\DD\cap\MM$ is finite. Let~$\delta_0$ be so small that, for each
$\zM\in\MM\cap\DD$, we have $\D_{\delta_0}(\zM)\subset\OO$,
$\overline{\D_{\delta_0}(\zM)}\cap\MM=\{\zM\}$ and the statements
in Lemma~\ref{lemma3.4} hold true for $\delta\le\delta_0$. Let
$\delta\in(0,\delta_0]$. We claim that if a curve $\CC\in\eusb C$
intersects the disc~$\D_\delta(\zM)$ for a $\zM\in\MM\cap\DD$,
then the restriction $\CC\cap\D_\delta(\zM)$ is a simple
curve connecting~$\zM$ to $\partial\D_\delta(\zM)$. Indeed, each
curve $\CC\in\eusb C$ terminates either on~$\partial\OO$ or
on~$\MM$. If~$\CC$ ``enters''~$\D_\delta(\zM)$ and does not
hit~$\zM$, our assumptions about~$\delta_0$ imply that~$\CC$
``leaves''~$\D_\delta(\zM)$ through the boundary. But
Lemma~\ref{lemma3.5} ensures that one of the phases coexisting
along~$\CC$ dominates in a small neighborhood on the ``left''
of~$\CC$, while the other dominates in a small neighborhood on the
``right'' of~$\CC$. The only way this can be made consistent with
the connectivity of the sets~$I_m^{(\delta)}$ in
Lemma~\ref{lemma3.4} is by assuming that
$I_m^{(\delta)}\ne\emptyset$ only for the two~$m$'s coexisting
along~$\CC$. But that still contradicts Lemma~\ref{lemma3.4}, by
which $I_m^{(\delta)}\ne\emptyset$ for at least \emph{three}
distinct~$m$. Thus, once a curve $\CC\in\eusb C$
intersects~$\D_\delta(\zM)$, it must terminate at~$\zM$.

Let $\DD_0=\DD\setminus\bigcup_{z\in\MM}\D_{\delta_0}(z)$
and let $\Delta\colon\DD_0\to[0,\infty)$ be a function given by
\begin{equation}
\Delta(z)=\inf\bigl\{\delta'\in(0,\delta_0)\colon
\D_{\delta'}(z)\subset\OO,\,
\D_{\delta'}(z)\cap{\textstyle\bigcup_{\CC\in\eusb C}}\,\CC
\text{ is disconnected}\bigr\}.
\end{equation}
We claim that~$\Delta$ is bounded from below by a positive constant.
Indeed,~$\Delta$ is clearly continuous and, since~$\DD_0$ is
compact,~$\Delta$ attains its minimum at some
$z\in\DD_0$. If $\Delta(z)=0$, then~$z$ is a limit point
of~$\bigcup_{\CC\in\eusb C}\CC$ and thus~$z\in\CC$
for some $\CC\in\eusb C$. Moreover, for infinitely many
$\delta'\in (0,\delta_0)$, the circle~$\partial\D_{\delta'}(z)$
intersects the set $\bigcup_{\CC\in\eusb C}\CC$ in at least three
different points.
Indeed, the curve $\CC\ni z$ provides two intersections; the third
intersection is obtained by adjusting the radius
$\delta'$ so that
$\D_{\delta'}(z)\cap{\textstyle\bigcup_{\CC\in\eusb C}}$ is
disconnected. Thus, we are (again) able to construct three
sequences~$(z_k)$,~$(z_k')$ and~$(z_k'')$ such that, without loss
of generality, $z_k,z_k',z_k''\in\CalS_a\cap\CalS_b$ for some
distinct~$a,b\in\RR$ (only two phases can exist in sufficiently small
neighborhoods of points in~$\DD_0$), $|z_k-\barz|=|z_k'-\barz|
=|z_k''-\barz|\to0$, but $z_k\ne z_k'\ne z_k''\ne z_k$ for
all~$k$. However, this contradicts Lemma~\ref{lemma3.2}, because
its part~(2) cannot hold simultaneously for all three pairs of
sequences~$(z_k,z_k')$,~$(z_k',z_k^{\prime\prime})$
and~$(z_k,z_k'')$.

Now we are ready to define the set of points $z_1,\dots,z_\ell$.
Let~$\epsilon$ be the minimum of
the function~$\Delta$ in~$\DD_0$ and let
$\delta=\min(\delta_0,\epsilon)$.
Consider the following collections  of open finite
discs:
\begin{equation}
\label{covers}
\begin{aligned}
\eusm S_1&=\bigl\{\D_\delta(z)\colon z\in\MM\cap\DD\bigr\},
\\
\eusm S_2&=\bigl\{\D_{\delta}(z)\colon
z\in\DD\cap{\textstyle\bigcup_{\CC\in\eusb C}\CC},\,
\dist(z,\textstyle{\bigcup_{\D\in\eusm S_1}}\,\D)>
\tfrac23\delta\bigr\},
\\
\eusm S_3&=\bigl\{\D_\delta(z)\colon z\in\DD,\,
\dist(z,\textstyle{\bigcup_{\D\in\eusm S_1\cup\eusm S_2}}\,\D)>
\tfrac23\delta\bigr\}.
\end{aligned}
\end{equation}
It is easy to check that the union of these discs covers~$\DD$.
Let $\eusm S= {\eusm S}_1\cup{\eusm S}_2\cup{\eusm S}_3$.
By compactness of~$\DD$,
we can choose a finite collection ${\eusm S}'\subset \eusm S$
still covering~$\DD$.
It remains to show that the sets
$\AA=\GG\cap\D$ for $\D\in\eusm S'$
will have the desired properties.
Let $\D\in\eusm S'$ and let~$z$ be the center of~$\D$.
If~$\D\in\eusm S_3$, then $\GG\cap\D=\emptyset$.
Indeed, if~$z'$ is a coexistence point, then
$\D_\delta(z')\in\eusm S_1\cup\eusm S_2$
and thus $\dist(z,z')>\delta+\tfrac23\delta$ and hence $z'\not\in\D$.
Next, if~$\D\in\eusm S_2$, then $z\in\GG$ and,
by the definition of~$\delta_0$ and~$\epsilon$,
the disc~$\D$ contains no multiple point and
intersects~$\GG$ only in one component.
This component is necessarily part of one of
the curves~$\CC\in\eusb C$.
Finally, if~$\D\in\eusm S_1$, then~$z$ is
a multiple point and, relying on our previous reasoning,
several curves $\CC\in\eusb C$ connect~$z$
to the boundary of~$\D$.
Since Lemma~\ref{lemma3.4} implies the existence
of exactly $|\QQ(z)|$ coexistence points on
$\partial\D$, there are exactly~$|\QQ(z)|$ such curves.
The proof is finished by noting that every multiple
point appears as the center of some disc $\D\in\eusm S'$,
because that is how the collections \eqref{covers}
were constructed.
\end{proofsect}

\section{Partition function zeros}
\noindent 
The goal of this section is to prove
Theorems~\ref{T:deg}-\ref{T:Mph}. The principal tool which enables
us to control the distance between the roots of~$Z_L^\per$ and the
solutions of equations \twoeqref{Req}{Imq} or \eqref{Meq}
is Rouch\'e's Theorem (see e.g.~\cite{Gamelin}). 
For reader's convenience, we transcribe the corresponding
statement here:

\begin{theorem}[Rouch\'e's Theorem]
Let~$\DD\subset\C$ be a bounded domain with piecewise smooth boundary~$\partial\DD$.
Let~$f$ and~$g$ be analytic on~$\DD\cup\partial\DD$. If~$|g(z)|<|f(z)|$ for
all~$z\in\partial\DD$, then~$f$ and~$f+g$ have the same number of zeros in~$\DD$,
counting multiplicities.
\end{theorem}

\noindent
More details on the use of this theorem and the corresponding bounds are stated in Sect.~\ref{sec4.2}
for the case of two-phase coexistence and in Sect.~\ref{sec4.3} for the case of multiple phase coexistence.

Root degeneracy will be controlled using a link between the non-degeneracy conditions from Assumption~B and certain Vandermonde determinants; cf Sect.~\ref{sec4.1}.
Throughout this section, we will use the shorthand
\begin{equation}
\label{S(Q)-def}
\CalS_\epsilon(\QQ)=\bigcap_{m\in \QQ}\CalS_\epsilon(m)
\end{equation}
to denote the set of points~$z\in\OO$ where all phases from a non-empty $\QQ\subset\RR$
are ``almost stable'' (as quantified by $\epsilon>0$).

\subsection{Root degeneracy}
\label{sec4.1}\noindent
In this section we will prove Theorem~\ref{T:deg}.
We begin with a claim about the
Vandermonde
matrix
defined in terms of the functions
\begin{equation}
\label{bm} b_m(z)=\frac{\partial_z
\zeta_m^{(L)}(z)}{\zeta_m^{(L)}(z)}, \qquad
z\in\CalS_{\kappa/L}(m),
\end{equation}
where the dependence of~$b_m$ on~$L$ has been suppressed in
the notation.
Let us fix a non-empty~$\QQ\subset\RR$ and let~$q=|\QQ|$. For
each~$z\in\CalS_{\kappa/L}(\QQ)$, we introduce the~$q\times
q$ Vandermonde matrix~$\M(z)$ with elements
\begin{equation}
\label{Matrix}
\M_{\ell,m}(z)=b_m(z)^\ell,
\qquad
m\in\QQ,\,\,\ell=0,1,\dots,q-1.
\end{equation}
Let~$\Vert\M\Vert$ denote the~$\ell^2(\QQ)$-norm of~$\M$ (again
without making the~$\QQ$-dependence of this norm notationally
explicit). Explicitly,~$\Vert\M\Vert^2$ is defined by the supremum
\begin{equation}
\Vert\M\Vert^2=\sup\biggl\{\sum_{\ell=0}^{q-1}\,\Bigl|\sum_{m\in\QQ}
\M_{\ell,m}\hatw_m\Bigr|^2\colon\sum_{m\in\QQ}|\hatw_m|^2=1\biggr\},
\end{equation}
where~$(\hatw_m)$ is a~$|\QQ|$-dimensional complex vector.

Throughout the rest of this section, the symbol $\Vert\cdot\Vert$
will refer to the (vector or matrix) $\ell^2$-norm
as specified above.
The only exceptions are the $\ell^p$-norms $\Vert\mathbf{q}\Vert_1$,
$\Vert\mathbf{q}\Vert_2$ and $\Vert\mathbf{q}\Vert_\infty$
of the $r$-tuple $(q_m)_{m\in\RR}$, which are defined in the usual way.

\begin{mylemma}
\label{lemma4.1}
Suppose that
Assumption~B3 holds and let~$\LB$ be
as in Assumption~B3. For each $\QQ\subset\RR$, there exists a
constant $K=K(\QQ)<\infty$ such that
\begin{equation}
\label{InvMbd} \bigl\Vert\M^{-1}(z)\bigr\Vert\le K,
\text{ for all }
z\in\CalS_{\kappa/L}(\QQ)
\text{ and } L\ge\LB.
\end{equation}
In particular,~$\M(z)$ is invertible for all~$z\in
\CalS_{\kappa/L}(\QQ)$ and~$L\ge \LB$.
\end{mylemma}

\begin{proofsect}{Proof}
Let~$\QQ\subset\RR$ and~$q=|\QQ|$. Let us choose a point~$z\in
\CalS_{\kappa/L}(\QQ)$ and let~$\M$ and~$b_m$,~$m\in\QQ$, be the
quantities~$\M(z)$ and~$b_m(z)$,~$m\in\QQ$. First we note that,
since~$\M$ is a Vandermonde matrix, its determinant can be
explicitly computed: $\det\M=\prod_{m<n}(b_n-b_m)$, where ``$<$''
denotes a complete order on~$\QQ$. In particular, Assumption~B3
implies that~$|\!\det\M|\ge{\tilde\alpha}^{q(q-1)/2}>0$
once~$L\ge \LB$.

To estimate the matrix norm of~$\M^{-1}$,
let~$\lambda_1,\dots,\lambda_q$ be the eigenvalues of the
Hermitian matrix~$\M\,\M^+$ and note that~$\lambda_\ell>0$ for
all~$\ell=1,\dots,q$ by our lower bound on~$|\!\det\M|$.
Now,~$\Vert\M^+\Vert^2$ is equal to the spectral radius of the
operator~$\M\,\M^+$, and $\Vert\M^{-1}\Vert^2$ is equal to the
spectral radius of the operator~$(\M\,\M^+)^{-1}$. By the
well-known properties of the norm we thus have
\begin{equation}
\Vert\M\Vert^2=\Vert\M^+\Vert^2=\max_{1\le\ell\le q}\lambda_\ell,
\end{equation}
while
\begin{equation}
\Vert\M^{-1}\Vert^2=\max_{1\le\ell\le q}\lambda_\ell^{-1}.
\end{equation}
Now~$|\!\det\M|^2=\det\M\,\M^+=\lambda_1\dots\lambda_q$ and a
simple algebraic argument gives us that
\begin{equation}
\label{M-1}
\Vert\M^{-1}\Vert\le\frac{\Vert
\M\Vert^{q-1}}{|\!\det\M|}.
\end{equation}
Using the lower bound on~$|\!\det\M|$, this implies that
$\Vert\M^{-1}\Vert\le {\tilde \alpha}^{-q\frac{q-1}2}\Vert
\M\Vert^{q-1}$. The claim then follows by invoking the
uniform boundedness of the matrix elements of~$\M$
(see the upper bound from Assumption~B3), which implies
that~$\Vert\M\Vert$ and hence also $\Vert\M^{-1}\Vert$ is uniformly
bounded from above throughout~$\CalS_{\kappa/L}(\QQ)$.
\end{proofsect}

Now we are ready to prove Theorem~\ref{T:deg}.
To make the reading easier, let us note that for $\QQ=\{m\}$,
the expression
\eqref{2.7} defining $\UU_\epsilon(\QQ)$
can be simplified to
\begin{equation}
\label{UU-single-m}
\UU_\epsilon(\{m\})
=\bigl\{z\in\OO\colon |\zeta_n(z)|<e^{-\epsilon/2}|\zeta(z)|
\text{ for all }n\neq m\bigr\},
\end{equation}
a fact already mentioned right after \eqref{2.7}.

\begin{proofsect}{Proof of Theorem~\ref{T:deg}}
Let~$m\in\RR$. Since the sets $\UU_{\kappa/L}(\QQ)$, $\QQ\subset\RR$,
cover~$\OO$, it suffices to prove that $Z_L^\per\ne0$ in
$\UU_{L^{-d}\omega_L}(\{m\})\cap \UU_{\kappa/L}(\QQ)$ for each
$\QQ\subset\RR$. In fact, since $z\in\UU_{L^{-d}\omega_L}(\{m\})$
implies that $m$ is stable, $|\zeta_m(z)|=\zeta(z)$, we may assume
without loss of generality that $m\in\QQ$, because otherwise
$\UU_{L^{-d}\omega_L}(\{m\})\cap \UU_{\kappa/L}(\QQ)=\emptyset$.
Thus, let ~$m\in\QQ\subset\RR$ and fix a
point~$z\in\UU_{L^{-d}\omega_L}(\{m\})\cap \UU_{\kappa/L}(\QQ)$.
By Assumption~B4, we have the bound
\begin{multline}
\label{ZLlb1} 
\qquad
\bigl|Z_L^\per(z)\bigr|\ge  \zeta(z)^{L^d}
\biggl(q_m\Bigl|\frac{\zeta_m^{(L)}(z)}{\zeta(z)}\Bigr|^{L^d}
\\-
\sum_{n\in\QQ\smallsetminus\{m\}}
q_n\Bigl|\frac{\zeta_n^{(L)}(z)}{\zeta(z)}\Bigr|^{L^d}
-C_0 L^{d}\Vert\mathbf{q}\Vert_1 e^{-\tau L}\biggr).
\qquad
\end{multline}
Since~$z\in\UU_{L^{-d}\omega_L}(\{m\})$, we have~$|\zeta_n(z)| <
\zeta(z)e^{-\frac12L^{-d}\omega_L}$ for~$n\neq m$. In conjunction
with Assumption~B2, this implies
\begin{equation}
\label{4.9a}
\Bigl|\frac{\zeta_n^{(L)}(z)}{\zeta(z)}\Bigr|^{L^d} \le e^{L^d
e^{-\tau L}} e^{-\frac12\omega_L},
\qquad n\ne m.
\end{equation}
On the other hand, we also have
\begin{equation}
\label{4.10a}
\Bigl|\frac{\zeta_m^{(L)}(z)}{\zeta(z)}\Bigr|^{L^d} \ge e^{-L^d
e^{-\tau L}},
\end{equation}
where we used that~$|\zeta_m(z)|=\zeta(z)$.
Since $\omega_L\to\infty$, \twoeqref{4.9a}{4.10a} show
that the right-hand side \eqref{ZLlb1} is dominated by
the term with index~$m$, which is bounded away from zero
uniformly in~$L$. Consequently,~$Z_L^\per\ne0$ throughout
$\UU_{L^{-d}\omega_L}(\{m\})\cap \UU_{\kappa/L}(\QQ)$,
provided~$L$ is sufficiently large.

Next we will prove the claim about the degeneracy of the roots.
Let us fix~$\QQ\subset\RR$ and let, as before,~$q=|\QQ|$.
Suppose that~$L\ge \LB$ and let~$z\in
\UU_{\kappa/L}(\QQ)$ be a
root of~$Z_L^\per$ that is at least~$q$-times degenerate.
Since~$Z_L^\per$
is analytic in a neighborhood of~$z$, we have
\begin{equation}
\label{degen}
\partial_z^\ell Z_L^\per(z)=0,\qquad
\ell=0,1,\dots,q-1.
\end{equation}
It will be convenient to introduce $q$-dimensional
vectors $\mathbf{x}=\mathbf{x}(z)$
and $\mathbf{y}=\mathbf{y}(z)$ such that \eqref{degen}
can be expressed as
\begin{equation}
\label{Mx=y}
\M(z) \mathbf{x} =\mathbf{y},
\end{equation}
with $\M(z)$ given by \eqref{bm} and \eqref{Matrix}.
Indeed, let $\mathbf{x}=\mathbf{x}(z)$
be the vector with components
\begin{equation}
x_m=q_m\Bigl(\frac{\zeta_m^{(L)}(z)}{\zeta(z)}\Bigr)^{L^d}, \qquad
m\in\QQ.
\end{equation}
Similarly, let~$\mathbf{y}=\mathbf{y}(z)$ be the
vector with components
$y_0,\dots,y_{q-1}$, where
\begin{equation}
\label{yell}
\begin{aligned}
y_\ell&=L^{-d\ell}\zeta(z)^{-L^d}\partial_z^\ell
\varXi_{\QQ,L}(z)
\\
&\qquad\qquad
-\sum_{m\in\QQ}q_m\,\zeta(z)^{-L^d}
\Bigl\{L^{-d\ell}
\partial_z^\ell \bigl[\zeta_m^{(L)}(z)\bigr]^{L^d}-
b_m(z)^\ell \bigl[\zeta_m^{(L)}(z)\bigr]^{L^d}\Bigr\}.
\end{aligned}
\end{equation}
Recalling the definition $\varXi_{\QQ,L}(z)$ from \eqref{ZLper},
it is easily seen that \eqref{Mx=y} is equivalent to
\eqref{degen}.

We will now produce appropriate bounds on
the~$\ell^2(\QQ)$-norms~$\Vert\mathbf{y}\Vert$
and~$\Vert\mathbf{x}\Vert$ which hold uniformly
in~$z\in\UU_{\kappa/L}(\QQ)$, and show that \eqref{Mx=y}
contradicts Lemma~\ref{lemma4.1}.
To estimate~$\Vert\mathbf{y}\Vert$, we first note that
there is a
constant~$A<\infty$, independent of~$L$, such that,
for all~$\ell=0,\dots,q-1$ and all~$z\in\UU_{\kappa/L}(\QQ)$,
\begin{equation}
\label{yell2}
\Bigl|L^{-d\ell}
\partial_z^\ell \bigl[\zeta_m^{(L)}(z)\bigr]^{L^d}-
b_m(z)^\ell \bigl[\zeta_m^{(L)}(z)\bigr]^{L^d}\Bigr|\le
AL^{-d}\zeta(z)^{L^d}.
\end{equation}
Here the leading order term from~$L^{-d\ell}\partial_z^\ell
[\zeta_m^{(L)}(z)]^{L^d}$ is exactly canceled by the term $b_m(z)^\ell
[\zeta_m^{(L)}(z)]^{L^d}$, and the remaining terms can be bounded
using \eqref{uplogderL}. Invoking \eqref{yell2} in \eqref{yell} and
applying \eqref{error}, we get
\begin{equation}
\Vert \mathbf{y}\Vert\le A\Vert\mathbf{q}\Vert_1\sqrt qL^{-d}
+\bigl(\,\max_{0\le\ell\le{q-1}}C_\ell\bigr)\Vert\mathbf{q}\Vert_1\sqrt qL^{d}
e^{-\tau L},
\end{equation}
where the
factor~$\sqrt q$
comes from the conversion of $\ell^\infty$-type
bounds \eqref{yell2} into a bound on the
$\ell^2$-norm~$\Vert \mathbf{y}\Vert$.
On the other hand, by \eqref{zetamL} and~$q_m\ge1$ we immediately
have
\begin{equation}
\Vert \mathbf{x}\Vert\ge e^{-e^{-\tau L}}.
\end{equation}
But~$\Vert\mathbf{x}\Vert\le\Vert\M^{-1}(z)\Vert\,\Vert
\mathbf{y}\Vert$, so once~$L$ is sufficiently large, this
contradicts the upper bound $\Vert\M^{-1}(z)\Vert\le K$ implied by
Lemma~\ref{lemma4.1}. Therefore, the root at~$z$ cannot be more
than~$(q-1)$-times degenerate after all.
\end{proofsect}

\subsection{Two-phase coexistence}
\label{sec4.2}\noindent Here we will prove Theorem~\ref{T:2ph} on
the location  of partition function zeros in the range of
parameter~$z$ where only two phases from~$\RR$ prevail. Throughout
this section we will assume that Assumptions~A and~B are satisfied
and use $\kappa$ and $\tau$ to denote the constants from
Assumption~B. We will also use $\delta_L(z)$ for the function
defined in \eqref{deltaLz}.

\smallskip
The proof of Theorem~\ref{T:2ph} is  based directly on three
technical lemmas, namely, Lemma \ref{f--eq}--\ref{lemmag} below,
whose proofs are deferred to Sect.~\ref{sec-lemmas}. The general
strategy is as follows: First, by Lemma~\ref{f--eq}, we will know
that the solutions to \twoeqref{Req}{Imq} are within an $O(e^{-\tau
L})$-neighborhood from the solutions of similar equations, where
the functions~$\zeta_m$ get replaced by their analytic
counterparts~$\zeta_m^{(L)}$. Focusing on specific indices~$m$
and~$n$, we will write these analytic versions of
\twoeqref{Req}{Imq} as $f(z)=0$, where~$f$ is the function defined by
\begin{equation}
\label{fz}
f(z)=q_m\zeta_m^{(L)}(z)^{L^d}+q_n\zeta_n^{(L)}(z)^{L^d},
\qquad
z\in\CalS_{\kappa/L}(\{m,n\}).
\end{equation}
The crux of the proof of Theorem~\ref{T:2ph} is then to show that
the solutions of $f(z)=0$ are located within an appropriate distance
from the zeros of~$Z_L^\per(z)$. This will be achieved by invoking
Rouch\'e's Theorem for the functions~$f$ and $f+g$, where~$g$ is
defined by
\begin{equation}
\label{gz}
g(z)=Z_L^\per(z)-f(z),
\qquad z\in\CalS_{\kappa/L}(\{m,n\}).
\end{equation}
To apply Rouch\'e's Theorem, we will need that $|g(z)|<|f(z)|$ on
boundaries of certain discs in~$\CalS_{\kappa/L}(\{m,n\})$; this
assumption will be verified by combining Lemma~\ref{lemmaf} (a
lower bound on $|f(z)|$) with Lemma~\ref{lemmag} (an upper bound
on $|g(z)|$). The argument is then finished by applying
Lemma~\ref{f--eq} once again to conclude that any two distinct
solutions of the equations \twoeqref{Req}{Imq}, and thus also any
two distinct roots of $Z_L^\per$, are farther than a
uniformly-positive constant times~$L^{-d}$. The actual
proof follows a slightly different path than indicated here
in order to address certain technical details.

\smallskip
We begin by stating the aforementioned technical lemmas. The first
lemma provides the necessary control over the distance between the
solutions of \twoeqref{Req}{Imq} and those of the equation~$f(z)=0$.
The function~$f$ is analytic and it thus makes sense to consider
the multiplicity of the solutions. For that reason we will prefer
to talk about the roots of the function~$f$.

\begin{mylemma}
\label{f--eq}
There exist finite, positive constants $B_1$, $B_2$, $\tilde C_1$ and
$L_1$, satisfying the bounds $B_1<B_2$ and $\tilde C_1e^{-\tau L} < {B_1}L^{-d}$ whenever
$L\ge L_1$, such that for all $L\ge L_1$, all $s\leq (B_1+B_2)
L^{-d}$ and all $z_0\in\CalS_{\kappa/(2L)}(\{m,n\})$ with
$\D_{s}(z_0)\subset\OO$, the disc $\D_{s}(z_0)$ is a subset of
$\CalS_{\kappa/L}(\{m,n\})$ and the following statements hold:
\settowidth{\leftmargini}{(111)}
\begin{enumerate}
\item[(1)]
If $s\leq B_1L^{-d}$, then
disc $\D_{s}(z_0)$ contains at most one solution
of the equations \twoeqref{Req}{Imq} and at most one
root of function $f$, which is therefore non-degenerate.
\item[(2)] If $s\geq\tilde C_1e^{-\tau L}$ and if~$z_0$ is a
solution of the equations \twoeqref{Req}{Imq}, then $\D_{s}(z_0)$
contains at least one root of $f$.
\item[(3)] If $s\geq\tilde C_1e^{-\tau L}$ and if~$z_0$ is a root
of the function $f$, then $\D_{s}(z_0)$ contains at least one solution
of the equations \twoeqref{Req}{Imq}.
\item[(4)] If $s=B_2L^{-d}$ and if both $m$ and $n$ are stable at
$z_0$, then $\D_{s}(z_0)$ contains at least  one solution
of the equations \twoeqref{Req}{Imq}.
\end{enumerate}
\end{mylemma}

The next two lemmas state  bounds on $|f(z)|$ and $|g(z)|$
that will be needed to apply Rouch\'e's Theorem.
First we state a lower bound on $|f(z)|$:

\begin{mylemma}
\label{lemmaf}
There exist finite, positive constants $\tilde c_2$ and~$\tilde C_2$ obeying~$\tilde c_2\le\tilde C_2$ and, for any $\tilde C\geq \tilde C_2$ and any sequence~$(\epsilon_L)$
of positive numbers satisfying
\begin{equation}
\label{4.10aa} \lim_{L\to\infty}L^d\epsilon_L=0,
\end{equation}
there exists a constant~$L_2<\infty$ such that for all~$L\ge L_2$ the
following is true:
If~$z_0$ is a point in~$\CalS_{\kappa/(4L)}(\{m,n\})\cap(\CalS_m\cup\CalS_n)$
and ${\D_{\tilde C\epsilon_L}(z_0)}\subset\OO$, then there exists
a number $s(z_0)\in\{\tilde c_2\epsilon_L,\tilde C_2\epsilon_L\}$
such that ${\D_{s(z_0)}(z_0)}\subset \CalS_{\kappa/(2L)}(\{m,n\})$
and
\begin{equation}
\label{2phbd-a}
\liminf_{s\uparrow s(z_0)}\,\,\inf_{z\colon|z-z_0|=s}|f(z)|
>\epsilon_LL^d\zeta(z_0)^{L^d}.
\end{equation}
Moreover, if $f$ has a root in ${\D_{\tilde c_2\epsilon_L}(z_0)}$,
then $s(z_0)$ can be chosen as $s(z_0)=\tilde C\epsilon_L$.
\end{mylemma}

The reasons why we write a limit in \eqref{2phbd-a} will be seen
in the proof of Theorem~\ref{T:2ph}. At this point let us just say
that we need to use Lemma~\ref{lemmaf} for the maximal choice
$s(z_0)=\tilde C\epsilon_L$ in the cases when we know that
$\D_{\tilde C\epsilon_L}(z_0)\subset\OO$ but do not know the same
about the closure of $\D_{\tilde C\epsilon_L}(z_0)$. In light of
continuity of $z\mapsto|f(z)|$, once $s(z_0)<\tilde C\epsilon_L$,
the limit is totally superfluous.

\smallskip
Now we proceed to state a corresponding upper bound on $|g(z)|$:

\begin{mylemma}
\label{lemmag} There exists a constant $A_3\in(0,\infty)$ and, for
each $C\in(0,\infty)$ and any sequence~$\gammaL$ obeying the
assumptions \eqref{gLass}, there exists a number $L_3<\infty$ such
that
\begin{equation}
\label{gbd}
\sup_{z\colon |z-z_0|<
C\delta_L(z_0) }|g(z)|\le A_3\delta_L(z_0)L^d\zeta(z_0)^{L^d}
\end{equation}
holds for any $L\ge L_3$ and any $z_0\in\UU_\gammaL$ with
$\D_{C\delta_L(z_0)}(z_0)\subset\OO$.
\end{mylemma}

With Lemmas~\ref{lemmaf}--\ref{lemmag} in hand,
the proof of Theorem~\ref{T:2ph} is rather straightforward.

\begin{proofsect}{Proof of Theorem~\ref{T:2ph}}
Let~$m$ and~$n$ be distinct indices from $\RR$ and let us abbreviate
$\UU_{\gammaL}=\UU_{\gammaL}(\{m,n\})$ and
$\CalS_\epsilon=\CalS_\epsilon(\{m,n\})$. Let $f(z)$ and~$g(z)$ be
the functions from \twoeqref{fz}{gz}. 
Let~$B_1$,~$B_2$, $\tilde C_1$,~$\tilde c_2$, 
$\tilde C_2$ and $A_3$ be the
constants whose
existence is guaranteed by Lemmas~\ref{f--eq}-\ref{lemmag} and
let~$L_1$ be as in Lemma~\ref{f--eq}. Since~$A_3$ appears on the
right-hand side of an upper bound, without loss of generality  we
can assume that
\begin{equation}
\label{A3constraint}
\tilde c_2A_3\ge\tilde C_1.
\end{equation}
Further, let us choose the constants~$C$ and $D$ such that
\begin{equation}
C=\tilde C_1+\tilde C_2A_3, \quad \text{and} \quad D=B_1+B_2.
\end{equation}
Next, let~$L_2$ be the constant for which Lemma~\ref{lemmaf} holds
for both $\tilde C=\tilde C_2$ and $\tilde C=C/A_3$ and for
both~$\epsilon_L=A_3 e^{-\tau L}$
and~$\epsilon_L=A_3L^de^{-\frac12 \gammaL L^d}$. Finally,
let~$L_3$ be the constant for which Lemma~\ref{lemmag} holds
with~$C$ as defined above.

The statement of Theorem~\ref{T:2ph} involves two additional
constants chosen as follows: First, a constant~$B$ for which we
pick a number from $(0,\tfrac2{\sqrt3}B_1)$ (e.g, $B_1/3$ will
do). Second, a constant~$L_0$ which we choose such that
$L_0\ge\max\{L_1,L_2,L_3\}$ and that the bounds
\begin{equation}
\label{C<B_1}
\gammaL\le\frac\kappa{4L},
\quad
e^{-\tau L}\le L^de^{-\frac12\gammaL L^d},
\quad
CL^de^{-\frac12\gammaL L^d}
+\tilde C_1e^{- \tau L}\le \frac{\sqrt3-1}2BL^{-d}
\end{equation}
hold true for all $L\ge L_0$. Fix $L\ge L_0$ and consider the set
\begin{equation}
\UU=\bigl\{z_0\in\UU_{\gamma_L}
\colon\D_{C\delta_L(z_0)}(z_0)\subset\OO\bigr\}.
\end{equation}
Notice that our choice of~$L_0$ guarantees that
$\UU\subset\UU_{\gamma_L}
\subset\CalS_{\kappa/(4L)}\cap(\CalS_m\cup\CalS_n)$,
while the fact that~$\tilde C\le C/A_3$ for both choices of
$\tilde C$ above ensures that for any $z_0\in\UU$, the disc
$\D_{\tilde CA_3\delta(z_0)}(z_0)$ is contained in~$\OO$. These
observations verify the assumptions of Lemma~\ref{lemmaf}---with
$\epsilon_L=A_3\delta_L(z_0)$ and~$\tilde C$ equal to
either~$\tilde C_2$ or~$C/A_3$---as well as of Lemma~\ref{lemmag},
for any~$z_0\in\UU$.

First, we will attend to the proof of claim~(2).
Let $z_0\in\Omega_L^\star\cap\UU$ be a root of
$Z_L^\per=f+g$.
Lemma~\ref{lemmaf} with $\tilde C=\tilde C_2$ and
$\epsilon_L=
A_3\delta_L(z_0)$ and Lemma~\ref{lemmag}
then imply the existence of a radius
$s(z_0)$
with $s(z_0)\le\tilde C_2 \epsilon_L=
\tilde C_2A_3\delta_L(z_0)< C\delta_L(z_0)$ such that
\begin{equation}
\label{f>g}
\bigl|f(z)\bigr|>\bigl|g(z)\bigr|,\qquad z\in\partial\D_{s}(z_0)
\end{equation}
holds for $s=s(z_0)$.
(Note that here the limit in \eqref{2phbd-a}
can be omitted.)
Hence, by Rouch\'e's Theorem,~$f$ and~$f+g$ have an equal number
of roots in~$\D_{s(z_0)}(z_0)$, including multiplicity. In
particular, the function $f$ has a root $z_1$ in $\D_{s(z_0)}(z_0)$
which by
Lemma~\ref{lemmaf} lies also in $\CalS_{\kappa/(2L)}$.
Since $s(z_0) +  \tilde C_1 e^{-\tau L}\le C\delta_L(z_0)$
by the definition of $C$ and the second bound in \eqref{C<B_1},
we may use Lemma \ref{f--eq}(3) to infer that
the equations
\twoeqref{Req}{Imq} have a solution
$z\in\D_{\tilde C_1 e^{-\tau L}}(z_1)\subset\D_{C\delta_L(z_0)}(z_0)$.
Moreover,
\eqref{C<B_1} implies that
$C\delta_L(z_0)\le B_1 L^{-d}$ so  by Lemma~\ref{f--eq}(1)
there is only one such
solution in the entire disc $\D_{C\delta_L(z_0)}(z_0)$.

Next, we will prove claim~(3).
Let~$z_0\in\Omega_L(\QQ)\cap\UU$ be
a solution to the equations \twoeqref{Req}{Imq}. By Lemma~\ref{f--eq}(2), there
exists a root 
$z_1\in\D_{\tilde C_1e^{-\tau L}}(z_0)
\subset\D_{C\delta_L(z_0)}(z_0)$ of
the function~$f$.
Lemma~\ref{f--eq}(1) then shows that~$z_1$ is in fact the only
root of~$f$ in $\D_{C\delta_L(z_0)}(z_0)$. Applying
Lemma~\ref{lemmaf} for the point~$z_0$ and the choices
$\epsilon_L=A_3\delta_L(z_0)$ and $\tilde C=C/A_3$ in conjunction
with Lemma~\ref{lemmag}, there exists a radius $s(z_0)$ such that
\eqref{f>g} holds true for any $s<s(z_0)$ sufficiently
near~$s(z_0)$. Moreover, by the bound
\eqref{A3constraint} we know that $z_1\in\D_{\tilde C_1 e^{-\tau
L}}(z_0)\subset \D_{\tilde c_2 \epsilon_L}(z_0)$ is a root of~$f$
within distance $\tilde c_2\epsilon_L$ from~$z_0$, and so the last
clause of Lemma~\ref{lemmaf} allows us to choose
$s(z_0)=C\delta_L(z_0)$. Let $s_0<s(z_0)$ be such that \eqref{f>g}
holds for $s\in(s_0,s(z_0))$ and pick an $s\in(s_0,s(z_0))$.
Rouch\'e's Theorem for the discs $\D_{s}(z_0)$ and the fact
that~$f$ has only one root in~$\D_{C\delta_L(z_0)}(z_0)$ imply the
existence of a unique zero~$z$ of $f(z)+g(z)=Z_L^\per(z)$
in~$\D_{s}(z_0)$. The proof is finished by taking the
limit~$s\uparrow C\delta_L(z_0)$.

Further, we will pass to claim (4).  Let~$z_1$ and~$z_2$ be two
distinct roots of $Z_L^\per$ in $\UU_{\gammaL}$ such that both
$\D_{BL^{-d}}(z_1)\subset\OO$ and $\D_{BL^{-d}}(z_2)\subset\OO$
are satisfied. We will suppose that $|z_1-z_2|<BL^{-d}$ and derive
a contradiction. Let $z=\frac12(z_1+z_2)$ be the middle point of
the segment between~$z_1$ and~$z_2$. Since $|z_1-z_2|<BL^{-d}$, a
simple geometrical argument shows that the disc of radius
$s=\tfrac{\sqrt3}2BL^{-d}$ centered at~$z$ is entirely contained
in $\D_{BL^{-d}}(z_1)\cup\D_{BL^{-d}}(z_2)\subset\OO$. Next, by
Lemmas~\ref{lemmaf}-\ref{lemmag}, there exist two roots~$z_1'$
and~$z_2'$ of~$f$ such that $z_1'\in\D_{C\delta(z_1)}(z_1)$ and
$z_2'\in\D_{C\delta(z_2)}(z_2)$. (We may have that $z_1=z_2$, in
which case $z_1=z_2$ would be a degenerate root of~$f$.) Now our
assumptions on~$B$ and~$L_0$ imply that
\begin{equation}
\frac{\sqrt3}2BL^{-d}\ge \frac B2L^{-d}+C\delta_L(z_1)
\ge|z-z_1|+|z_1-z_1'|\ge|z-z_1'|,
\end{equation}
and similarly for~$z_2'$. Consequently, both~$z_1'$ and~$z_2'$ lie
in~$\D_s(z)$. But this contradicts Lemma~\ref{f--eq} and the bound
$\tfrac{\sqrt3}2B<B_1$, implying that $\D_s(z_0)$ contains at most one
non-degenerate root of~$f$.

Finally, we will prove claim~(1).
Let $z_0\in\GG\cap\UU_\gammaL(\QQ)$ with $\D_{DL^{-d}}(z)\subset\OO$.
According to Lemma~\ref{f--eq}(4), the disc~$\D_{B_2L^{-d}}(z)$
contains at least one one solution $z_1$
of the equations \twoeqref{Req}{Imq}. Checking that
$B_2L^{-d}+C\delta_L(z_1)\le (B_2+B_1)L^{-d}$
in view of \eqref{C<B_1} and the definition of $B$,
we know that $\D_{C\delta(z_1)}(z_1)\subset\OO$ and we can use
already proven claim (3) to get the existence of a root of $Z_L^\per$ in
$\D_{C\delta_L(z_1)}(z_1)$, which is a subset of~$\D_{DL^{-d}}(z_0)$.
\end{proofsect}

This concludes the proof of Theorem~\ref{T:2ph} subject to the
validity of Lemmas~\ref{f--eq}-\ref{lemmag}. The proofs of these
lemmas have been deferred to Sect.~\ref{sec-lemmas}.

\subsection{Proof of Proposition~\ref{cor2.1}}
Fix distinct indices $m,n\in\RR$. Our strategy is to first prove
the claim for the density of the solutions of the equations
\twoeqref{Req}{Imq},
\begin{equation}
\label{til-adens}
\tilde\rho_{m,n}^{(L,\epsilon)}(z)=\frac1{2\epsilon
L^d}\bigl|\Omega_L(\{m,n\})\cap\D_\epsilon(z)\bigr|,
\end{equation}
and then to argue that the density $\rho_{m,n}^{(L,\epsilon)}$ yields
the same limit.

Let $z_0\in\GG(\{m,n\})\setminus\MM$, where $\MM$ is the set of
all multiple points.
By Theorem~\ref{T:PD} and Assumptions~A1-A2,
there exists an $\epsilon>0$ such that, throughout the disc
$\D_\epsilon=\D_\epsilon(z_0)\subset \OO$, we have
$Q(z)\subset\{m,n\}$ and the the function
$F_{m,n}(z)=\zeta_m(z)/\zeta_n(z)$ is twice continuously
differentiable and nonvanishing.  Clearly, all solutions of
the equations \twoeqref{Req}{Imq} in $\D_\epsilon$
must lie in the set 
\begin{equation}
\GG^{(L)}=\bigl\{z\in\D_\epsilon\colon
|F_{m,n}(z)|=(q_n/q_m)^{1/L^d}\bigr\}.
\end{equation}
Denoting the set
$\GG(\{m,n\}\cap\D_\epsilon$ by $\GG^{(\infty)}$, we now
claim that for sufficiently
small $\epsilon$, the sets $\GG^{(\infty)}$
and $\GG^{(L)}$ can be viewed as
differentiable parametric curves $\gamma\colon(t_-,t_+)\to \D_\epsilon$
and $\gamma^{(L)}\colon  (t_-^{(L)},t_+^{(L)})\to \D_\epsilon$
for which
\settowidth{\leftmargini}{(1111)}
\begin{enumerate}
\item[(1)]
$t_-^{(L)}\to t_-$ and $t_+^{(L)}\to t_+$
\item[(2)]
$\gamma^{(L)}\to\gamma$ uniformly on  $\in(t_-,t_+)$
\item[(3)]
$\hatv_L\to\hatv$ uniformly on   $(t_-,t_+)$
\end{enumerate}
hold true as $L\to\infty$.
Here $\hatv_L(t)=\tfrac \textd{\textd t}\gamma^{(L)}(t)$
and $\hatv(t)=\tfrac \textd{\textd t}\gamma(t)$ denote
the tangent~vectors to~$\gamma^{(L)}$ and~$\gamma$, respectively.

We will construct both curves as solutions to the differential equation
\begin{align}
\label{diff-equ}
\frac{\textd z(t)}{\textd t} =
\texti\frac{\partial_z \phi_{m,n}(z(t))}{|\partial_z \phi_{m,n}(z(t))|}
\end{align}
with $\phi_{m,n}(z)=\log|F_{m,n}(z)|$ (note that for $\epsilon$
small enough, the right hand side is a well defined, continuously
differentiable function of $z(t)\in\D_\epsilon$ by
Assumptions~A1-A2 and the fact that $|\partial_z
\phi_{m,n}(z_0)|\geq \alpha/2$ according to Assumption~A3).
In order to define the curves $\gamma^{(L)}(\cdot)$ and $\gamma(\cdot)$
we will choose a suitable starting point at $t=0$.
For $\gamma(\cdot)$, this will just be the point $z_0$, while for
$\gamma^{(L)}(\cdot)$ we will choose a point $z_0^{(L)}\in\D_\epsilon$
which obeys the conditions $\phi_{m,n}(z_0^{(L)})=\eta_L$
and $|z_0-z_0^{(L)}|\leq 3\alpha^{-1} \eta_L $, where
$\eta_L=L^{-d}\log(q_n/q_m)$.
To construct the point $z_0^{(L)}\in\D_\epsilon$,
we use again the smoothness of $\phi_{m,n}$.
Namely, by Assumption~A1-2, the function
$\phi_{m,n}(x+\texti y)=\log|F_{m,n}(x+\texti y)|$ is twice continuously
differentiable on $\D_\epsilon$ if $\epsilon$ is sufficiently
small, and by Assumption~A3 we either have
$|\partial\phi_{m,n}(x+\texti y)/\partial x|\geq\alpha/3$, or
$|\partial\phi_{m,n}(x+\texti y)/\partial y|\geq\alpha/3$.  Assuming,
without loss of generality, that
$|\partial\phi_{m,n}(x+\texti y)/\partial y|\geq\alpha/3$ on all of
$\D_\epsilon$, we then define $z_0^{(L)}$ to be the unique
point for which $\RE z_0^{(L)}=\RE z_0$ and
$\phi_{m,n}(z_0^{(L)})=\eta_L$.  By the assumption
$|\partial\phi_{m,n}(x+\texti y)/\partial y|\geq\alpha/3$,
we then have $|z_0-z_0^{(L)}|\leq 3\alpha^{-1} \eta_L$,
as desired.

Having chosen $z_0^{(L)}$,
the desired curves $\gamma^{(L)}\colon (t_-^{(L)},t_+^{(L)})\to
\D_\epsilon$
and $\gamma\colon (t_-,t_+)\to \D_\epsilon$
are obtained as the solutions of the equation
\eqref{diff-equ} with initial condition $\gamma^{(L)}(0)=z_0^{(L)}$
and $\gamma(0)=z_0$, respectively. Here $t_-^{(L)}$, $t_+^{(L)}$,
$t_-$, and $t_+$ are determined by the condition that $t_-^{(L)}$
and $t_-$ are the largest values $t<0$ for which
$\gamma^{(L)}(t)\in\partial\D_\epsilon$ and
$\gamma(t)\in\partial\D_\epsilon$, respectively, and $t_+^{(L)}$
and $t_+$ are the smallest values $t>0$ for which
$\gamma^{(L)}(t)\in\partial\D_\epsilon$ and
$\gamma(t)\in\partial\D_\epsilon$, respectively.
Since the right-hand side of \eqref{diff-equ} has modulus one,
both curves are parametrized by the arc-length.
Moreover, decreasing~$\epsilon$ if necessary, the functions
$\gamma^{(L)}$ can be
extended to all $t\in(t_-,t_+)$.
To see that the limits in (1-3) above hold, we just refer
to the Lipschitz continuity of the right hand side of \eqref{diff-equ}
and the fact that, by definition,
$|\gamma^{(L)}(0)-\gamma(0)|=O(L^{-d})$.
Let $K$ be the Lipschitz constant of the right-hand side of
\eqref{diff-equ}
in a neighborhood containing $\gamma^{(L)}(t)$ for all $t\in(t_-,t_+)$.
Choosing~$\epsilon$ so small that both $t_+-t_-$ and
$t^{(L)}_+-t^{(L)}_-$
are less than, say, $1/(2K)$, integrating \eqref{diff-equ} and invoking
the Lipschitz continuity, we~get
\begin{equation}
\sup_{t_-<t<t_+}|\gamma^{(L)}(t)-\gamma(t)|
\leq |\gamma^{(L)}(0)-\gamma(0)|
+\tfrac12\sup_{t_-<t<t_+}|\gamma^{(L)}(t)-\gamma(t)|.
\end{equation}
This shows that $\gamma^{(L)}(t)\to\gamma(t)$ uniformly in
$t\in(t_-,t_+)$.
Using Lipschitz continuity once more, we get a similar bound on the
derivatives.
But then also the arc-lengths corresponding to $\gamma^{(L)}$ must
converge to the arc-length of $\gamma$, which shows that also
$t_+^{(L)}\to t_+$ and $t_-^{(L)}\to t_-$.

Consider now the curve $\gamma(t)$.
Given that $|F_{m,n}(z)|$ is constant along $\gamma$, we have
\begin{equation}
\label{ARG}
\frac{\textd\Arg F_{m,n}(\gamma(t))}{\textd t}
=\frac1\texti\frac{\textd\log F_{m,n}(\gamma(t))}{\textd t}
=-\texti  \partial_z  \log F_{m,n}(z)\bigr|_{z=\gamma(t)}
\hatv(t).
\end{equation}
Referring to Assumption~A3 and the fact that $|\hatv(t)|=1$, we find
that the modulus of the left-hand side is bounded below by $\alpha$.
Using continuity of the derivative
$\tfrac\textd{\textd t}\Arg F_{m,n}$ in $\D_\epsilon$,
we observe that one of the two
alternatives occurs on all the interval  $(t_-^{(L)},t_+^{(L)})$:
\begin{equation}
\label{alternative}
\text{either } \qquad
\frac{\textd\Arg F_{m,n}(\gamma^{(L)}(t))}{\textd t}\ge \frac\alpha 2
\qquad\text{ or}\qquad
\frac{\textd\Arg F_{m,n}(\gamma^{(L)}(t))}{\textd t}\le -\frac\alpha 2,
\end{equation}
provided $\epsilon$ is sufficiently small.
By Lemma~\ref{f--eq}, the disc $\D_\epsilon$ contains a finite number
$k=2\epsilon L^d \tilde\rho_{m,n}^{(L,\epsilon)}(z_0)$
of solutions of the equations
\eqref{Req} and \eqref{Imq} which in the present notation read
\begin{align}
{}&|F_{m,n}(z)|=\Bigl(\frac{q_n}{q_m}\Bigr)^{1/L^d},\\
{}&L^d\Arg F_{m,n}(z)= \pi\operatorname{ mod }2\pi.
\end{align}
Assuming, without loss of generality, that the former alternative
in \eqref{alternative} takes place, and ordering all the solutions
consecutively along the curve $\gamma^{(L)}$, i.e., letting
$z_1=\gamma^{(L)}(t_1)$,\dots, $z_k=\gamma^{(L)}(t_k)$
where $t_-^{(L)}\le t_1<\dots<t_k\le t_+^{(L)}$, we have
\begin{equation}
\label{F_{m,n}_j}
\Arg F_{m,n}(z_{j+1})-\Arg F_{m,n}(z_j)=2\pi L^{-d}
\end{equation}
for any $j=1,\dots,k-1$, as well as
\begin{equation}
\Arg F_{m,n}(z_{1})-\Arg F_{m,n}(z_-)\le 2\pi L^{-d}
\end{equation}
and
\begin{equation}
\Arg F_{m,n}(z_{+})-\Arg
F_{m,n}(z_k)\le 2\pi L^{-d}.
\end{equation}
In view of the first equality in \eqref{ARG} rephrased for
$\gamma^{(L)}$, the left hand side of
\eqref{F_{m,n}_j} can be rewritten~as
\begin{equation}
\Arg F_{m,n}(z_{j+1})-\Arg F_{m,n}(z_j)=
\int_{t_j}^{t_{j+1}}
\Bigl|\frac{\textd  \log F_{m,n}(\gamma^{(L)}(t))}{\textd t}\Bigr|
\textd t
\end{equation}
and thus
\begin{equation}
\label{dzk}
\biggl| \int_{t_-^{(L)}}^{t_+^{(L)}}
\Bigl|\frac{\textd \log F_{m,n}(\gamma^{(L)}(t))}{\textd t}\Bigr|
\textd t-2k\pi L^{-d}\biggr|
\le
2\pi L^{-d}.
\end{equation}
Let us divide the whole expression by~$L^d$ and take the limit
$L\to\infty$.
Now~$\gamma^{(L)}$ converge to~$\gamma$ along with their first
derivatives,
uniformly in $t\in(t_-,t_+)$,
and the limits $t_\pm^{(L)}$ converge to $t_\pm$.
The Bounded Convergence Theorem then shows that the integral in
\eqref{dzk}
converges to a corresponding integral over $\gamma$.  Recalling that
$\tilde\rho_{m,n}^{(L,\epsilon)}(z_0)= k/({2\epsilon L^d})$, we thus get
\begin{equation}
\label{integral}
\begin{aligned}
\lim_{L\to\infty}\tilde\rho_{m,n}^{(L,\epsilon)}(z_0)
&=\frac 1{4\pi\epsilon}
\int_{t_{-}}^{t_{+}}
\Bigl|\frac{\textd  \log F_{m,n}(\gamma_0(t))}{\textd t}\Bigr| \textd t
\\
&=\frac 1{4\pi\epsilon} \int_{\gamma_0}\bigl|\partial_z
\log F_{m,n}(z)
\bigr| |d z|
\end{aligned}
\end{equation}
where the last integral denotes the integration with respect to the
arc length.
Taking into
account the Lipschitz continuity of $|\partial_z \log F_{m,n}(z)|$,
the last integral in \eqref{integral} can be
approximated by $\bigl(\bigl|\partial_z  \log F_{m,n}(z_0)\bigr| +
O(\epsilon)\bigr) |\gamma| $.  By the smoothness of the curve
$\gamma$, we estimate its
length by $|\gamma|=2\epsilon\bigl(1 + O(\epsilon)\bigr)$, so that
\begin{equation}
\label{limitatildytrho}
\lim_{\epsilon\downarrow 0}\lim_{L\to\infty}
\tilde\rho_{m,n}^{(L,\epsilon)}(z_0)
=\frac{1}{2\pi} \bigl|\partial_z  \log F_{m,n}(z_0)\bigr|
=\frac{1}{2\pi} \Bigl|\frac{\partial_z\zeta_m(z_0)}{\zeta_m(z_0)}
-\frac{\partial_z\zeta_n(z_0)}{\zeta_n(z_0)}\Bigr|.
\end{equation}

To finish the proof, we need to show that
$\rho_{m,n}^{(L,\epsilon)}(z_0)$
will converge to the same limit.
According to Theorem~\ref{T:2ph}, we have
\begin{equation}
\bigl||\Omega_L^*\cap\D_\epsilon(z)|-
|\Omega_L(\{m,n\})\cap\D_\epsilon(z)|\bigr|\le2
\end{equation}
for all $z\in\GG(m,n)$ such that~$|\QQ(z)|=2$ and $\epsilon$
sufficiently small. Hence
\begin{equation}
\bigl| \rho_{m,n}^{(L,\epsilon)}(z)
-\tilde \rho_{m,n}^{(L,\epsilon)}(z)\bigr|\le
\frac1{\epsilon L^{d}},
\end{equation}
and the claim of the proposition  follows by \eqref{limitatildytrho}.
\qed


\subsection{Multiple phase coexistence}
\label{sec4.3}\noindent
In this section we will prove
Theorem~\ref{T:Mph}, which deals with the zeros of~$Z_L^\per$ in
the vicinity of multiple points.
Let $\zM\in\OO$ be a multiple point and let
$\QQ=\QQ(\zM)$.
For each~$m\in\QQ$, let~$\phi_m(L)$
and~$v_m$ be as in \eqref{phi-v}.
Define the functions
\begin{equation}
\label{445}
\tilde f(z)=\sum_{m\in\QQ}q_m\,e^{\texti\phi_m(L)+v_m (z-\zM)L^d},
\end{equation}
\begin{equation}
\tilde g(z)=Z_L^\per(z)\zeta(\zM)^{-L^d}-f(z),
\end{equation}
and
\begin{equation}
\label{447}
\xi(z)=\exp\bigl\{\max_{m\in\QQ}\RE(v_m(z-\zM))\bigr\}.
\end{equation}
As in the case of two-phase
coexistence, the proof uses Rouch\'e's Theorem for the functions
$\tilde f$ and $\tilde f+\tilde g$.
For this we will need a lower bound on~$|\tilde f|$
and an upper bound on~$|\tilde g|$.

\begin{mylemma}
\label{lemma4.4} Suppose Assumptions~A and~B hold. Given
$\QQ\subset \RR$ with $|\QQ|\ge 3$ and abbreviating~$q=|\QQ|$ and
$R_L=L^{-d(1+1/q)}$, let~$(\epsilon_L)$ be a sequence of positive
numbers such that
\begin{equation}
\label{epscond}
\lim_{L\to\infty}L^{2d}\epsilon_L=\infty\qquad\text{but}\qquad
\lim_{L\to\infty}L^{2d-d/q}\epsilon_L=0.
\end{equation}
Then there is a constant~$L_5<\infty$ such that for
any~$z_0\in\C$ and any~$L\ge L_5$ there
exists~$s(z_0)\in[R_L/q,R_L]$ for which the bound
\begin{equation}
\label{Mphbd}
\inf_{z\colon|z-z_0|=s(z_0)}\,
\bigl|f(z)\bigr|>L^d\epsilon_L\, \xi(z_0)^{L^d}
\end{equation}
holds.
\end{mylemma}

\begin{mylemma}
\label{lemma4.10} Let $\zM\in\OO$ be a multiple point, let
$\QQ=\QQ(\zM)$, $q=|\QQ|$, and $R_L=L^{-d(1+1/q)}$. There exists a
constant
${A_6}\in(0,\infty)$ and, for each sequence~$(\rho_L)$ of
positive numbers obeying \eqref{rholim}, a number~$L_6<\infty$
such that if $L\ge L_6$ then
$\D_{\rho_L'}(\zM)\subset\UU_{\kappa/L}(\QQ)$, where
$\rho_L'=\rho_L+R_L$.  Furthermore, we have
\begin{equation}
\label{4.50}
\sup_{z\colon|z-z_0|\leq R_L}\bigl|\tilde g(z)\bigr|
\le {A_6}\rho_L^2 L^d{\xi}(z_0)^{L^d}
\end{equation}
whenever~$z_0\in\D_{\rho_L}(\zM)$.
\end{mylemma}

With these two lemmas we can  proceed directly to the proof of
Theorem~\ref{T:Mph}.

\begin{proofsect}{Proof of Theorem~\ref{T:Mph}}
The proof is close in spirit to the proof of Theorem~\ref{T:2ph}.
Let~$\zM$ be a multiple point and let~$\QQ=\QQ(\zM)$. Consider a
sequence~$(\rho_L)$ of positive numbers such that \eqref{rholim}
holds. Choosing~$\epsilon_L={A_6}\rho_L^2$, where ${A_6}$ is the
constant from Lemma~\ref{lemma4.10}, we note that the conditions
\eqref{epscond} are satisfied due to our conditions on~$\rho_L$
from \eqref{rholim}. We will then prove Theorem~\ref{T:Mph} with
$L_0=\max\{L_5,L_6\}$, where $L_5$ and $L_6$ are the constants
from Lemma~\ref{lemma4.4} and~\ref{lemma4.10}, respectively. The
proof again boils down to a straightforward application of
Rouch\'e's Theorem.

Indeed, let $L\ge L_0$ and note that by Lemmas~\ref{lemma4.4}
and~\ref{lemma4.10}, for each~$z_0\in\D_{\rho_L}(\zM)$ there is
an~$s(z_0)\in[R_L/q,R_L]$ such~that on $\D_{s(z_0)}(z_0)$, we have
\begin{equation}
\label{4.115}
\bigl|\tilde f(z)\bigr|>\bigl|\tilde g(z)\bigr|.
\end{equation}
Consider the set of these discs $\D_{s(z_0)}(z_0)$---one for
every~$z_0\in\D_{\rho_L}(\zM)$. These discs cover the closure of
$\D_{\rho_L}(\zM)$, so we can choose a finite subcover~$\eusm S$.
Next we note that \eqref{4.115} implies that neither $\tilde f$
nor $\tilde f+\tilde g$ have more than finitely many zeros in
$\D_{\rho_L}(\zM)$ (otherwise, one of these functions would be
identically zero). Without loss of generality, we can thus assume
that the discs centered at the zeros of $\tilde f$ and $\tilde
f+\tilde g$ in $\D_{\rho_L}(\zM)$ are included in $\eusm S$.
Defining $\UU=\bigcup_{\D\in\eusm S}\D$, we clearly have
$\D_{\rho_L}(\zM)\subset\UU\subset\D_{\rho_L'}(\zM)$.

Let now $\eusm K$ be the set of all components of
$\UU\setminus\bigcup_{\D\in\eusm S}\partial\D$. Let $\KK\in\eusm
K$ be one such component. By \eqref{4.115} we know that $|\tilde
f(z)|>|\tilde g(z)|$ on the boundary of $\KK$ and Rouch\'e's
Theorem then guarantees that~$\tilde f$ has as many zeros in~$\KK$
as~$\tilde f+\tilde g$, provided we count multiplicity correctly.
Moreover, both functions $\tilde f$ or $\tilde f+\tilde g$ have no
zeros on $\bigcup_{\D\in\eusm S}\partial\D$. Since~$\tilde
f(z)+\tilde g(z) =Z_L^\per(z)\zeta(\zM)^{-L^d}$
and~$\zeta(\zM)^{-L^d}>0$, the zeros of~$\tilde f+\tilde g$ are
exactly those of~$Z_L^\per$. The above construction of $\UU$ and
$\eusm S$ then directly implies the desired correspondence of the
zeros. Namely, in each $\KK\in\eusm K$, both $\tilde f$ and
$Z_L^\per$ have the same (finite) number of zeros, which can
therefore be assigned to each other. Now $\tilde f$ and $Z_L^\per$
have no zeros in $\UU\setminus\bigcup_{\KK\in\eusm K}\KK$, so
choosing one such assignment in each $\KK\in\eusm K$ extends into
a one-to-one assignment of $\Omega_L^\star\cap\UU$ and
$\Omega_L(\QQ)\cap\UU$. Moreover, if $z\in\Omega_L^\star\cap\KK$ and
$\tilde z\in\Omega_L(\QQ)\cap\KK$ for some $\KK\in\eusm K$ (which
is required if $z$ and $\tilde z$ are the corresponding roots),
then $z$ belongs to the disc $\widetilde\D\in\eusm S$ centered at
$\tilde z$ and $\tilde z$ belongs to the disc $\D\in\eusm S$
centered at~$z$. Consequently, $z$ and $\tilde z$ are not farther
apart than $R_L=L^{-d(1+1/q)}$. This completes the proof.
\end{proofsect}

\subsection{Proof of Proposition~\ref{prop2.6}}
\label{sec4.3a}\noindent
Assuming that $L^{-d}\omega_L\le \gamma_L$, it clearly suffices to
ascertain that
\begin{equation}
\label{staci}
\bigcup_{\QQ\colon |\QQ|\ge 3}\CalS_{\gamma_L}(\QQ)\cap\DD
\subset\bigcup_{ \zM\in\DD\cap\MM}\D_{\rho_L}(\zM).
\end{equation}
To this end let us first observe that continuity of the functions $\zeta_m$ implies
\begin{equation}
\label{limS}
\lim_{L\to\infty}\CalS_{\gamma_L}(\QQ)=\bigcap_{m\in \QQ}\CalS_m
\end{equation}
since $\gamma_L\to 0$.  The set $\DD\cap\MM$ is finite  according
to Theorem~\ref{T:PD}. Hence, there exists a constant $\delta_0>0$
and, for each $\delta\in(0,\delta_0]$, a constant
$L_0=L_0(\delta)$, such that the discs $\D_{\delta}(\zM)$,
$\zM\in\DD\cap\MM$, are mutually disjoint,
\begin{equation}
\QQ(z)\subset \QQ(\zM) \qquad\text{whenever}\qquad z\in \D_{\delta}(\zM),
\end{equation}
and
\begin{equation}
\label{tezstaci}
\bigcup_{\QQ\colon |\QQ|\ge 3}  \CalS_{\gamma_L}(\QQ)\cap\DD
\subset
\bigcup_{\zM\in\DD\cap\MM}\D_{\delta}(\zM)
\end{equation}
whenever $0<\delta\leq\delta_0$ and $L\ge L_0(\delta)$. It is
therefore enough to show that there exist constants $\chi>0$ and
$\delta\in(0,\delta_0)$ such that for any multiple point
$\zM\in\DD$, we have
\begin{equation}
\label{jestestaci}
\D_{\delta}(\zM)\cap\CalS_{\gamma_L}(\QQ(\zM))
\subset\D_{\rho_L}(\zM)
\end{equation}
once $\rho_L\ge \chi \gamma_L$ and $L\geq L_0(\delta)$.

We will prove \eqref{jestestaci} in two steps: First we will show
that there is a constant $\chi>0$ such that for any multiple
point $\zM$, any $z\neq\zM$, and any $n\in\QQ(\zM)$, there exists
$m\in\QQ(\zM)$  for which
\begin{equation}
\label{uhel}
\RE\bigl[(z-\zM)(v_n(\zM)-v_m(\zM))\bigr]\ge 2\chi  |z-\zM|,
\end{equation}
and then we will show that \eqref{uhel} implies
\eqref{jestestaci}. To prove \eqref{uhel}, we first refer to the
fact that we are dealing with a finite number of strictly convex
polygons with vertices $\{v_k(\zM)\colon k\in\QQ(\zM)\}$ according
to Assumption~A4 and thus, given $z$ and $n$, the label $m$ can be
always chosen so that the angle between  the complex numbers
$z-\zM$ and $v_n(\zM)-v_m(\zM)$ is not smaller than a given fixed
value. Combining this fact with the lower bound from
Assumption~A3, we get~\eqref{uhel}.

We are left with the proof of \eqref{jestestaci}. Let us thus
consider a multiple point $\zM\in\DD$ with $\QQ(\zM)=\QQ$, and a
point $z\in \D_{\delta}(\zM)\setminus\D_{\rho_L}(\zM)$. We will
have to show that there exists an  $m\in\QQ$ with $z\notin
\CalS_{\gammaL}(m)$.  Recalling that $\QQ(z')\subset\QQ$ for all
$z'\in\D_{\delta}(\zM)$, let $n\in\QQ$ be such that
$|\zeta_n(z)|=\zeta(z)$. Choosing   $m\in\QQ(\zM)$ so that
\eqref{uhel} is satisfied  and using, as in the proof of
Lemma~\ref{lemma3.1}, $F_{n,m}(z)$ to denote the function
$F_{n,m}(z)=\zeta_n(z)/\zeta_m(z)$, we apply, as in \eqref{Tay},
the Taylor expansion to $\log|F_{n,m}(z)|$ to get
\begin{multline}
\qquad
\log|F_{n,m}(z)|=\RE\bigl[(z-\zM)(v_n(\zM)-v_m(\zM))\bigr]
+O(|z-\zM|^2)
\\\ge \chi  |z-\zM|\geq \chi\rho_L.
\qquad
\end{multline}
Here we also used that $|F_{n,m}(\zM)|=1$ and assumed that $\delta$
was chosen small enough to guarantee that the error term is smaller
than $\chi  |z-\zM|$. As a result, we get
\begin{equation}
\label{}
|\zeta_m(z)|\le e^{-\chi \rho_L}\zeta(z)\le e^{-\gamma_L}\zeta (z)
\end{equation}
implying that $z\not\in\CalS_{\gamma_L}(m)$. Thus, the inclusion
\eqref{jestestaci} is verified and \eqref{staci} follows. \qed

\section{Technical lemmas}
\label{sec5}\noindent The goal of this section is to provide the proofs of
Lemmas~\ref{f--eq}-\ref{lemma4.10}. We will begin with some
preparatory statements concerning Lipschitz continuity of
the~$\zeta_m$ and~$\zeta$.

\subsection{Lipschitz properties of the functions 
$\log|\zeta_m|$ and $\log\zeta$}
\label{sec4.aux}\noindent
In this section, we prove two auxiliary lemmas needed for the
proofs of our main theorems.
For any $z_1,z_2\in\C$, we will use $[z_1,z_2]$ to denote the closed
segment
\begin{equation}
[z_1,z_2]=\bigl\{tz_1+(1-t)z_2\colon t\in[0,1]\bigr\}.
\end{equation}
The following Lipschitz bounds are (more or less) a direct
consequence of formulas \eqref{zetamL} and \eqref{uplogderL} in
Assumption~B.

\begin{mylemma}
\label{lemma-ratio} Suppose Assumptions~A and~B hold and
let~$\kappa$,~$\tau$, and~$M$ be as in Assumption~B.
Let~$m\in\RR$, and let~$z_1,z_2\in \CalS_{\kappa/L}(m)$ be such
that~$[z_1,z_2]\subset\CalS_{\kappa/L}(m)$. Then
\begin{equation}
\label{ratio1} \Bigl|\frac{\zeta_m(z_1)}{\zeta_m(z_2)}\Bigr|
\le e^{2e^{-\tau L} +M |z_1-z_2|}.
\end{equation}
Moreover, for all~$z_1,z_2\in \OO$ such that~$[z_1,z_2]\subset\OO$,
we have
\begin{equation}
\label{ratio2} \frac{\zeta(z_1)}{\zeta(z_2)}\le e^{M  |z_1-z_2|}.
\end{equation}
\end{mylemma}

\begin{remark}
Since~$z\mapsto|\zeta_m(z)|$ are all twice continuously differentiable 
and hence Lipschitz throughout~$\OO$, so is their maximum~$z\mapsto\zeta(z)$.
The reason why we provide a (rather demanding) proof of \eqref{ratio2}
is that we need this bound to hold uniformly throughout~$\OO$ and the 
constant~$M$ from Assumption~B(3) to appear explicitly on the right-hand side.
The first part of the lemma underlines what is hard about the second part:
On the basis of Assumption~B, the uniform Lipchitz bound in \eqref{ratio1} can 
be guaranteed only in the region where~$m$ is ``almost stable.''
\end{remark}

\begin{proofsect}{Proof of Lemma~\ref{lemma-ratio}}
Let~$[z_1,z_2]\subset\CalS_{\kappa/L}(m)$. The
bound \eqref{ratio1} is directly proved by combining
\eqref{zetamL} with the estimate
\begin{equation}
\label{ration3a}
\bigl|\log|\zeta_m^{(L)}(z_1)|-\log|\zeta_m^{(L)}(z_2)|\bigr|\le
M|z_1-z_2|,
\end{equation}
implied by \eqref{uplogderL}.
Indeed, introducing $\varphi(t)=\zeta_m^{(L)}(z_1+t(z_2-z_1))$, we have
\begin{equation}
\Bigl|\frac{\textd}{\textd t}\log|\varphi(t)|\Bigr|
=\Bigl|\frac1{\varphi(t)}\frac{\textd|\varphi(t)|}{\textd t}\Bigr|
\le \Bigl|\frac1{\varphi(t)}\Bigr|
\Bigl|\frac{\textd\varphi(t)}{\textd t}\Bigr|\le M|z_2-z_1|
\end{equation}
implying \eqref{ration3a}.
By passing to the limit $L\to\infty$, we conclude that
\begin{equation}
\label{ratio3}
\bigl|\log\zeta(z_1)-\log\zeta(z_2)\bigr|\le M|z_1-z_2|
\end{equation}
holds provided $[z_1,z_2]\subset\CalS_m$.

To prove  \eqref{ratio2}, let~$z_1,z_2\in\OO$ with
$[z_1,z_2]\subset\OO$. If the segment $[z_1,z_2]$ intersects the
coexistence set~$\GG$ only in a finite number of points, then
\eqref{ratio2} is an easy consequence of \eqref{ratio3}. However,
this may not always be the case and hence we need a more general
argument. Note that continuity of both sides requires us to prove
\eqref{ratio2} only for a dense set of points~$z_1$ and~$z_2$.
This and the fact that each compact subset of~$\OO$ contains only
a finite number of multiple points from~$\MM=\{z\in\OO\colon
|\QQ(z)|\ge3\}$ permit us to assume that $z_1, z_2\notin\GG$ and
that the segment $[z_1,z_2]$ does not contain a multiple point,
i.e., $[z_1,z_2]\cap\MM=\emptyset$.

Suppose now that the bound \eqref{ratio2} fails. We claim that
then there exist a point $\bar x\in[z_1,z_2]$, with $\bar x\neq
z_1,z_2$, and two sequences $(x_n)$ and $(y_n)$ of points from $
[z_1,\bar x]\cap\GG$ and $ [\bar x,z_2]\cap \GG$, respectively,
such that the following holds:
\settowidth{\leftmargini}{(1111)}
\begin{enumerate}
\item[(1)]
$x_n\ne y_n$ for all~$n$ and
$\lim_{n\to\infty}x_n=\lim_{n\to\infty}y_n=\bar x$.
\item[(2)]
There exists a number $M'>M$ such that
\begin{equation}
\label{M'}
\Bigl|\log\frac{\zeta(x_n)}{\zeta(y_n)}\Bigr|>M'|x_n-y_n|
\end{equation}
 for all $n$.
\end{enumerate}
The proof of these facts will be simplified by introducing the
\emph{Lipschitz ratio}, which for any pair of distinct numbers
$x,y\in[z_1,z_2]$ is defined by the formula
\begin{equation}
R(x,y)=\frac{|\log\zeta(x)-\log\zeta(y)|}{|x-y|}.
\end{equation}
The significance of this quantity stems from its behavior under
subdivisions of the interval. Namely, if~$x$ and~$y$ are distinct
points and $z\in(x,y)$, then we have
\begin{equation}
\label{maxR}
R(x,y)\le\max\bigl\{R(x,z),R(z,y)\bigr\},
\end{equation}
with the inequality being strict unless $R(x,z)=R(z,y)$.

To prove the existence of sequences satisfying (1) and (2) above,
we need  a few observations: First, we note that
$M'=R(z_1,z_2)>M$ from our assumption that \eqref{ratio2} fails.
Second, whenever  $x,y\in [z_1,z_2]$ are such that $R(x,y)>M$,
then \eqref{ratio3} implies the  existence of~$x',y'\in[x,y]$ such
that $x',y'\in\GG$ and $R(x',y')\ge R(x,y)$. Indeed, we choose
$x'$ to be the nearest point to $x$ from the closed set
$[x,y]\cap\GG$, and similarly for $y'$. The fact that the Lipschitz
ratio increases in the process is a direct consequence of
\eqref{maxR}. Finally, if distinct $x,y\in [z_1,z_2]\cap\GG$
satisfy $R(x,y)>M$, then there exists a pair of distinct points
$x',y'\in [x,y]\cap\GG$ such that $|x'-y'|\le \frac12 |x-y|$ and
$R(x',y')\ge R(x,y)$. To prove this we use \eqref{maxR} with
$z=\frac12(x+y)$ to choose the one of the segments $[x,z]$ or $[z,y]$
that has the Lipschitz ratio not smaller than $R(x,y)$ and then
use the preceding observation on the chosen segment.

Equipped with these observations, we are ready to prove the
existence of the desired sequences. Starting with the second
observation above applied for $x=z_1$ and $y=z_2$, we get
$x_1,x_2\in[z_1,z_2]\cap\GG$ such that $R(x_1,x_2)>M'$. Notice
that $x_1\neq z_1$ and $x_2\neq z_2$ since $z_1,z_2\notin \GG$.
Next, whenever the pair $x_n,y_n$ is chosen, we use the third
observation to construct the pair $x_{n+1},y_{n+1}\in
[x_n,y_n]\cap\GG$ of points such that $|x_{n+1}-y_{n+1}|\le
\frac12 |x_n-y_n|$ and $R(x_{n+1},y_{n+1})\ge R(x_n,y_n)\ge M'$.
Clearly, the sequences $(x_n)$ and $(y_n)$  converge to a common
limit $\bar x\in [x_1,y_1]$, which is distinct from~$z_1$
and~$z_2$.

We will now show that \eqref{M'} still leads to a contradiction
with \eqref{ratio2}. First we note that the point $\bar x$, being
a limit of points from $\GG\setminus\MM$, is a two-phase
coexistence point and so Theorem~\ref{T:PD}(2) applies in a disc
$\D_\epsilon(\bar x)$ for $\epsilon>0$ sufficiently small. Hence,
there is a unique smooth coexistence curve~$\CC$ connecting $\bar
x$ to the boundary of $\D_\epsilon(\bar x)$ and, since $(x_n)$ and
$(y_n)$ eventually lie on~$\CC$, its tangent vector at $\bar x$ is
colinear with the segment $[z_1,z_2]$. Since in $\D_\epsilon(\bar
x)$, the coexistence curve is at least twice continuously
differentiable, the tangent vector to $\CC$ has a bounded
derivative throughout~$\D_\epsilon(\bar x)$. As a consequence, in
the disc  $\D_\delta(\bar x)$ with $\delta\le\epsilon$, the curve
$\CC$ will not divert from the segment $[z_1,z_2]$ by more than
$C\delta^2$, where $C=C(\epsilon)<\infty$.

Now we are ready to derive the anticipated contradiction: Fix~$n$
and let $\delta_n$ be the maximum of $|x_n-\bar x|$ and $|y_n-\bar
x|$. Let $\hate$ be a unit vector orthogonal to the segment
$[z_1,z_2]$ and consider the shifted points
$x_n'=x_n+2C\delta_n^2\hate$ and $y_n'=y_n+2C\delta_n^2\hate$.
Then we can write
\begin{equation}
\frac{\zeta(x_n)}{\zeta(y_n)}=\frac{\zeta(x_n)}{\zeta(x_n')}
\frac{\zeta(x_n')}{\zeta(y_n')}\frac{\zeta(y_n')}{\zeta(y_n)}.
\end{equation}
Assuming that $n$ is sufficiently large to ensure that
$\delta_n\sqrt{1+4C^2\delta_n^2}\le\epsilon$, the  segment
$[x_n',y_n']$ lies in~$\D_\epsilon(\bar x)$ entirely on one
``side'' of~$\CC$ and is thus contained in $\CalS_m$ for some
$m\in\RR$. On the other hand, given the bounded derivative of the
tangent vector to $\CC$, each segment $[x_n,x_n']$ and
$[y_n,y_n']$ intersects the curve $\CC$ exactly once, which in
light of $x_n,y_n\in\GG$ happens at the endpoint. This means that
also $[x_n,x_n']\subset\CalS_m$ and $[y_n,y_n']\subset\CalS_m$ for
the same~$m$. Consequently, all three ratios can be estimated
using \eqref{ratio2}, yielding
\begin{equation}
R(x_n,y_n)\le
M\,\frac{|x_n-x_n'|+|x_n'-y_n'|+|y_n'-y_n|}{|x_n-y_n|}
\le M+4MC\delta_n,
\end{equation}
where we used that $|x_n'-y_n'|=|x_n-y_n|$ and $|x_n-y_n|\ge
\delta_n$. But $\delta_n\to0$ with $n\to\infty$ and thus the ratio
$R(x_n,y_n)$ is eventually strictly less than~$M'$, in
contradiction with~\eqref{M'}. Hence, \eqref{ratio2} must have
been true after~all.
\end{proofsect}

The previous lemma will be particularly useful in terms of the
following corollary.

\begin{mycorollary}
\label{cor-D} Suppose that Assumptions~A and~B hold and let
$0<\tilde\kappa\leq\kappa$, where $\kappa$ is the constant from
Assumption~B. Then there exist constants $c<\infty$ and
$L_4<\infty$ such that the following is true for all $L\ge L_4$
and all $s\leq c/L$:
\settowidth{\leftmargini}{(111)}
\begin{enumerate}
\item[(1)] For  $m\in\RR$ and $z\in\CalS_{\tilde\kappa/(2L)}(m)$
with $\D_s(z)\subset \OO$, we have
\begin{equation}\label{4.53}
\D_{s}(z)\subset\CalS_{\tilde\kappa/L}(m).
\end{equation}
\item[(2)]
For  $z\in\OO$ with $\D_{s}(z)\subset\OO$, the set
\begin{equation}
\label{4.23}
\QQ'=\bigl\{m\in\RR\colon\D_{s}(z)
\subset\CalS_{\tilde\kappa/L}(m)\bigr\}
\end{equation}
in non-empty and
\begin{equation}
\label{D-in}
\D_{s}(z)\subset\UU_{\tilde\kappa/L}(\QQ').
\end{equation}
\item[(3)]
For $\gammaL\leq\tilde\kappa/(2L)$,  $\QQ\subset\RR$ and
$z\in\UU_\gammaL(\QQ) \cap\UU_{2\tilde\kappa/L}(\QQ)$
with $\D_{s}(z)\subset\OO$, we have
\begin{equation}
\label{D0-in}
\D_{s}(z)\subset\UU_{\tilde\kappa/L}(\QQ).
\end{equation}
\end{enumerate}
\end{mycorollary}

\begin{proofsect}{Proof}
Let $M$ be as in Assumption~B. We then choose $c>0$ sufficiently
small and $L_4<\infty$ sufficiently large to ensure that for
$L\geq L_4$ we have
\begin{equation}\label{L4-cond}
\frac{\tilde\kappa}{8M}-\frac1MLe^{-\tau L}\geq 2c.
\end{equation}
First,  we will show that the claims (1), (2), and (3) above
reduce to the following statement valid for each $m\in\RR$:
If~$z,z'$ are complex numbers such that the bound $|z-z'|\leq
2c/L$, the inclusion $[z,z')\subset\OO$, and
$z\in\OO\setminus\CalS_{\tilde\kappa/L}(m)$ hold, then also
\begin{equation}
\label{D1incl}
[z,z')\subset\OO\setminus \overline{\CalS_{\tilde\kappa/(2L)}(m)}.
\end{equation}
We proceed with the proof of~(1-3) given this claim; the
inclusion \eqref{D1incl} will be established at the end of this
proof.

\smallskip\noindent
\textit{Ad~(1):} Let  $z\in\CalS_{\tilde\kappa/(2L)}$ with
$\D_s(z)\subset \OO$ and assume that \eqref{4.53} fails.  Then
there exist some $z'\in\OO\setminus \CalS_{\tilde\kappa/L}(m)$
with $|z-z'|<s$ and $[z,z']\subset \OO$. But by \eqref{D1incl},
this implies
$[z',z)\cap\overline{\CalS_{\tilde\kappa/(2L)}(m)}=\emptyset$,
which means that
$[z',z]\cap{\CalS_{\tilde\kappa/(2L)}(m)}=\emptyset$. This
contradicts the fact that $z\in\CalS_{\tilde\kappa/(2L)}(m)$.

\smallskip\noindent
\textit{Ad~(2):} Let $z\in\OO$ with $\D_{s}(z)\subset\OO$. By the
definition of stable phases, there is at least one $m\in\RR$ such
that $z\in\CalS_m\subset\CalS_{\tilde\kappa/(2L)}(m)$. Combined
with \eqref{4.53}, this proves that the set $\QQ'$ is non-empty.
To prove \eqref{D-in}, it remains to show that
$\D_s(z)\subset\OO\setminus\overline{\CalS_{\tilde\kappa/(2L)}(m)}$
whenever $m\notin\QQ'$. By the definition of $\QQ'$, $m\notin\QQ'$
implies that there exists a $z'\in\D_s(z)$ such that
$z'\in\OO\setminus\CalS_{\tilde\kappa/L}(m)$.  Consider  an
arbitrary $z''\in \partial\D_s(z)$.  For such a $z''$, we have
that  $|z'-z''|\leq 2c/L$ and $[z',z'')\subset\OO$, so by
\eqref{D1incl}, we conclude that
$[z',z'')\subset\OO\setminus\overline{\CalS_{\tilde\kappa/(2L)}(m)}$.
Since this is true for all $z''\in\partial\D_s(z)$, we get the
desired statement
$\D_s(z)\subset\OO\setminus\overline{\CalS_{\tilde\kappa/(2L)}(m)}$.

\smallskip\noindent
\textit{Ad~(3):} Let $\QQ\subset\RR$,
$z\in\UU_\gammaL(\QQ)\cap\UU_{2\tilde\kappa/L}(\QQ)$ and
$\D_{s}(z)\subset\OO$. If $m\in\QQ$, then
$z\in\CalS_{\gammaL}(m)\subset\CalS_{\tilde\kappa/(2L)}(m)$ by the
definition of $\UU_\gammaL(\QQ)$ and the condition that
$\gammaL\leq \tilde\kappa/(2L)$.  With the help of \eqref{4.53},
this implies that $\D_{s}(z)\subset\CalS_{\tilde\kappa/L}(m)$ for
all $m\in\QQ$. Recalling the definition of
$\UU_{\tilde\kappa/L}(\QQ)$, we are left with the proof that
$\D_{s}(z)\subset\OO\setminus\overline{\CalS_{\tilde\kappa/(2L)}(m)}$
whenever $m\notin\QQ$.  But if $m\notin\QQ$, then
$z\in\OO\setminus\CalS_{\tilde\kappa/L}(m)$ because we assumed
that $z\in\UU_{2\tilde\kappa/L}(\QQ)$. By \eqref{D1incl} we
conclude that $[z,z')\subset\OO\setminus
\overline{\CalS_{\tilde\kappa/(2L)}(m)}$ whenever
$z'\in\partial\D_s(z)$, which proves
$\D_{s}(z)\subset\OO\setminus\overline{\CalS_{\tilde\kappa/(2L)}(m)}$.

\smallskip
We are left with the proof of \eqref{D1incl}, which will be done
by contradiction. Assume thus that $m\in\RR$ and let $z,z'$ be two
points such that $|z-z'|\leq 2c/L$, $[z,z')\subset\OO$ and
$z\in\OO\setminus\CalS_{\tilde\kappa/L}(m)$ hold, while
\eqref{D1incl} fails to hold, so that
$[z,z')\cap\overline{\CalS_{\tilde\kappa/(2L)}(m)}\neq\emptyset$.
Let~$z_1\in[z,z')\cap\overline{\CalS_{\tilde\kappa/(2L)}(m)}$.
Since $[z,z')\subset\OO$, we have in particular that
$[z_1,z]\subset\OO$. Let $z_2$ be defined as the nearest point to
$z_1$ on the linear segment $[z_1,z]$ such that
$z_2\not\in\CalS_{3\tilde\kappa/(4L)}(m)$. By continuity of the
functions $\zeta_k$, we have $[z_1,z_2] \subset
\CalS_{\tilde\kappa/L}(m)\subset\CalS_{\kappa/L}(m)$ so that the
bounds in Lemma~\ref{lemma-ratio} are at our disposal. Putting
\twoeqref{ratio1}{ratio2} together, we have
\begin{equation}
\label{4.26}
\Bigl|\frac{\zeta_m(z_1)}{\zeta(z_1)}\Bigr|\,
\Bigl|\frac{\zeta(z_2)}{\zeta_m(z_2)}\Bigr|
\le e^{2e^{-\tau L}+2M|z_1-z_2|}.
\end{equation}
Now, since~$z_1\in\CalS_{\tilde\kappa/(2L)}(m)$
and~$z_2\not\in\CalS_{3\tilde\kappa/(4L)}(m)$, we can infer that
the left-hand side is larger than~$e^{\tilde\kappa/(4L)}$. Hence,
we must have
\begin{equation}
|z_1-z_2|\ge\frac\kappa{8ML}-\frac1Me^{-\tau L} \ge\frac{2c}L,
\end{equation}
where the last inequality is a consequence of \eqref{L4-cond}. Now
$z_1,z_2\in[z,z')$ implies $|z_1-z_2|<|z-z'|$, which contradicts
the assumption that $|z-z'|\leq 2c/L$ and thus proves
\eqref{D1incl}.
\end{proofsect}

\subsection{Proofs of Lemmas~\ref{f--eq}-\ref{lemmag}}
\label{sec-lemmas}\noindent 
Here we will establish the three
technical lemmas on which the proof of Theorem~\ref{T:2ph} was based.
Throughout this section, we fix  distinct $m,n\in\RR$ and
introduce the abbreviations
$\CalS_\epsilon=\CalS_\epsilon(\{m,n\})$ and
$\UU_\epsilon=\UU_\epsilon(\{m,n\})$. We will also let $f$ and $g$
be the  functions defined in \twoeqref{fz}{gz}.

\smallskip
First we will need to establish a few standard facts concerning the local
inversion of analytic maps and its behavior under perturbations by
continuous functions. The proof is based on Brouwer's Fixed Point Theorem, 
see e.g. \cite[Chapter~2]{Milnor}.

\begin{mylemma}
\label{lemma-inv}
Let  $z_0\in\C$, $\epsilon>0$, and let $\phi\colon\D_\epsilon(z_0)\to\C$
be an analytic map for which
\begin{equation}
\label{1/2-bound}
|\phi'(z_0)|^{-1} \bigl|\phi'(z)-\phi'(z_0)\bigr|\le \frac12
\end{equation}
holds for all $z\in\D_\epsilon(z_0)$.
Let $\delta\le\epsilon|\phi'(z_0)|/2$.
Then, for every $w\in\D_\delta(\phi(z_0))$, there exists a unique point
$z\in\D_\epsilon(z_0)$ such that $\phi(z)=w$.

In addition, let  $\eta\in[0,\delta/2)$ and let
$\theta\colon\D_\epsilon(z_0)\to\C$ be a continuous  map
satisfying
\begin{equation}
\label{etabd} |\theta(z)|\le\eta, \qquad  z\in\D_\epsilon(z_0).
\end{equation}
Then for each $z\in\D_\epsilon(z_0)$ with $\phi(z)\in\D_\eta(\phi(z_0))$
there exists a point $z'\in\D_\epsilon(z_0)$ such  that
\begin{equation}
\label{fifi}
\phi(z')+\theta(z')=\phi(z).
\end{equation}
Moreover,  $|z'-z|\le 2\eta|\phi'(z_0)|^{-1}$.
\end{mylemma}

\begin{proofsect}{Proof}
Following standard proofs of the theorem about local inversion of
differentiable maps (see, e.g., \cite{Federer}, Sect.~3.1.1), we
seek the inverse of~$w$ as a fixed point of the (analytic)
function $z\mapsto\psi(z)=z+\phi'(z_0)^{-1}(w-\phi(z))$. The
condition \eqref{1/2-bound} guarantees that $z\mapsto\psi(z)$ is a
contraction on~$\D_\epsilon(z_0)$. Indeed, for every
$z\in\D_\epsilon(z_0)$ we have
\begin{equation}
\label{E:Dpsi}
|\psi'(z)|=\bigl|1-\phi'(z_0)^{-1}\phi'(z)\bigr|
\le |\phi'(z_0)|^{-1} \bigl|\phi'(z)-\phi'(z_0)\bigr|\le \tfrac12,
\end{equation}
which implies that $|\psi(z)-\psi(z')|\le\tfrac12|z-z'|$  for all
$z,z'\in\D_\epsilon(z_0)$. The actual solution to $\phi(z)=w$ is
obtained as the limit $z=\lim_{n\to\infty} z_n$ of iterations
$z_{n+1}=\psi(z_n)$ starting at $z_0$. In view of the above
estimates, we have $|z_{n+1}-z_n|\le\tfrac12|z_n-z_{n-1}|$ and,
summing over $n$, we get
$|z_n-z_0|\le2|z_1-z_0|\le2|\phi'(z_0)|^{-1} |w-\phi(z_0)|$. Since
$|w-\phi(z_0)|<\delta$, we have that $z_n$ as well as its
limit belongs to $\D_\epsilon(z_0)$.

Next we shall attend to the second part of the claim.
The above argument allows us to define the  left inverse of $\phi$ as
the function $\phi^{-1}\colon\D_\delta(\phi(z_0))\to\D_\epsilon(z_0)$
such that $\phi^{-1}(w)$ is the  unique value $z\in\D_\epsilon(z_0)$
for which $\phi(z)=w$.
Let  $\eta\in[0,\delta/2)$ and let $z\in\D_\epsilon(z_0)$ be such that
$\phi(z)\in\D_\eta(\phi(z_0))$.
Consider the function  $\Psi\colon\D_\delta(\phi(z_0))\to\C$ defined  by
\begin{equation}
\Psi(w)=\phi(z)-\theta(\phi^{-1}(w)).
\end{equation}
By our choice of $z$ and \eqref{etabd}, we have
$|\Psi(w)|\le2\eta$ for any $w\in\D_\delta(\phi(z_0))$. Thus,
$\Psi$ maps the closed disc $\overline{\D_{2\eta}(\phi(z_0))}$
into itself and, in light of continuity of~$\Psi$, Brouwer's
Theorem implies that~$\Psi$ has a fixed point $w'$  in
$\overline{\D_{2\eta}(\phi(z_0))}$. From the relation
$\Psi(w')=w'$ we then  easily show that  \eqref{fifi} holds for
$z'=\phi^{-1}(w')$. To control the  distance between $z$ and $z'$,
we just note that the above Lipschitz bound  on~$\psi$ allows us
to conclude that $|z'-z|\le2|\phi'(z_0)|^{-1}|\phi(z')- \phi(z)|$.
Applying \eqref{fifi} and \eqref{etabd}, the right-hand side is
bounded by $2\eta|\phi(z_0)|^{-1}$.
\end{proofsect}

Now we are ready to start proving Lemmas~\ref{f--eq}-\ref{lemmag}.
The first claim to prove concerns the relation of the solutions
of \twoeqref{Req}{Imq} and the roots of the function~$f$
defined in \eqref{fz}.

\begin{proofsect}{Proof of Lemma~\ref{f--eq}}
Let  $\tilde\alpha$, $M$ and $\tau$ be the constants from
Assumption~B. Let~$c$  and~$L_4$ be the constants from
Corollary~\ref{cor-D} with $\tilde\kappa=\kappa$. The proof will
be carried out for the constants $B_1$, $\tilde C_1$ and $L_1$
chosen as follows: We  let
\begin{equation}
B_1=\frac1{4M},
\quad
B_2=\frac{16+4|\log(q_n/q_m)|}{\tilde\alpha}
\quad\text{and}\quad\tilde  C_1=\frac{10}{\tilde\alpha},
\end{equation}
and assume that $L_1$ is so large that $L_1\ge L_4$ and for all
$L\ge L_1$, we have $\tilde C_1e^{-\tau L}<B_1L^{-d}$ and the bounds:
\begin{equation}
\label{5.26}
(B_1+B_2)L^{-d}\leq\frac cL\le\frac1{4M},\quad
2e^{-\tau L}+\frac\kappa L\le \frac14,
\end{equation}
\begin{equation}
\label{5.26a}
\frac 2{\tilde\alpha}(M+M^2)(B_1+B_2)L^{-d}\le\frac12,
\end{equation}
and also
\begin{equation}
\label{5.27}
2e^{-\tau L}+2M
B_1L^{-d}\le L^{-d},\quad
\tilde\alpha>2\sqrt2e^{-\tau L},
\end{equation}
\begin{equation}
\label{5.27a}
\pi L^{-d}+2e^{-\tau L}
< 4L^{-d}
\quad\text{and}\quad
\tilde C_1 e^{-\tau L}\leq\tfrac 12 B_2L^{-d}.
\end{equation}
Let us fix a value $L\ge L_1$ and choose a point
$z_0\in\CalS_{\kappa/(2L)}$ and a number $s\leq (B_1+B_2)L^{-d}$ such
that $\D_s(z_0)\subset\OO$. Corollary~\ref{cor-D}(1) combined with
the first bound in \eqref{5.26} implies that
$\D_s(z_0)\subset\CalS_{\kappa/L}$.

We will apply Lemma~\ref{lemma-inv} for suitable choices of $\phi$
and $\theta$ defined in terms of the functions
$F_{m,n}\colon\D_s(z_0)\to\C$ and
$F_{m,n}^{(L)}\colon\D_s(z_0)\to\C$ defined by
\begin{equation}
\label{FLdef}
F_{m,n}(z)=\frac{\zeta_m(z)}{\zeta_n(z)} \quad \text{and } \quad
F_{m,n}^{(L)}(z)=\frac{\zeta_m^{(L)}(z)}{\zeta_n^{(L)}(z)}.
\end{equation}
We will want to define $\phi(z)$ as the logarithm of
$F_{m,n}^{(L)}(z)$, and $\theta(z)$ as the logarithm of
the ratio
$F_{m,n}^{(L)}(z)/F_{m,n}(z)$, but in order to do so, we
will have to specify the branch of the complex logarithm
we are using.  To this end, we will first analyze the
image of the functions $F_{m,n}^{(L)}(z)$ and
$F_{m,n}^{(L)}(z)/F_{m,n}(z)$.

According to Assumption~B2, for any
$z\in\D_s(z_0)\subset\CalS_{\kappa/L}$, we have
$|F_{m,n}^{(L)}(z)|\in(2/3,3/2)$ in view of the second bound in
\eqref{5.26} with the observation that $\frac14<\log\frac32$.
A simple calculation and the bound \eqref{uplogderL} show that
$\Arg F_{m,n}^{(L)}(z)$ and $\Arg F_{m,n}^{(L)}(z_0)$ differ
by less than $2M(B_1+B_2)L^{-d}\leq\frac12$.
Indeed,
the difference  $\Arg F_{m,n}^{(L)}(z)-\Arg F_{m,n}^{(L)}(z_0)$ is
expressed in terms of the integral of
$\partial_zF_{m,n}^{(L)}/F_{m,n}^{(L)}$ along any path in
$\D_s(z_0)$ connecting~$z_0$ and~$z$. The latter logarithmic
derivative is bounded uniformly by $2M$ throughout $\D_s(z_0)$.
Consequently, $z\mapsto F_{m,n}^{(L)}(z)$ maps $\D_s(z_0)$ into
the open set of complex numbers
$\{\rho e^{\texti\omega}\colon\rho\in(\frac23,\frac32),
|\omega-\omega_0|<\frac1{2}\}$,
where $\omega_0=\Arg F_{m,n}(z_0)$.
The function $z\mapsto F_{m,n}^{(L)}(z)/F_{m,n}(z)$,
on the other hand, maps $\D_s(z_0)$ into
the open set of complex numbers
$\{\rho e^{\texti\omega}\colon\rho\in(\frac23,\frac32),
|\omega|<\frac1{4}\}$,
as can be easily inferred from Assumption B2 and the second bound
in \eqref{5.26}.  Given these observations, we  choose
the branch of the complex logarithm with cut along the ray
$\{re^{-\texti\omega_0/2}\colon r>0\}$, and define
\begin{equation}
\phi(z)=\log F_{m,n}^{(L)}(z)
\end{equation}
and
\begin{equation}
\theta(z)=\log\frac{F_{m,n}^{(L)}(z)}{F_{m,n}(z)}.
\end{equation}
Having defined the functions $\phi$ and $\theta$,  we note that,
by Assumptions~A and~B, $\phi$ is  analytic while $\theta$ is
twice continuously differentiable throughout $\D_s(z_0)$.
Moreover, these functions are directly related to the equations
$f(z)=0$ and \twoeqref{Req}{Imq}. Indeed, $f(z)=0$ holds for some
$z\in\D_s(z_0)$ if and only if $F_{m,n}^{(L)}(z)$ is an~$L^d$-th
root of $-(q_n/q_m)$, i.e.,
$\phi(z)=(\log(q_n/q_m)+\texti \pi(2k+1))L^{-d}$ for some integer $k$.
Similarly, $z\in\D_s(z_0)$ is a  solution of \twoeqref{Req}{Imq}
if and only if $\phi(z)+\theta(z)$ is of the form
$(\log(q_n/q_m)+\texti \pi(2k+1))L^{-d}$ for some integer $k$.
Furthermore, these functions obey the bounds
\begin{equation}
\label{5.30}
\tilde\alpha\le|\phi'(z)|\le2M,\qquad
|\phi'(z)-\phi'(z_0)|\le 2(M+M^2)(B_1+B_2)L^{-d},
\end{equation}
and
\begin{equation}
\label{5.31}
|\theta(z)|\le 2e^{-\tau  L},\qquad
|\theta(z)-\theta(z')|\le 2\sqrt2e^{-\tau L}|z-z'|
\end{equation}
for all  $z,z'\in\D_s(z_0)$.  Here the first three bounds are
obvious consequences of Assumption~B, while the third follows from
Assumption~B by observing that the derivative matrix $D\theta(z)$
is bounded in norm by $2\sqrt 2$ times the right hand side of
\eqref{der-zetamL}. Note that, in light of \twoeqref{5.26}{5.26a}, these
bounds directly verify the assumptions \eqref{1/2-bound} and
\eqref{etabd} of Lemma~\ref{lemma-inv} for $\eta=2e^{-\tau L}$
and any $\epsilon\le s$.
We proceed  by applying Lemma~\ref{lemma-inv} with different
choices of $\epsilon$ to give the  proof of (2-4) of
Lemma~\ref{f--eq}, while part (1) turns out to be a direct
consequence of the bounds \twoeqref{5.30}{5.31}.

Indeed, let us first show that for $s\leq B_1L^{-d}$ the disc
$\D_s(z_0)$ contains at most one solution to \twoeqref{Req}{Imq}
and at most one root of the equation $f(z)=0$.  We will prove both
statements by contradiction.  Starting with the solutions to
\twoeqref{Req}{Imq}, let us thus assume that $z_1,z_2\in\D_s(z_0)$
are two distinct solutions to the equations \twoeqref{Req}{Imq}.
Setting $w_1=\phi(z_1)+\theta(z_1)$ and
$w_2=\phi(z_2)+\theta(z_2)$ this means that $w_1-w_2$ is an
integer multiple of $2\pi\texti L^{-d}$. However,  the bounds
\eqref{5.30} and \eqref{5.31} combined with the first bound in
\eqref{5.27} guarantee that $|w_1-w_2|\le 4e^{-\tau
L}+4MB_1L^{-d}\le 2 L^{-d}$ and thus $w_1=w_2$. But then the bound
$|\phi(z_1)-\phi(z_2)|\ge\tilde\alpha|z_1-z_2|$ implies that
$|\theta(z_1)-\theta(z_2)|\ge\tilde\alpha|z_1-z_2|$, which, in
view of the second bound in \eqref{5.27}, contradicts the second
bound in \eqref{5.31}. Hence, we must have had $z_1=z_2$ in the
first place.  Turning to the equation $f(z)=0$, let us now assume that
$z_1$ and $z_2$ are two different roots of this equation. Setting
$w_1=\phi(z_1)$ and $w_2=\phi(z_2)$, we again have $w_1=w_2$, this
time by the first bound in \eqref{5.30} and the very definition of
$B_1$, which implies that $4MB_1=1$.  But once we have $w_1=w_2$,
we must have $z_1=z_2$ since
$|\phi(z_1)-\phi(z_2)|\ge\tilde\alpha|z_1-z_2|$ by our lower bound
on $\phi'(z)$, implying that there exists at most one
$z\in\D_s(z_0)$ that solves the equation $f(z)=0$. If such a
solution~$z$ exists, Assumption~B immediately implies that
$f'(z)\ne0$, and so~$z$ is a non-degenerate root of~$f$.

Next, we will show that within a
$\tilde  C_1e^{-\tau L}$-neighborhood
of each solution $z_0$ of the equations \twoeqref{Req}{Imq}
there is a  root of~$f$. Indeed, let
$\epsilon=\tilde C_1 e^{-\tau L}$ and $\delta=5e^{-\tau L}$.
By the first bound in \eqref{5.30}
and our choice of $\tilde C_1$, we then have
$\delta\leq\epsilon|\phi'(z_0)|/2$, so the first part of
Lemma~\ref{lemma-inv} is at our disposal.
Since~$z_0$ is assumed to be a solution to
\twoeqref{Req}{Imq}, we have that  $\phi(z_0)+\theta(z_0)$ is
of the form $(\log(q_n/q_m)+\texti \pi(2k+1)) L^{-d}$, where
$k$ is an integer. In light of the bound
$|\theta(z_0)|\le2 e^{-\tau L}$, the disc $\D_\delta(\phi(z_0))$
contains the point  $w=\phi(z_0)+\theta(z_0)$. By the first part
of Lemma~\ref{lemma-inv}, there  exists a point
$z\in\D_\epsilon(z_0)$ such that $\phi(z)=w$, implying
that~$z$ is a root of $f$.

As a third step we will prove that if~$z_0$ is a root of $f$, then there
exists a solution to \twoeqref{Req}{Imq} in $\D_{\tilde
C_1e^{-\tau L}}(z_0)$.
By the relation between~$f$ and~$\phi$ we now
know that $\phi(z_0)$ is of the form $(\log(q_n/q_m)+\texti \pi(2k+1)) L^{-d}$
 for some integer $k$.  We again set $\epsilon=\tilde C_1 e^{-\tau
L}$ and $\delta=5e^{-\tau L}$.  Choosing $\eta=2e^{-\tau L}$ and
noting that $2\eta<\delta$, we  apply the second part of
Lemma~\ref{lemma-inv} to conclude that there must be a point
$z'\in\D_\epsilon(z_0)$ such that
$\phi(z')+\theta(z')=\phi(z_0)=(\log(q_n/q_m)+\texti \pi(2k+1)) L^{-d}$,
which means that~$z'$ is  a solution to \twoeqref{Req}{Imq}.

Finally, we will show that if~$z_0\in\CalS_m\cap\CalS_n$, then
there exists a solution to \twoeqref{Req}{Imq} in the disc
$\D_{B_2L^{-d}}(z_0)$.
To this end, we first note that
$z_0\in\CalS_m\cap\CalS_n$ implies that
$\phi(z_0)+\theta(z_0)$ is purely imaginary.
Combined with the first bound in \eqref{5.31} we conclude
that
within a distance of at most
$(|\log(q_m/q_n)|+\pi)L^{-d}+2e^{-\tau L}$
from $\phi(z_0)$,
there exists a point of the form $w=(\log(q_n/q_m)+\texti \pi(2k+1)) L^{-d}$
for some integer $k$.
We now set $\epsilon=B_2L^{-d}/2$ and
$\delta=(|\log(q_m/q_n)|+4)L^{-d}$.  By the first condition
in \eqref{5.27a}, we then have
$|\phi(z_0)-w|<\delta$, while the first bound in \eqref{5.30}
together with the definition of $B_2$ implies that
$\delta\leq \epsilon|\phi'(z_0)|/2$.  We therefore can use
the first part of Lemma~\ref{lemma-inv}
to conclude that there must be a point
$z'\in\D_\epsilon(z_0)$ such that $\phi(z')=w$,
implying that $z'$ is a root of $f(z')=0$.
Finally, by the already proven statement (3) of the lemma,
there must be a solution of the equations \twoeqref{Req}{Imq}
within a distance strictly less than $\tilde C_1 e^{-\tau}$
from $z'$.  Since $\epsilon+\tilde C_1 e^{-\tau}\leq B_2L^{-d}$
by the second condition in \eqref{5.27a}, this gives the desired solution
of the equations \twoeqref{Req}{Imq} in the disc
$\D_{B_2L^{-d}}(z_0)$.
\end{proofsect}

Next we will prove Lemma~\ref{lemmaf} which provides a
lower bound on~$f(z)$ on the boundary of certain discs.

\begin{proofsect}{Proof of Lemma~\ref{lemmaf}}
Let~${\tilde \alpha}$ and~$M$ be
as in Assumption~B3, let $\tilde\kappa=\kappa/2$, and
let~$c$ and~$L_4$ be the
constants from Corollary~\ref{cor-D}.
We will prove the claim with
\begin{equation}
{\tilde c_2}=(2eM\Vert\mathbf{q}\Vert_\infty)^{-1}
\qquad\text{and}\qquad
\tilde C_2=
\max\{{\tilde c_2},
{22 e}{\tilde\alpha}^{-1}\}
\end{equation}
and, given $\tilde C\ge\tilde C_2$, with $L_2$ defined by
the condition
that $L_2\geq L_4$ and
\begin{equation}
\label{Lcond}
\tilde C\epsilon_L
\le c/L,
\quad
L^de^{-\tau L}\le 1,
\quad
e^{\tilde C M L^d \epsilon_L}\leq 2
\end{equation}
and
\begin{equation}
\label{LcondB}
2e(M+M^2)\Vert\mathbf{q}\Vert_\infty \tilde C^2L^d\epsilon_L\leq 1
\end{equation}
hold whenever $L\geq L_2$.

Fix~$L\ge L_2$ and choose a point $z_0\in
\CalS_{\kappa/(4L)}\cap(\CalS_m\cup\CalS_n)$ with
${\D}_{\tilde{C}\epsilon_L}(z_0)\subset\OO$. Let
$s<\tilde C\epsilon_L$ and note that
by \eqref{Lcond} we have $s<c/L$.
Applying Corollary~\ref{cor-D}(1) to the disc $\D_s(z_0)$ we find
that $\D_s(z_0)\subset\CalS_{\kappa/(2L)}\subset\CalS_{\kappa/L}$.
In particular, the bounds of Assumption~B are at our disposal
whenever $z\in{\D_{\tilde C\epsilon_L}(z_0)}$. The proof will
proceed by considering two separate cases depending (roughly) on
whether $|f(z_0)|$ is ``small'' or ``large.'' We will first
address the latter situations.
Let us therefore suppose that
$|f(z_0)|> 4L^d\epsilon_L\,\zeta(z_0)^{L^d}$.
In this case,
we will show that \eqref{2phbd-a} holds with
$s(z_0)={\tilde c_2}\epsilon_L$. (Note that
$s(z_0)\leq \tilde C_2\epsilon_L\leq\tilde C\epsilon_L$
by our definition of $\tilde C_2$.)
A crucial part of the proof consists
of the derivation of an appropriate estimate on the derivative
of~$f$. Let $s<\tilde C\epsilon_L$ and let $z$ be such that
$|z-z_0|\le s$. Recalling the definition \eqref{bm} of $b_m(z)$
and using Assumptions~B2-B3, the second and third bound in
\eqref{Lcond} and the fact that one of the values $|\zeta_m(z_0)|$
and $|\zeta_n(z_0)|$ must be equal to $\zeta(z_0)$, we have
\begin{equation}
\label{f'bound}
\begin{aligned}
|f'(z)\bigr|
&= L^d\Bigl|q_mb_m(z)\zeta_m^{(L)}(z)^{L^d}
+q_nb_n(z)\zeta_n^{(L)}(z)^{L^d}\Bigr|
\\
&\leq
L^d
\Bigl[q_m M|\zeta_m(z_0)|^{L^d}+q_nM|\zeta_n(z_0)|^{L^d}\Bigr]
e^{M|z-z_0|L^d+L^de^{-\tau L}}
\\
&\leq
4eM\Vert\mathbf{q}\Vert_\infty
L^d\zeta(z_0)^{L^d}
\end{aligned}
\end{equation}
whenever $z\in \CalS_{\kappa/L}$. As argued above,
$z\in\D_{\tilde C\epsilon_L}(z_0)$ implies that
$[z_0,z]\subset \CalS_{\kappa/L}$, so by the Fundamental Theorem
of Calculus we have
\begin{equation}
\label{4.77}
\bigl|f(z)\bigr|\ge\bigl|f(z_0)\bigr| -
4eM\Vert\mathbf{q}\Vert_\infty L^d\zeta(z_0)^{L^d}
s
\ge
4L^d\,\zeta(z)^{L^d}
\Bigl(\epsilon_L-\frac s{2\tilde c_2}\Bigr)
\end{equation}
for all $z\in\D_s(z_0)$.
The bound \eqref{2phbd-a} now follows by letting
$s\uparrow\tilde c_2\epsilon_L$.

Next we will address the cases with~$|f(z_0)|\leq
4L^d\epsilon_L\zeta(z_0)^{L^d}$.
Let $s<\tilde C\epsilon_L$ and pick $z$ such that
$|z-z_0|=s$.
This point belongs to the disc $\D_{\tilde C\epsilon_L}(z_0)$
which
we recall is a subset of~$\CalS_{\kappa/L}$.
The second-order expansion formula
\begin{equation}
f(z)=f(z_0)+f'(z_0)(z-z_0)
+(z-z_0)^2\int_0^1\textd  t\int_0^{t}\textd \tilde t
\,f^{\prime\prime}\bigl(\tilde tz+(1-\tilde t)z_0\bigr)
\end{equation}
then yields the estimate
\begin{equation}
\label{4.17}
\bigl|f(z)\bigr|\ge\bigl|f(z_0)+(z-z_0)f'(z_0)\bigr|-
{\widetilde K}\bigl(\tilde C \epsilon_L\bigr)^2L^{2d}\zeta(z_0)^{L^d}
\end{equation}
where
\begin{equation}
\widetilde
K=\frac 12\zeta(z_0)^{-L^d}L^{-2d}
\sup\bigl\{|f^{\prime\prime}(z)|\colon z\in
\UU, \,|z-z_0|< \tilde C\epsilon_L\bigr\}.
\end{equation}
Proceeding as in the bound \eqref{f'bound}, we easily get
\begin{equation}
\widetilde K\le 2e\Vert\mathbf{q}\Vert_{\infty}
\bigl[M^2(1-L^{-d}) + ML^{-d}\bigr],
\end{equation}
which implies that
$\widetilde K\le2e\Vert\mathbf{q}\Vert_{\infty} [M^2 +M]$.

It remains to estimate the absolute value on the right-hand side of
\eqref{4.17}.
Abbreviating~$b_m=b_m(z_0)$ and~$b_n=b_n(z_0)$, we can write
\begin{equation}
\begin{aligned}
f'(z_0)&=L^d\bigl(b_mq_m\zeta_m^{(L)}(z_0)^{L^d}
+b_nq_n\zeta_n^{(L)}(z_0)^{L^d}\bigr)\\
&=L^d(b_m-b_n)q_m\zeta_m^{(L)}(z_0)^{L^d}+b_nL^df(z_0).
\end{aligned}
\end{equation}
Without loss of generality, let us suppose
that~$|\zeta_m(z_0)|\ge|\zeta_n(z_0)|$
and, consequently, $|\zeta_m(z_0)|=\zeta(z_0)$,
because~$z_0\in\CalS_m\cup\CalS_n$.
Applying Assumption~B3 together
with the assumed upper bound on~$|f(z_0)|$, we get
\begin{equation}
\label{4.20}
\bigl|(z-z_0)f'(z_0)+f(z_0)\bigr|\ge \bigl({\tilde
\alpha} q_mse^{-L^de^{-\tau L}} -4\epsilon_L\,(1+s L^d
M)\bigr)L^d\zeta(z_0)^{L^d},
\end{equation}
where we recalled that $|z-z_0|=s$.
Since~$s\le\tilde C\epsilon_L$, the third inequality in
\eqref{Lcond} gives that~$s L^d M\le\tilde CML^d\epsilon_L<1$.
Let now $s$ be so large that $s\geq\frac12\tilde C\epsilon_L$.
Using this bound in
the first term in \eqref{4.20} and using
the second inequality in \eqref{Lcond}
we thus get
\begin{equation}
\label{4.20a}
\bigl|(z-z_0)f'(z_0)+f(z_0)\bigr|\ge
\bigl(\tfrac12\tilde\alpha\tilde C_2e^{-1} -8\bigr)
L^d\epsilon_L\zeta(z_0)^{L^d}
\geq 3L^d\epsilon_L\zeta(z_0)^{L^d}.
\end{equation}
Moreover, using the above bound on~$\widetilde K$ and the
inequality in \eqref{LcondB}, the last term on the right-hand side
of~\eqref{4.17} can be shown not to exceed~$L^d\epsilon_L
\zeta(z_0)^{L^d}$.  Putting \eqref{4.17} and \eqref{4.20a}
together with these estimates, we have~$|f(z)|\geq 2 L^d\epsilon_L
\zeta(z_0)^{L^d}$ for all $z\in\D_{\tilde C\epsilon_L}(z_0)$
such that $s=|z-z_0|$
satisfies $\frac12\tilde C\epsilon_L\le s<\tilde C\epsilon_L$.
The proof is
finished by taking $s\uparrow\tilde C\epsilon_L$.

The last statement of the lemma is an immediate consequence of the
fact that whenever the above procedure picks $s(z_0)=\tilde
c_2\epsilon_L$ and $\tilde c_2<\tilde C$, then
the argument \twoeqref{f'bound}{4.77} implies the stronger bound
\begin{equation}
\label{2phbd-b}
\inf_{z\colon|z-z_0|<s(z_0)}|f(z)|\geq 2L^d\epsilon_L\zeta(z_0)^{L^d}.
\end{equation}
Now, if $f$ has a root in $\D_{\tilde c_2\epsilon_L}(z_0)$, then
this bound shows
that we could not have chosen $s(z_0)=\tilde c_2\epsilon_L$.
Therefore, $s(z_0)$
must be equal to the other possible value, i.e., we must have
$s(z_0)=\tilde C\epsilon_L$.
\end{proofsect}

\begin{proofsect}{Proof of Lemma~\ref{lemmag}}
We will prove
\eqref{gbd} with $A_3=2C_0\Vert\textbf{q}\Vert_1$,
where $C_0$ is as in \eqref{error} for $\ell=0$.
Let~$\tilde L_0$ and~$M$ be as in Assumption~B and
let~$L_4$ and~$c$ be as in Corollary~\ref{cor-D}.
Let $C\in(0,\infty)$ and let us
choose $L_3\geq \max\{L_4,\tilde L_0\}$
in such a way that
\begin{equation}
\label{L3-cond}
\max\bigl\{Ce^{-\tau L},CL^de^{-\frac 12 L^d\gammaL}\bigr\}
\leq
\frac cL,
\quad
MCL^de^{-\tau L}
\leq\log 2,
\end{equation}
\begin{equation}
\label{L3-cond2}
\frac12L^d\gammaL+MCL^{2d}e^{-\frac 12L^d\gammaL}\leq\tau L,
\end{equation}
and
\begin{equation}
\label{L3-cond3}
\gamma_L\leq\frac\kappa{2L}
\quad\text{and}\quad
MCL^{2d}e^{-\frac 12L^d\gammaL}
+L^de^{-\tau L}
\leq 2d\log L+\log C_0
\end{equation}
hold for all $L\geq L_3$.

We will treat separately the cases
$z_0\in\UU_\gammaL\cap\UU_{2\kappa/L}(z_0)$ and
$z_0\in\UU_\gammaL\setminus\UU_{2\kappa/L}(z_0)$. Let us first
consider the former case, so
that~$\delta_L(z_0)=e^{-\tau L}$.
The first condition in
\eqref{L3-cond}, the fact that
${\D_{C\delta_L(z_0)}(z_0)}\subset\OO$ and
$\gammaL\leq\kappa/(2L)$
therefore
allow us to use Corollary~\ref{cor-D}(3), from which we conclude that
$\D_{C\delta_L(z_0)}(z_0)\subset\UU_{\gammaL}$. For
$z\in{\D_{C\delta_L(z_0)}(z_0)}$ we may thus apply
the~$\ell=0$ version of \eqref{error} to the
function~$g(z)=\varXi_{\{m,n\},L}(z)$. Combined with the bound
\eqref{ratio2}, the second condition in \eqref{L3-cond} and our
definition of $A_3$ this immediately gives the desired
bound \eqref{gbd}.

Next we will attend to the cases
when~$z_0$ lies in~$\UU_\gammaL\setminus\UU_{2\kappa/L}$, so
that~$\delta_L(z_0)=L^de^{-\frac 12L^d\gammaL}$.
Let us define $\QQ'$ as in \eqref{4.23} with $s=C\delta_L(z_0)$,
i.e.,
\begin{equation}
\QQ'=\bigl\{k\in\RR\colon\D_{C\delta_L(z_0)}\subset\CalS_{\kappa/L}(k)\bigr\}.
\end{equation}
By Corollary~\ref{cor-D}(2), the set $\QQ'$ is non-empty and
${\D_{C\delta_L(z_0)}(z_0)}\subset\UU_{\kappa/L}(\QQ')$.
Let $z\in{\D_{C\delta_L(z_0)}(z_0)}$ and let us
estimate~$g(z)$. We will proceed
analogously to the preceding case; the only
difference is that this time we have
\begin{equation}
g(z)=\varXi_{\QQ',L}(z)+h(z),
\end{equation}
where the extra
term~$h(z)$
is given by
\begin{equation}
h(z)=\sum_{k\in\QQ'\smallsetminus\{m,n\}}q_k
\bigl[\zeta_k^{(L)}(z)\bigr]^{L^d}.
\end{equation}
Now~$|\varXi_{\QQ',L}(z)|$ is estimated as before: Using
that $z\in\UU_{\kappa/L}(\QQ')$, the bounds \eqref{error} and
\eqref{ratio2} immediately yield that $|\varXi_{\QQ',L}(z)| \le
C_0\|\mathbf{q}\|_1L^d\delta_L(z_0) \zeta(z_0)^{L^d}$.
(Here we used that the term
$e^{ML^dC\delta_L(z_0)}e^{-\tau L}$ is bounded by
$e^{-\frac 12L^d\gammaL}\leq\delta_L(z_0)$
as follows from \eqref{L3-cond2}.)

Therefore, we just need to produce an
appropriate bound on~$|h(z)|$.
To that end, we note that,
since
$[z_0,z]\subset\UU_{\kappa/L}(\QQ')$
and $|z-z_0|\leq C\delta_L(z_0)$,
we have from \eqref{ration3a} and Assumption~B2
that
\begin{equation}
\bigl|{\zeta_k^{(L)}(z)}\bigr|^{L^d}
\le
\bigl|{\zeta_k^{(L)}(z_0)}\bigr|^{L^d}
e^{MCL^d\delta_L(z_0)}
\le
\bigl|{\zeta_k(z_0)}\bigr|^{L^d}
e^{MCL^d\delta_L(z_0)+L^de^{-\tau L}}
\end{equation}
whenever $k\in\QQ'$.  Since
$z_0\in \UU_{\gammaL}$, which implies
$|\zeta_k^{(L)}(z)|\leq {\zeta(z_0)}e^{-\gammaL/2}$
whenever $k\notin\{m,n\}$,
we thus have
\begin{equation}
\bigl|{\zeta_k^{(L)}(z)}\bigr|^{L^d}
\le
e^{MCL^d\delta_L(z_0)+L^de^{-\tau L}}
e^{-\frac 12\gammaL L^d}{\zeta(z_0)}^{L^d}
\end{equation}
for every~$k\in\QQ'\setminus\{m,n\}$. Using the last bound in
\eqref{L3-cond3}, we conclude that $|h(z)|$
is bounded by
$C_0\Vert\mathbf{q}\Vert_1L^d\delta_L(z_0)
\zeta(z_0)^{L^d}$.
From here
\eqref{gbd} follows.
\end{proofsect}

\subsection{Proof of Lemmas~\ref{lemma4.4} and ~\ref{lemma4.10}}
\label{sec-multilemmas}\noindent
Here we will establish the two technical lemmas on which
the proof of Theorem~\ref{T:Mph} was based.
Throughout this section we will assume that a multiple point
$\zM\in\OO$ is fixed and that $\QQ=\QQ(\zM)$.
We will also use $\tilde f$, $\tilde g$ and $\xi$ to denote the
functions defined in \twoeqref{445}{447}.

\smallskip
Lemma~\ref{lemma4.4} is an analogue of Lemma~\ref{lemmaf} from
Sect.~\ref{sec4.2}  the corresponding proofs are also analogous.
Namely, the proof of Lemma~\ref{lemmaf} was
based on the observation that either~$|f(z)|$ was itself large in
a neighborhood of $z_0$, or it was small, in which case we knew
that~$|f'(z)|$ was large. In Lemma~\ref{lemma4.4}, the function
$\tilde f(z)$ is more complicated; however, a convenient
reformulation in terms of Vandermonde matrices allows us to
conclude that at least one among its  first $(q-1)$ derivatives is
large. This is enough to push the argument through.

\begin{proofsect}{Proof of Lemma~\ref{lemma4.4}}
Abbreviating~$q=|\QQ|$ and using  $A(q)=2q^{3q(q+1)/2}q!\sqrt q  $
and the constants $K=K(\QQ)$ and $\tilde L_0$ from
Lemma~\ref{lemma4.1}  and $M$ from Assumption~B, let
$\epsilon=1/(3K)$ and $L_5\geq \tilde L_0$
be such that
\begin{equation}
\label{L5-2}
e^{ML^dR_L}\le2,
\quad
2\Vert q\Vert_1 M^q \le L^{2d}
\epsilon_L
\quad \text{ and } \quad
A(q) L^{2d-d/q}\epsilon_L\le \epsilon/\sqrt q
\end{equation}
for all $L\ge L_5$. A choice of $L_5$ yielding \eqref{L5-2} is
possible in view of \eqref{epscond}.

Choosing~$z_0\in\C$, we use~$F(z)$ to
denote the function
$F(z)=\tilde f(z) \xi(z_0)^{-L^d}$.
First, we claim that  if \eqref{Mphbd} fails to hold for
some~$L\ge L_5$, then
we have
\begin{equation}
\label{derepsbd} \bigl|F^{(\ell)}(z_0)\bigr|\le
\frac{\epsilon}{\sqrt q}L^{d\ell},\qquad \ell=0,\dots,q-1.
\end{equation}
Indeed, let us
observe that, if \eqref{Mphbd} fails to hold, then there must
exist a collection of points~$z_k$, with~$k=1,\dots,q$, such that
\begin{equation}
|z_k-z_0|=\tfrac{k}{q}R_L\quad\text{and}\quad
\bigl|F(z_k)\bigr|\le L^d\epsilon_L,
\end{equation}
for all~$k=1,\dots,q$. Further, notice that, for~$|z-z_0|\le R_L$,
 we have the bound
\begin{equation}
\label{4.37a}
\bigl|e^{v_m(z-\zM)L^d} \xi(z_0)^{-L^d}\bigr|
\le
e^{\RE(v_m(z-z_0))L^d}
\le e^{ML^dR_L}, \qquad m\in\QQ,
\end{equation}
implying~$|F^{(q)}(z)|\le 2\sum_{m\in\QQ}q_m\,|v_m|^q L^{dq}$ in
view of the first condition in \eqref{L5-2}. In particular, we
have~$|F^{(q)}(z)|R_L^{q}\le 2\Vert\mathbf{q}\Vert_1M^q L^{-d}$
for all~$z$ in the~$R_L$-neighborhood of~$z_0$. With help of the
second condition in \eqref{L5-2},  Taylor's theorem yields
\begin{equation}
\label{Tbd}
\Bigl|\sum_{\ell=0}^{q-1}\frac{F^{(\ell)}(z_0)}{\ell!}(z_k-z_0)^\ell
\Bigr|\le 2L^d\epsilon_L,\qquad k=1,\dots,q.
\end{equation}

Now we will write \eqref{Tbd} in
vector notation and use our
previous estimates on Vandermonde matrices to derive \eqref{derepsbd}.
Let~$\mathbf{x}=(x_0,x_1,\dots,x_{q-1})$ be the vector
with components
\begin{equation}
x_\ell=R_L^\ell\,\frac{F^{(\ell)}(z_0)}{\ell!}
\Bigl(\frac{z_k-z_0}{|z_k-z_0|}\Bigr)^\ell, \qquad
\ell=0,1,\dots,q-1,
\end{equation}
and let~$\N=(\N_{k,\ell})$ be the $q\times q$-matrix with
elements~$\N_{k,\ell}=|z_k-z_0|^\ell R_L^{-\ell}=(k/q)^\ell$. The
bound \eqref{Tbd} then implies that the vector $\N\mathbf{x}$ has
each component bounded by $2L^d\epsilon_L$ and so
$\Vert\N\mathbf{x}\Vert\le 2\sqrt qL^d\epsilon_L$. On the other
hand, since~$\N$ is a Vandermonde matrix, the norm of its inverse
can be estimated as in~\eqref{M-1}. Namely, using the
inequalities~$|\det\N|\ge q^{-q(q-1)/2}$ and~$\Vert\N\Vert\le q$,
we get
\begin{equation}
\Vert\N^{-1}\Vert\le\frac{\Vert\N\Vert^{q-1}}{|\det\N|} \le
q^{q(q-1)/2+q(q-1)}.
\end{equation}
But
then~$\Vert\mathbf{x}\Vert\le\Vert\N^{-1}\Vert\Vert\N\mathbf{x}\Vert
\le q^{3q(q-1)/2}2
\sqrt qL^d\epsilon_L$ implying
\begin{equation}
\label{4.29} L^{-d\ell}|F^{(\ell)}(z_0)|\le
\ell!(L^dR_L)^{-\ell}\Vert\mathbf{x}\Vert\le
A(q)L^{2d-d/q}\epsilon_L,
\end{equation}
where we used that $L^d(L^dR_L)^{-\ell}$ is maximal for
$\ell=q-1$, in which case it equals $L^{2d-d/q}$. With the help of
the last condition in \eqref{L5-2}, the claim \eqref{derepsbd}
follows for all~$L\ge L_5$.

Having proved \eqref{derepsbd}, we will now
invoke the properties of Vandermonde matrices once again
to show that \eqref{derepsbd}
contradicts Lemma~\ref{lemma4.1}.
Let~$\mathbf{y}$ be the $q$-dimensional vector with components
\begin{equation}
y_m=q_me^{\texti\phi_m(L)+v_m(z-\zM)L^d}\xi(z_0)^{-L^d}, \qquad
m\in\QQ.
\end{equation}
Let~$\BbbO=(\BbbO_{\ell,m})$ be the~$q\times q$ matrix with
matrix elements~$\BbbO_{\ell,m}=v_m^\ell$. (Here~$\ell$ takes values
between~$0$ and~$q-1$, while~$m\in\QQ$.) Recalling the definition
of~$F(z)$, the bound \eqref{derepsbd} can be rewritten as
$|[\BbbO\mathbf{y}]_\ell|\leq \epsilon/\sqrt q$.  It therefore
implies that
\begin{equation}
\label{xuma} \Vert\BbbO\mathbf{y}\Vert\le\epsilon.
\end{equation}
The matrix $\BbbO$ corresponds to the $L\to\infty$ limit of the
matrix~$\M$ in \eqref{Matrix} evaluated at~$\zM$. In particular,
since $\zM\in\CalS_{\kappa/L}(m)$ for all~$L$ and all
$m\in\QQ(\zM)$ and in view of the second bound in Assumption~B2,
the bound \eqref{InvMbd} applies to $\BbbO$ as well. Having
thus~$\Vert\BbbO^{-1}\Vert\le K$ with the constant~$K$ from
Lemma~\ref{lemma4.1}, we can conclude that
\begin{equation}
\label{108}
\Vert\mathbf{y}\Vert\le \Vert \BbbO^{-1}\Vert
\Vert\BbbO\mathbf{y}\Vert\le K\Vert\BbbO\mathbf{y}\Vert
\le K\epsilon\le\frac13
\end{equation}
using our choice~$\epsilon=1/(3K)$.
On the other hand, let~$m$ be an index for which the
maximum in the definition of~$\xi(z_0)$ is attained.
Then we have
\begin{equation}
\bigl|e^{v_m(z-\zM)L^d}\xi(z_0)^{-L^d}\bigr|=
e^{\RE(v_m(z-z_0))L^d}\ge e^{-ML^dR_L}\ge\frac12, \qquad m\in\QQ,
\end{equation}
according to the first condition in \eqref{L5-2}. Moreover,
$q_m\ge 1$ and thus $\Vert y\Vert\ge \frac12$ in contradiction to
\eqref{108}. Thus, \eqref{Mphbd} must hold for
some~$s(z_0)\in[R_L/q,R_L]$ once~$L$ exceeds~$L_5$.
\end{proofsect}

Lemma~\ref{lemma4.10} is also quite similar to the
corresponding statement (Lemma~\ref{lemmag}) from two-phase
coexistence.

\begin{proofsect}{Proof of Lemma~\ref{lemma4.10}}
We will prove the Lemma for
${A_6}=2e(C_0+3)(M+M^2)\Vert\mathbf{q}\Vert_1$,
where $M$ and $C_0$ are the constants from Assumption~B.

Let $c$ and $L_4$ be the constants from Corollary~\ref{cor-D} for
$\tilde\kappa=\kappa$. Since $\zM\in\OO$ is a multiple point with
$\QQ(\zM)=\QQ$, we clearly have that $\zM\in\UU_\epsilon(\QQ)$
whenever $\epsilon$ is small enough.  Since~$\OO$ is open, we also
have that $\D_s(\zM)\subset\OO$ whenever $s$ is sufficiently
small.  As a consequence, there is a constant $\tilde L_6=\tilde
L_6(\zM)$ such that
$\zM\in\UU_{2\kappa/L}(Q)\cap\UU_{\kappa/2L}(Q)$ and
$\D_{c/L}(\zM)\subset\OO$ whenever $L\geq \tilde L_6$.
Using Corollary~\ref{cor-D}, we reach the conclusion that
$\D_{s}(\zM)\subset\UU_{\kappa/L}(\QQ)$ whenever $L\geq
\max\{\tilde L_6,L_4\}$ and $s\leq c/L$.
We now choose $L_6\geq\max\{\tilde L_6,L_4\}$ in such a way that
\begin{equation}
\label{L6-cond}
\begin{aligned}
\rho'_L\leq c/L,
\qquad
\rho_L'\leq 2\rho_L,
\qquad
&(1+2\rho_L)e^{-\tau L}\leq
(M+M^2)\rho_L^2,
\\
4(M+M^2)\rho_L^2 L^d\leq 1,
&\quad
e^{ML^dR_L}\leq 2
\end{aligned}
\end{equation}
whenever $L\geq L_6$.  By the above conclusion and the first
condition in \eqref{L6-cond}, we then have
$\D_{\rho_L'}(\zM)\subset\UU_{\kappa/L}(\QQ)$ whenever $L\geq
L_6$.

To prove \eqref{4.50}, let us recall the definition of
$\varXi_{\QQ,L}(z)$ in formula \eqref{ZLper} from Assumption~B4.
Then we can write~$\tilde g(z)$ as
$\varXi_{\QQ,L}(z)\zeta(\zM)^{-L^d}+h(z)$, where
\begin{equation}
h(z)=\sum_{m\in\QQ}q_m\,
\biggl[\Bigr(\frac{\zeta_m^{(L)}(z)}{\zeta(\zM)}\Bigr)^{L^d}
-e^{\texti\phi_m(L)+v_m (z-\zM)L^d}\biggr].
\end{equation}
Our goal is to show that both $\varXi_{\QQ,L}(z)\zeta(\zM)^{-L^d}$
and $h(z)$ satisfy a bound of the type \eqref{4.50}.

We will begin with the bound on $h(z)$.
First we  recall the definition of $\phi_m(L)$ to
write
\begin{equation}
\Bigr(\frac{\zeta_m^{(L)}(z)}{\zeta(\zM)}\Bigr)^{L^d}=
\Bigr(\frac{\zeta_m^{(L)}(z)}{\zeta_m^{(L)}(\zM)}\Bigr)^{L^d}
\Bigr(\frac{\zeta_m^{(L)}(\zM)}{\zeta_m(\zM)}\Bigr)^{L^d}
e^{\texti\phi_m(L)}.
\end{equation}
The first term on the right-hand side is to the
leading order equal to $e^{b_m(z-\zM)L^d}$, which is
approximately equal to $e^{v_m(z-\zM)L^d}$.  To control
the difference between these two terms, and to
estimate the
deviations from the leading order behavior, we combine the bound
\eqref{der-zetamL}  with the second-order Taylor formula
and \eqref{uplogderL} to show
that, for all~$z\in\D_{\rho'_L}(\zM)$ and all $m\in\QQ$,
\begin{equation}\label{4.52a}
\bigl|\log\bigl(\zeta_m^{(L)}(z)/\zeta_m^{(L)}(\zM)\bigr)
-v_m(z-\zM)\bigr|
\le
e^{-\tau L}\rho_L'+
\frac 12(M+M^2)(\rho'_L)^2,
\end{equation}
where we have chosen the principal branch of the complex logarithm.
Combining this estimate with the second and third condition in
\eqref{L6-cond} and the bound \eqref{zetamL}
from Assumption~B2, we get
\begin{equation}
\label{4.52aa}
\bigl|L^d\log\bigl(\zeta_m^{(L)}(z)/\zeta^{(L)}(\zM)\bigr)
-v_m(z-\zM)L^d-\texti \phi_m(L)\bigr|
\le
3(M+M^2)\rho_L^2L^d.
\end{equation}
Using the fourth condition in \eqref{L6-cond}
and the fact that $|e^w-1|\leq e|w|$ whenever $|w|\leq 1$,
we get
\begin{equation}
\bigl|h(z)\bigr|\le
3e(M+M^2)\Vert\mathbf{q}\Vert_1\,L^d\rho_L^2{\xi}(z)^{L^d}\,.
\end{equation}
Now
${\xi}(z)^{L^d}\le{\xi}(z_0)^{L^d}e^{ML^d R_L}\le
2{\xi}(z_0)^{L^d}$ by the fifth condition in
\eqref{L6-cond}, so we
finally have the bound $|h(z)|\le
A{\xi}(z_0)^{L^d}L^d\rho_L^2$, with~$A$ given
by~$A=6e(M+M^2)\Vert\mathbf{q}\Vert_1$.

It remains to prove a corresponding bound for~$\varXi_{\QQ,L}(z)
\zeta(\zM)^{-L^d}$. First we recall our previous observation
that~$\D_{\rho'_L}(\zM)\subset\UU_{\kappa/L}(\QQ)$, so we have
Assumption~B4 at our disposal.
Then \eqref{error} yields
\begin{equation}
\label{4.54a}
\bigl|\varXi_{\QQ,L}(z) \zeta(\zM)^{-L^d}\bigr|\le
C_0L^d\Vert\mathbf{q}\Vert_1 e^{-\tau L}
\Bigl[\frac{\zeta(z)}{\zeta(\zM)}\Bigr]^{L^d}, \qquad
z\in\D_{\rho'_L}(\zM).
\end{equation}
Also, by the definition of
$\UU_{\kappa/L}(\QQ)$, we have that
$\zeta(z)=\min_{m\in\QQ}|\zeta_m(z)|$
whenever $z\in\D_{\rho'_L}(\zM)$.
For $z\in\D_{\rho'_L}(\zM)$, we can therefore
find a index $m\in\QQ$ such that~$|\zeta_m(z)|=\zeta(z)$.
With the help of \eqref{ratio2} and the bound \eqref{zetamL}
from Assumption~B, we thus get
\begin{equation}
\Bigl[\frac{\zeta(z)}{\zeta(\zM)}\Bigr]^{L^d}
\leq
\Bigr|\frac{\zeta_m(z_0)}{\zeta(\zM)}\Bigr|^{L^d}
\Bigr|\frac{\zeta_m(z)}{\zeta_m(z_0)}\Bigr|^{L^d}
\le
\Bigr|\frac{\zeta_m^{(L)}(z_0)}{\zeta(\zM)}\Bigr|^{L^d}
e^{MR_LL^d}e^{L^de^{-\tau}}.
\end{equation}
Combined with the estimate \eqref{4.52aa} for $z=z_0$,
and the last three conditions in \eqref{L6-cond},
this gives
\begin{equation}
\Bigl[\frac{\zeta(z)}{\zeta(\zM)}\Bigr]^{L^d}
\leq e^{MR_LL^d}e^{L^de^{-\tau}}
e^{3(M+M^2)\rho_L^2L^d}
\xi(z_0)^{L^d}
\leq 2e\xi(z_0)^{L^d}.
\end{equation}
Using the third condition in \eqref{L6-cond} one last time, we can
bound the right-hand side \eqref{4.54a}
by $2eC_0\Vert\mathbf{q}\Vert_1 (M+M^2)L^d\rho_L^2{\xi}(z_0)^{L^d}$.
Combined with the above bound on $|h(z)|$, this finally proves
\eqref{4.50}.
\end{proofsect}


\begin{acknowledgement}
The authors would like to thank A.~Sokal for his interest in
this work, R.~Shrock for bringing the
references~\cite{Chang-Shrock1,Chang-Shrock2,Matveev-Shrock1,%
Matveev-Shrock2,Shrock,Shrock-Tsai} to our attention,  and
A.~van Enter for pointing out the
reference~\cite{Saarloos-Kurtze}. M.B.~and R.K.~would like to
acknowledge the hospitality of Microsoft Research in Redmond,
where large parts of this work were carried out. The research of
R.K.~was partly supported by the grants GA\v{C}R~201/00/1149,
201/03/0478, and MSM~110000001. The authors also wish to thank 
anonymous referees for helpful suggestions concerning the presentation.
\end{acknowledgement}

\end{document}